\documentclass[footinbib,a4paper,aps,prx,superscriptaddress,reprint,twocolumn,preprintnumbers,amsmath,amssymb,nobalancelastpage,10pt]{revtex4-1}
\usepackage{amsmath}
\usepackage{verbatim}
\usepackage{graphicx}
\usepackage{color}
\usepackage{tikz}
\usepackage{subfigure}
\usepackage{float}
\usepackage{bigints}

\usepackage{mathptmx}
\usepackage{amssymb}
\usepackage{amsmath}
\usepackage{amsfonts}
\usepackage{array}

\usepackage{bm}
\usepackage{xcolor}
\graphicspath{{figures/}}

\usepackage{tikz-feynman}
\tikzfeynmanset{compat=1.0.0}

\usepackage{braket}

\definecolor{mygrey}{gray}{0.35}
\definecolor{myblue}{rgb}{0.2,0.2,0.8}
\definecolor{myzard}{cmyk}{0,0,0.05,0}
\definecolor{mywhite}{rgb}{1,1,1}
\definecolor{myred}{rgb}{0.644679,0.000000,0.065473}

\DeclareMathAlphabet{\mathpzc}{OT1}{pzc}{m}{it}

\usepackage[colorlinks=true,citecolor=myblue,linkcolor=myred]{hyperref}

 \def\ee{\mathord{\rm e}}
 
 \def\ii{\mathord{\rm i}}
\def\min{\mathord{\rm min}}

\def\half{\textstyle\frac{1}{2}}
\def\third{\textstyle\frac{1}{3}}

\renewcommand{\ii}{{\rm i}}
\renewcommand{\ee}{{\rm e}}
\def\beq{\begin{equation}}
\def\eeq{\end{equation}}
\def\barray{\begin{eqnarray}}
\def\earray{\end{eqnarray}}

\begin{document}

\title{Long-range Ising interactions mediated by $\lambda\phi^4$  fields:  \\
probing the renormalisation of sound  in crystals of trapped ions}

\author{G. Mart\'in-V\'azquez}
\affiliation{Departamento de F\'{i}sica Te\'{o}rica, Universidad Complutense, 28040 Madrid, Spain}
\affiliation{Nano and Molecular Systems Research Unit, University of Oulu, 90014 Oulu, Finland}
\author{G. Aarts}
\affiliation{Department of Physics, Swansea University, Singleton Campus, SA2 8PP Swansea, United Kingdom}
\affiliation{European Centre for Theoretical Studies in Nuclear Physics and Related Areas (ECT*) and Fondazione Bruno Kessler Strada delle Tabarelle 286, 38123 Villazzano (TN), Italy}
\author{M. M\"{u}ller}
\affiliation{Institute for Theoretical Nanoelectronics (PGI-2), Forschungszentrum J\"{u}lich, 52428 J\"{u}lich, Germany}
\affiliation{JARA-Institute for Quantum Information, RWTH Aachen University, 52056 Aachen, Germany}
\author{A. Bermudez}
\affiliation{Departamento de F\'{i}sica Te\'{o}rica, Universidad Complutense, 28040 Madrid, Spain}

\begin{abstract}
The generating functional of a self-interacting scalar quantum field theory (QFT), which contains all the relevant information about  real-time dynamics and scattering experiments, can be mapped onto a collection of multipartite-entangled two-level sensors via an interferometric protocol that exploits a specific set of source functions. Although one typically focuses on impulsive delta-like sources, as these give direct access to $n$-point Feynman propagators, we show in this work that using always-on harmonic sources can simplify substantially the sensing protocol.  In a specific regime, the effective real-time  dynamics of the quantum sensors can be described by a  quantum Ising model with long-range couplings, the range and strength of which contains all the relevant information about the renormalisation of the QFT, which can now be extracted in the absence of multi-partite entanglement. We present a detailed analysis of how this sensing protocol can be relevant to characterise the long-wavelength QFT that describes quantised sound waves of  trapped-ion crystals in the vicinity of a structural phase transition, opening a new route to characterise the associated renormalisation of sound.
 
\end{abstract}

\maketitle

\setcounter{tocdepth}{2}
\begingroup
\hypersetup{linkcolor=black}
\tableofcontents
\endgroup

  \section{\bf Introduction}

 Large-scale fault-tolerant quantum computers have the potential to solve  relevant computational problems  in ways that  surpass the capabilities of any classical device~\cite{nielsen_chuang_2010}. To achieve this  long-term goal, these  devices must process information by taking explicit advantage of the laws of quantum mechanics, while they also correct for the errors that occur during the computation due to imperfections and  noise. In this way, one  actively battles   a possible accumulation of the errors~\cite{RevModPhys.87.307}, as has already been demonstrated in various small-scale processors~\cite{PhysRevLett.81.2152,Chiaverini2004,Reed2012,Schindler1059,Nigg302,Kelly2015,Linkee1701074,Stricker2020,egan2021faulttolerant,Erhard2021}, including atomic, molecular and optical (AMO)  technologies such as trapped  ions~\cite{primer_qc_ions,Schindler_2013,doi:10.1063/1.5088164}.

Recent progress with noisy intermediate-scale quantum technologies~\cite{Preskill2018quantumcomputingin} has shown  that, even if these errors are not actively corrected for, and can thus   accumulate, the level of the overall noise is sufficiently low  that current  noisy devices can already solve  certain computational tasks that no classical computer could solve  in any feasible amount of time~\cite{Arute2019,Zhong1460}. This has allowed to demonstrate  the so-called  quantum advantage/supremacy~\cite{Harrow2017}. 
In  theoretical physics, some of the most complicated problems appear in the study of quantum systems with a large, sometimes even infinite, number of coupled degrees of freedom. These systems are typically formulated in terms  of quantum field theories (QFTs)~\cite{Peskin:1995ev,fradkin_2013}, some of which pose problems whose solution  has remained elusive for decades, in spite of the availability of large computational  resources.  With some notable exceptions~\cite{Jordan1130}, determining their specific complexity, or proving rigorously that  no classical computer will ever be able to solve them in a feasible amount of time, are  very delicate matters. In any case, given the  limitations of current computational approaches in QFTs, e.g. real time evolution or finite fermion densities,  and their importance in understanding Nature at its most fundamental level~\cite{doi:10.1080/00107510110063843}, any possible progress using alternative methods can be of great value. In this context, we find that the prospects of using  noisy intermediate-scale quantum technologies to address open questions in QFTs is fascinating~\cite{Feynman_1982,Cirac2012}. We also find particularly enticing that these quantum technologies may offer novel ways to experimentally realise and probe paradigmatic QFTs which, although not  finding counterparts in Nature, e.g. reduced dimensionality, contain key fundamental aspects that shape our understanding of it. We will emphasise below the results of the present work in this respect.

Although the initial focus of  such, so-called, quantum simulators (QSs)~\cite{Feynman_1982,Lloyd1073} has been on condensed-matter models, such as the bosonic~\cite{PhysRevLett.81.3108,Greiner2002} and fermionic~\cite{PhysRevLett.89.220407,Jordens2008,Schneider1520} Hubbard models with cold atoms~\cite{Bloch2012,JAKSCH200552}, or models of interacting spins~\cite{PhysRevLett.92.207901,Friedenauer2008,Islam2011,Lanyon57}
with trapped ions~\cite{Blatt2012,monroe2020programmable}; QSs may also find numerous applications for high-energy physics~\cite{Banuls2020}.
Trapped-ion systems have already been used to explore the predictions of the Dirac equation~\cite{PhysRevLett.98.253005,Gerritsma2010,PhysRevLett.106.060503}. Similarly,  cold Fermi gases in hexagonal optical lattices~\cite{PhysRevLett.98.260402,Tarruell2012} serve to  realise experimentally  the  fermion doubling in lattice discretisations of Dirac QFTs~\cite{gattringer_lang_2010}. One can also explore other discretisations where topology plays a key role~\cite{PhysRevLett.105.190404}, which connects to  a very fruitful line of research  in cold-atom QSs~\cite{Goldman2016}. The characteristic level of  microscopic control of these QSs allows to explore  QFTs with additional interactions, the strength of which can be tuned independently, connecting to paradigmatic  four-Fermi models~\cite{PhysRevLett.105.190403,PhysRevX.7.031057,BERMUDEZ2018149,ziegler2020correlated} or self-interacting scalar fields~\cite{Jordan1130,PhysRevX.7.041012,PhysRevA.99.052335}. In connection to our most fundamental theories of Nature~\cite{PhysRev.96.191}, QSs have also been proposed to target gauge theories~\cite{PhysRevA.73.022328,PhysRevLett.95.040402} with  both Abelian~\cite{Weimer2010} and 
non-Abelian~\cite{Tagliacozzo2013} groups. Proposals to include fermionic matter coupled to such gauge fields~\cite{PhysRevLett.109.125302,PhysRevLett.109.175302}, together with initial experimental progress along these lines~\cite{Martinez2016,Gorg2019,Schweizer2019,Mil1128,PhysRevX.10.021041,Yang2020,Kokail2019,2102.08920,PhysRevA.98.032331,PhysRevLett.121.170501}, have made it a very active and rapidly-evolving research direction~\cite{PhysRevD.100.034518,PhysRevD.102.094515} (see~\cite{Banuls2020} for a recent review) with interesting connections to topological phases of matter and topological order~\cite{PhysRevD.99.014503,PhysRevB.100.115152,PhysRevX.10.041007,borla2021quantum}.

In this article, we focus on one of the aspects emphasised above: the possibility of  realising  paradigmatic QFTs that can be probed  in  unprecedented ways. In the context of high-energy physics, one typically probes  these QFTs via scattering experiments where the particles, described as the fundamental excitations of the fields,  collide against  each other in large accelerators. In the context of AMO quantum technologies, there are  other techniques to probe a quantum field, which can be traced back to the origins of quantum optics and  the theory of photo-detection of  the electromagnetic  field~\cite{PhysRev.130.2529}. Historically,  there have been various models of the  photo-detector, such as  harmonic-oscillator probes~\cite{PhysRev.168.1896},  an ensemble of ancillary atoms with a single groundstate and a continuum of   excited levels~\cite{PhysRev.179.368}, or a simple two-level atom or qubit~\cite{PhysRevA.52.1525}, all of which  get excited by the absorption of photons from the electromagnetic field. By placing these probes at different  spatial locations, and detecting delayed photon coincidences, one gets information about  the retarded correlation functions of this gauge field. These correlation functions play a key role in the quantum theory of optical coherence~\cite{PhysRev.130.2529}, and form the basis of numerous groundbreaking experiments~\cite{BROWN1956,HANBURYBROWN1956} that have shaped the progress of AMO quantum technologies~\cite{PhysRevD.14.870}. Similar ancillary probes are referred to as particle detectors in  QFTs on curved spacetimes ~\cite{birrell_davies_1999}, such as a  quantum particle held in a box that moves with constant acceleration in the background of a  scalar field~\cite{PhysRevD.14.870}. The excitation rate of this accelerated particle, which is non-zero  even when the field is in the vacuum,  probes the so-called Unruh effect~\cite{RevModPhys.80.787}. Modified versions of this particle detector, such as a two-level atom~\cite{dewitt_2010,PhysRevD.29.1047} or a harmonic oscillator~\cite{PhysRevD.46.3271}, are commonly referred to as Unruh-DeWitt detectors which, if correctly switched on/off, can give information about the so-called Wightman correlation function of the scalar field~\cite{PhysRevD.98.105011}.  

In the context of contemporary applications of quantum technologies,  in particular quantum sensing~\cite{RevModPhys.89.035002}, these photon and particle detectors can be categorised as Rabi-type quantum sensors, as they rely on the excitation of the  probe to gather information about the quantum field.
There is, however,  a different kind of quantum sensor where, rather than focusing on its excitation probability,  one monitors its coherence  via Ramsey interferometry~\cite{PhysRev.78.695}. Ramsey-type sensors are 
 actually used in high-precision measurements, and exploited as frequency standards in the so-called  atomic clocks~\cite{RevModPhys.87.637}, which include both trapped-ion and cold-atom  technologies. In the context of QSs of condensed-matter models, Ramsey sensors have also been considered as probes for  quantum many-body properties, both in equilibrium~\cite{Bruderer_2006,PhysRevLett.121.130403} to probe equal-time correlation functions~\cite{PhysRevA.93.043612,PhysRevA053634}, and out of equilibrium~\cite{doi:10.1063/1.531672,PhysRevLett.96.106801,PhysRevLett.111.040601,PhysRevLett.110.230601,PhysRevLett.110.230602}, where noise and fluctuations can give crucial information about transport phenomena.

In the context of  QSs of high-energy physics, to the best of our knowledge, Ramsey-type probes have been much less studied. In~\cite{Zohar_2013}, these probes were used to characterise the Wilson loop operator of a  lattice gauge theory.  In~\cite{PhysRevX.7.041012}, Ramsey-type probes coupled to self-interacting  Klein-Gordon fields have been explored for trapped-ion QSs.
Although the real Klein-Gordon field~\cite{Klein1926,Gordon1926}  is an archetype QFT~\cite{Peskin:1995ev}, and the inclusion of self-interactions, e.g.  $\lambda\phi^4$ terms, shapes our understanding of crucial  aspects of QFTs such as  renormalisation~\cite{WILSON197475}; there is no fundamental particle in Nature  described by this real scalar field.   It is thus quite interesting that QSs have the potential to realise    this paradigmatic QFT in real experiments and, actually, explore different effective dimensionalities. Moreover, as put forth in~\cite{PhysRevX.7.041012}, one can devise a  Ramsey-type protocol to probe this field theory, turning key QFT concepts  such as the generating functional~\cite{Peskin:1995ev} into experimentally  measurable observables. In that work, we showed how one can devise an interferometric protocol that maps the information of such a generating functional onto a  multi-partite entangled state of ancillary qubits~\cite{PhysRevX.7.041012}, which are coupled to the scalar field via a generalisation of the so-called Schwinger sources~\cite{Schwinger452,Schwinger455}. It is interesting that, in contrast to the previous probing protocols that  infer either equal-time, retarded, or Wightman correlation functions, by switching  these sources on/off appropriately in the impulsive (i.e. anti-adiabatic) regime~\cite{PhysRevX.7.041012}, these entangled Ramsey sensors give access to any  $n$-point Feynman propagator. Such time-ordered functions form the basis of  current approaches to QFTs, and measuring them  is thus a  way of extracting all relevant  information about equilibrium or real-time dynamics of the QFT,  providing an alternative to scattering experiments.

In this article, we modify this sensing protocol and dispense with the requirement of initialising the qubits in a   multi-partite entangled state. We show that, by abandoning the impulsive regime to focus instead on harmonic Ising-Schwinger sources, the  dynamics can be described in terms of   inter-qubit Ising-type unitaries, which are  connected to the underlying generating functional evaluated at these specific   source functions. In a   regime in which these  sources are off-resonant with the massive $\lambda\phi^4$ QFT, we show that the qubits evolve under an effective quantum Ising model with long-range couplings that are fully controlled by a dimensionally-reduced Euclidean propagator of the self-interacting  Klein-Gordon field. Since this propagator contains relevant information about various aspects of  the renormalisation of the scalar field, our results show that these matters   can be directly probed by monitoring  the dynamics of the ancillary qubits which, we remark again, need not be prepared in a multi-partite entangled state. Additionally, in contrast to the impulsive protocol~\cite{PhysRevX.7.041012}, the sources need not be switched on/off in various combined experiments, but are simply adiabatically switched on during the whole probing sequence. 

We show that these results naturally connect to a long-wavelength description of the aforementioned QSs of spin models in trapped ions~\cite{PhysRevLett.92.207901,Friedenauer2008,Islam2011,monroe2020programmable}, if one works in the vicinity of a structural phase transition~\cite{PhysRevB.77.064111}. As discussed in detail below, the role of the real Klein-Gordon field is played by the transverse sound waves of the  ions which, for sufficiently low temperatures, describe the vibrations of the ions around the equilibrium positions of a  linear Coulomb crystal. In contrast to sound waves in solid-state materials, a detailed long-wavelength theory shows that these vibrational excitations, the so-called phonons, are massive particles and thus move inside an effective light-cone determined by the transverse  speed of sound, which would be fully tunable in an experiment. As one modifies the ratio of the trap frequencies, approaching the aforementioned structural phase transition, the bare mass of the effective Klein-Gordon field reduces, whereas the role of quartic non-linearities becomes more important. We show that,  by introducing additional lasers,  one can couple  the transverse motion of the ions to the internal atomic structure, which can be described in terms of qubits,  such that the  low-energy description  of this system is a scalar-sigma QFT of a self-interacting Klein-Gordon field coupled to a constrained sigma field that represents the underlying  qubits. By exploiting the connection to high-energy physics, and in particular to the concept of generating functionals and the renormalisation group, we show that the self-interacting scalar field mediates long-range Ising interactions between the qubits due to the exchange of virtual phonons which, by virtue of the additional    quartic self-interactions, can now scatter along the way. This allows us to predict that, in comparison to the standard phonon-mediated Ising interactions in harmonic trapped-ion crystals~\cite{PhysRevLett.92.207901,Friedenauer2008,Islam2011,monroe2020programmable}, by approaching the structural phase transition, the intensity and range of the Ising couplings get contributions from these  scattering events of the virtually-excited bosons, and can be neatly described in terms of the mass and wavefunction renormalisation of the scalar field.  In essence, the role of the quartic interactions leads to a renormalisation of transverse sound in the ion crystal, which can be probed by monitoring the real-time dynamics of an effective long-range Ising model. We believe that the results presented in this work are at reach of several current trapped-ion experiments.

Since the topic of this article is multidisciplinary in nature, we have made a special effort to present  our results in a way that is accessible to  two communities, those working in lattice field theories for high-energy physics and those more familiarised with quantum optics and AMO quantum technologies. We have thus included additional material in appendixes to make this work self-contained, and to present concepts in a way that will hopefully be accessible for both communities.

\section{\bf Klein-Gordon scalar  fields coupled to  $\mathbb{Z}_2$  fields}
\label{sec:KG_ising}
  
  In this section, we use canonical-quantisation techniques  to describe two different  sensing protocols for the generating functional of the massive Klein-Gordon field $\mathsf{Z}_0[J]$. In Appendix~\ref{sec:KG_Z_0}, the reader can find details about the derivation of  $\mathsf{Z}_0[J]$ for this  field theory, which we now build upon  to present an alternative take on the quantum sensing protocol  introduced in~\cite{PhysRevX.7.041012}. As described in Sec.~\ref{sec:Z0_impulsive}, the exact form of the unitary evolution operation sheds light on the need of  multi-partite entangled probes to extract   $\mathsf{Z}_0[J]$ within this scheme. This discussion also clarifies, as described in Sec.~\ref{sec:KG_Ising_spins}, that the mapping of $\mathsf{Z}_0[J]$ onto the probes relies on interactions mediated by the virtual exchange of Klein-Gordon bosons, which in turn suggests a simpler sensing protocol based  on the use of off-resonant harmonic sources, as introduced in Sec.~\ref{sec:KG_Ising_spins}. By measuring the dynamics of a pair of unentangled probes to infer the characteristics of an effective long-range Ising Hamiltonian, one can reconstruct the 2-point propagator and with it, also the free generating functional $\mathsf{Z}_0[J]$ in a simpler manner.

 \subsection{Impulsive  sensors  of the generating functional}
\label{sec:Z0_impulsive}

\begin{figure*}[t]
 \begin{centering}
  \includegraphics[width=2\columnwidth]{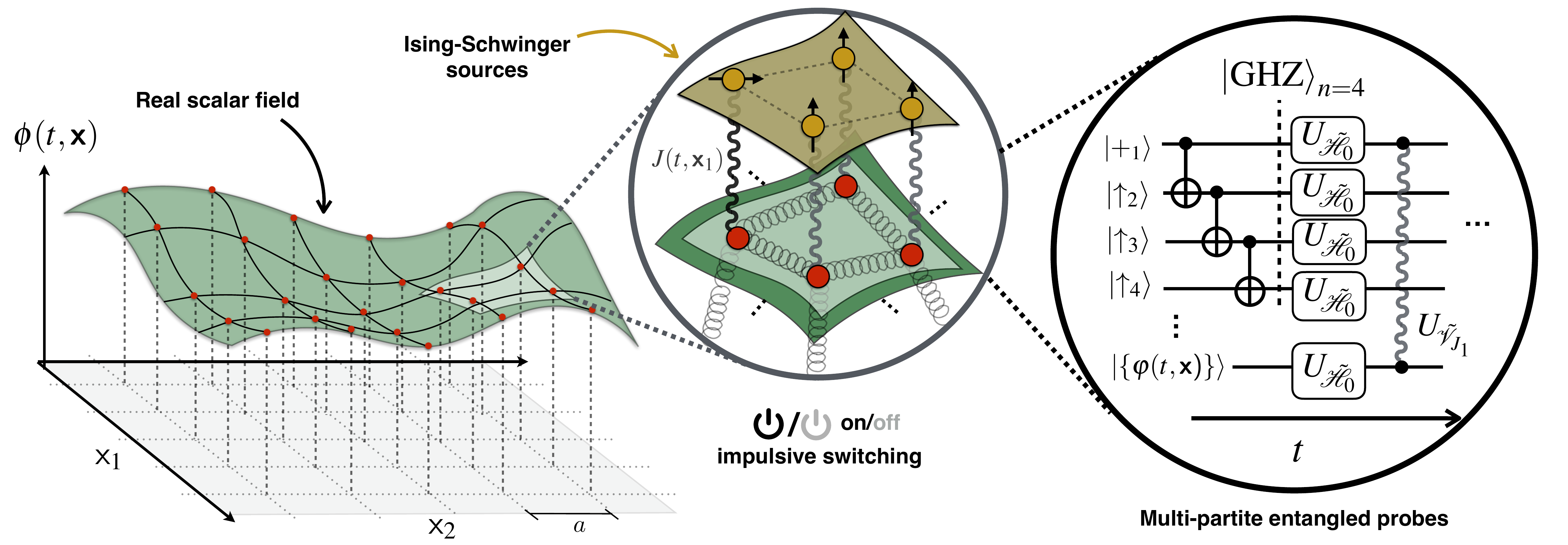}\\
  \caption{\label{Fig:scheme_impulsive_sensor} {\bf Scheme of the impulsive sensors for the generating functional:} A real scalar field $\phi(t,\textsf{\textbf{x}})$ is coupled locally by Ising-Schwinger sources of strength  $J(t,\textsf{\textbf{x}})$ to an ancillary Ising field which, in principle, also resides at every spacetime point $\boldsymbol{\sigma}(t,\textsf{\textbf{x}})$. In practice, the QFT will be regularised on a lattice of spacing $a$, and the discretised Ising field  need only be placed at a reduced set of spatial locations, which can be interpreted as a multi-partite sensor. By switching on/off  the Ising-Schwinger sources are impulsively, as discussed in the text,  one  maps the information of the generating functional of the field $\mathsf{Z}[J]$ onto the Ising spins (see the first inset). The full time evolution, including the initialisation of the Ising spins in a multi-partite entangled state, can be represented as a quantum circuit (see the second inset). }
\end{centering}
\end{figure*}

  In the framework of  canonical quantisation, the real scalar field evolving under a  Klein-Gordon  equation~\cite{Klein1926,Gordon1926} can be described by the following  Hamiltonian
\beq
\label{eq:KG_field}
H_0\!=\!\!\int\!\!{\rm d}^d\!\mathsf{x}\mathcal{H}_0,\hspace{1.2ex}\mathcal{H}_0=\half\pi^2(x)+ \half(\boldsymbol{\nabla}\phi(x))^2+\half{m_0^2}\phi^2(x),
\eeq 
where $\phi(x),\pi(x)$ are the field operator and its conjugate  momentum, both defined on a $D=d+1$ Minkowski spacetime $x=(t,\textsf{\textbf{x}})$, and satisfying equal-time  bosonic commutation relations    $[\phi(t,\textsf{\textbf{x}}),\pi(t,\textsf{\textbf{x}}')]=\ii\delta^d\!(\textsf{\textbf{x}}-\textsf{\textbf{x}}')$. Here, $m_0$ stands for the bare mass of the scalar bosons, $\boldsymbol{\nabla}$  contains only spatial derivatives, and we use natural units $\hbar=c=1$.

The generating functional $\mathsf{Z}_0[J]$ depends on the so-called Schwinger sources $J(x)$~\cite{Schwinger452,Schwinger455}, which are classical fields that introduce local perturbations  in the original QFT~\eqref{eq:KG_field}, modifying the number of excitations of the field, and thus creating or annihilating scalar bosons (see App.~\ref{sec:KG_Z_0}). 
We consider the promotion of these classical  sources  to quantum-mechanical  degrees of freedom~\cite{PhysRevX.7.041012}, by introducing an Ising field $\boldsymbol{\sigma}(x)$  expressed in terms of the identity and  Pauli matrices \beq
\sigma^0=\mathbb{I}_2,\hspace{1ex}\sigma^1=X,\hspace{1ex}\sigma^2=Y,\hspace{1ex}\sigma^3=Z,
\eeq
 in the corresponding basis $\{\ket{0_{\textsf{\textbf{x}}}}=\ket{\uparrow_{\textsf{\textbf{x}}}}, \ket{1_{\textsf{\textbf{x}}}}=\ket{\downarrow_{\textsf{\textbf{x}}}}\}$.
 With this notation, the Ising fields can be interpreted as qubits~\cite{nielsen_chuang_2010} that will serve as two-level quantum sensors located at different  spacetime points (see Fig.~\ref{Fig:scheme_impulsive_sensor}). Accordingly, the Hilbert space becomes a tensor product $\mathpzc{H}=\mathpzc{H}_{\phi}\otimes\mathpzc{H}_{\sigma}$, and one can exploit the coupling between the scalar and Ising fields to define a measurement protocol for $\mathsf{Z}_0[J]$. This becomes particularly interesting when including interactions, as the same measurement scheme  applies, but the generating functional  now contains non-trivial information about the interacting fields, such as   renormalisation and the underlying fixed points~\cite{WILSON197475,hollowood_2013}.

As discussed in~\cite{PhysRevX.7.041012}, the Schwinger sources are promoted to $J(x)\to \boldsymbol{J}(x)\cdot\boldsymbol{\sigma}(x)$, which becomes  interesting for $\boldsymbol{J}_\alpha(x)=J(x)(\delta_{\alpha,0}-\delta_{\alpha,3})/2$, as the scalar field  then couples to 
\beq
\boldsymbol{J}(x)\cdot\boldsymbol{\sigma}(x)=J(x)P(x),
\eeq
 where we have introduced the  orthogonal Ising projector  
 \beq
 P(x)=\ket{\downarrow_{\textsf{\textbf{x}}}}\!\!\bra{\downarrow_{\textsf{\textbf{x}}}}=\half(1-Z(x)).
 \eeq
 The new Hamiltonian density  $\tilde{\mathcal{H}}_0+\tilde{\mathcal{V
}_J}$ with these Ising-Schwinger sources is
\beq
\label{eq:Ising_source_coupling}
\tilde{\mathcal{H}}_0=\mathcal{H}_0+\delta\epsilon(x) Q(x),\hspace{1.5ex} \tilde{\mathcal{V
}_J}=-J(x)\phi(x)P(x),
\eeq
where we have introduced the remaining orthogonal projector 
\beq
Q(x)=\ket{\uparrow_{\textsf{\textbf{x}}}}\!\!\bra{\uparrow_{\textsf{\textbf{x}}}}=\half(1+Z(x)),
\eeq
 and $\delta\epsilon(x)$ is the energy density  for the Ising field. Since $[P(x),Q(y)]=0$, it is straightforward to see that the time-evolution operator $U(t_{\rm f},t_0)=U_{\!\!\tilde{\mathcal{H}}_0}U_{\tilde{\mathcal{V}}_J}$ can be expressed in terms of  two unitaries  \beq
 \label{eq:unitaries_spin}
  U_{\tilde{\mathcal{H}}_0}=\ee^{-\ii\int\!{\rm d}^Dx {\mathcal{\tilde{H}}}_0},\hspace{2ex}
 U_{\tilde{\mathcal{V}}_J}=\!\mathsf{T}\!\left\{\ee^{+\ii\int\!{\rm d}^Dx J(x)\phi_H(x)P(x)}\!\right\},
 \eeq
 in complete analogy to the situation for the standard Schwinger sources in Eq.~\eqref{eq:unitaries}.  
 Here, $\mathsf{T}\{\cdot\}$ is the time-ordering operator,  the scalar  fields $\phi_H(x)=(U_{\!\!\tilde{\mathcal{H}}_0})^\dagger \phi(x)U_{ \!\!\tilde{\mathcal{H}}_0}$ evolve in the Heisenberg picture with respect to the unsourced  Klein-Gordon Hamiltonian~\eqref{eq:KG_field}, whereas the Ising projectors do not change since $U_{\!\!\tilde{\mathcal{H}}_0}P(x)(U_{\!\!\tilde{\mathcal{H}}_0})^\dagger=P(x)$. By using the Magnus expansion~\cite{https://doi.org/10.1002/cpa.3160070404,Blanes_2010} described in Eq.~\eqref{eq:propagator} of the appendix, and carrying out the  steps in  analogy to the Klein-Gordon QFT with classical sources of Appendix~\ref{sec:KG_Z_0}, we find
 \beq
 \label{eq:int_U_spin}
 U_{\tilde{\mathcal{V}}_J}\!=U_N\hspace{0.2ex}\ee^{-\frac{1}{2}\!\!\!\bigintssss\!\!{\rm d}^{D}x_1\bigintssss\!\!{\rm d}^{D}x_2P(x_1)J(x_1)\Delta_{m_0}\!(x_1-x_2)J(x_2)P(x_2)}\!\!.
 \eeq
 Here, we have introduced the normal-ordered unitary 
 \beq
 \label{eq:normal_ordered_part}
 U_N=:\ee^{-\ii\!\!\bigintssss\!\!{\rm d}^Dx J(x)\phi_{\rm H}(x)P(x)}\!:,
 \eeq
 and the  time-ordered Feynman propagator of the scalar field
  \beq
  \label{eq:feynman_propagator}
 \Delta_{m_0}\!(x)=\int_k \tilde{\Delta}_{m_0}\!(k)\ee^{-\ii kx},\hspace{2ex} \tilde{\Delta}_{m_0}\!(k)=\frac{\ii}{k^2-m_0^2+\ii\epsilon},
 \eeq
 where     $k=(\omega,\textsf{\textbf{k}})$, and we define $\int_k=\int_{\phantom{k}}\!\!\! {\rm d}^Dk/(2\pi)^D$, and $\epsilon\to0^+$.
The structure of  equation~\eqref{eq:int_U_spin} already suggests that it may be possible to map all the  information of the generating functional~\eqref{eq:free_Z} of the massive Klein-Gordon field
   \beq
    \label{eq:free_Z}
  \mathsf{Z}_0[J]=\ee^{-\frac{1}{2}\!\!\bigintssss\!\!{\rm d}^Dx_1\!\!\bigintssss\!\!{\rm d}^Dx_2\hspace{0.2ex}J(x_1)\Delta_{m_0}\!(x_1-x_2)J(x_2)},
  \eeq
  onto the dynamics of the qubits by an appropriate protocol. Note that $\mathsf{Z}_{0}[0]=1$, so we are referring to the free normalised generating functional, or the full one $\mathsf{Z}[0]=1$,  in this article.

The key idea underlying the mapping of the generating-functional  onto the Ising probes,  presented in~\cite{PhysRevX.7.041012} using a different approach, is that the full time-evolution $U(t_{\rm f},t_0)=U_{\tilde{\mathcal{H}}_0}U_{\tilde{\mathcal{V}}_J}$ can map this  information  into the amplitudes of a specific initial state.  One starts by preparing  a GHZ state for the qubits through the  quantum circuit displayed in   Fig.~\ref{Fig:scheme_impulsive_sensor}, which applies concatenated  CNOT gates to a product input state, and should be extended to all the locations of the Ising spins. After these gates~\cite{nielsen_chuang_2010},  and right before the Ising-Schwinger sources are  switched on, the state is $\ket{\psi(t_0)}=\ket{0}\otimes\ket{\rm GHZ}$, where $\ket{0}$ is the Klein-Gordon vacuum, and $\ket{\rm GHZ}=(\Pi_{\textsf{\textbf{x}}}\ket{\uparrow_{\textsf{\textbf{x}}}}+\Pi_{\textsf{\textbf{x}}}\ket{\downarrow_{\textsf{\textbf{x}}}})\sqrt{2}$ is a multi-partite entangled state. We  now  let the scalar and Ising fields  couple through the Ising-Schwinger sources, and obtain  the time-evolved state $\ket{\psi(t_{\rm f})}=U_{\tilde{\mathcal{H}}_0}U_{\tilde{\mathcal{V}}_J}\ket{\psi(t_0)}$, where the unitaries are described in Eq.~\eqref{eq:unitaries_spin}. Due to trivial action of the normal-ordered part~\eqref{eq:normal_ordered_part} on  the Klein-Gordon vacuum, and the  action of the Ising projectors on the corresponding qubit states,  only the second part of the entangled state   evolves non-trivially, yielding
\beq
\label{eq:parity_oscillations}
\ket{\psi(t_{\rm f})}=\frac{1}{\sqrt{2}}\ket{0}\otimes\left(\!\Pi_{\textsf{\textbf{x}}}\ket{\uparrow_{\textsf{\textbf{x}}}}+\mathsf{Z}_0[J]\ee^{\ii\int{\rm d}^Dx\delta\epsilon(x)}\Pi_{\textsf{\textbf{x}}}\ket{\downarrow_{\textsf{\textbf{x}}}}\!\!\right).
\eeq 
Here, we readily find that the free generating functional~\eqref{eq:free_Z} appears in the relative amplitude of the  time-evolved spin state, after  
neglecting an irrelevant global phase factor that oscillates with the  zero-point energy of the scalar  field. 
The signal can be extracted by measuring the global parities  
  \beq
  \label{eq:parity_generating_functional}
  \begin{split}
  \mathsf{P}_1[J]=\bra{\psi(t_{\rm f})}\Pi_{\textsf{\textbf{x}}}X\!(\textsf{\textbf{x}})\ket{\psi(t_{\rm f})}={\rm Re}\{\mathsf{Z}_0[J]\ee^{\ii\int{\rm d}^Dx\delta\epsilon(x)}\},\\
    \mathsf{P}_2[J]=\bra{\psi(t_{\rm f})}\Pi_{\textsf{\textbf{x}}}Y\!(\textsf{\textbf{x}})\ket{\psi(t_{\rm f})}={\rm Im}\{\mathsf{Z}_0[J]\ee^{\ii\int{\rm d}^Dx\delta\epsilon(x)}\}.
    \end{split}
  \eeq

After  introducing the generating functional of connected propagators, which in the non-interacting case simply reads
\beq
\mathsf{W}_0[J]=-\ii\log Z_0[J]=\frac{\ii}{2}\!\!\bigintssss\!\!{\rm d}^Dx_1\!\!\!\!\bigintssss\!\!\!\!{\rm d}^Dx_2\hspace{0.2ex}J(x_1)\Delta_{m_0}\!(x_1-x_2)J(x_2),
\eeq
one can see that this functional is encoded in the relative phase of the multi-partite entangled state~\eqref{eq:parity_oscillations}, which is
which is the typical outcome of Ramsey interferometry~\cite{PhysRev.78.695},  one of the key methods in quantum sensing~\cite{RevModPhys.89.035002}.   
The global parities are thus a multi-partite generalization of a pair of  Ramsey signals 
   \beq
  \label{eq:parity_generating_functional_bis}
  \begin{split}
  \mathsf{P}_1[J]=\cos\left(\mathsf{W}_0[J]+\int{\rm d}^Dx\delta\epsilon(x)\right),\\
    \mathsf{P}_2[J]=\sin\left(\mathsf{W}_0[J]+\int{\rm d}^Dx\delta\epsilon(x)\right).
    \end{split}
  \eeq
  Selecting specific timings $(t_0,t_{\rm f})$ would allow one to reconstruct the desired generating functional for any particular set of sources $J(x)$ from experimental data. This result  teaches us an important lesson, namely that previously-considered  mathematical tools in QFTs, such as the generating functional, can indeed become observable quantities when combining ideas from high-energy physics and AMO quantum technologies.  will  be a generalisation of Ramsey interferometry that 
    As discussed  in~\cite{PhysRevX.7.041012}, the qubits need not be attached to every spacetime point when one is only interested in recovering a specific $n$-point propagator~\eqref{eq:vpa}. In that case, it suffices to use  $n$ qubits located at the desired spatial locations $\delta\epsilon(x)=\omega_0\sum_{i=1}^n\delta^d\!(\textsf{\textbf{x}}-\textsf{\textbf{x}}_i)$, where $\omega_0$ is the transition frequency between the two levels, and switch on and off the sources impulsively (i.e. non-adiabatically with respect to any other timescale in the problem) at the corresponding lapses of time $J(x)=\sum_i \mathsf{J}_i\delta^D\!(x-x_i)$, as indicated in Fig.~\ref{Fig:scheme_impulsive_sensor}. One would then  gather  the interferometric experimental   data $\mathsf{P}_1,\mathsf{P}_2$ for various combinations of the impulsive sources~\cite{PhysRevX.7.041012}, and infer the corresponding functional derivatives that lead to the desired $n$-point propagator. In the simplest situation, inferring $\Delta_{m_0}\!(x_1-x_2)$ requires repeating the interferometric protocol four times using two qubits  at each of the spatial locations of interest $\textsf{\textbf{x}}_1,\textsf{\textbf{x}}_2$, and switching on and off the  sources at the corresponding times $t_1,t_2$ of the spacetime points $x_1,x_2$. Together with the requirement of using  multi-partite entangled states for the initialisation of the Ramsey probe,  maintaining the coherence of macroscopically-distinct states, and repeating the whole experimental sequence to extract the quantum-mechanical statistics, this  sensing protocol turns out to be quite challenging from an experimental perspective. In the following section, we shall describe a simpler alternative.
  
\subsection{Harmonic  sensors and long-range Ising models}
  \label{sec:KG_Ising_spins}

The above simple calculation for the free Klein-Gordon field  unveils an interesting fact, which was partially hidden under the general formalism for interacting fields~\cite{PhysRevX.7.041012}. Equation~\eqref{eq:int_U_spin} shows that the $\mathsf{Z}_0[J]$ mapping relies on unitaries describing pairwise  interactions   $ U_{\tilde{\mathcal{V}}_J}=U_N\Pi_{x1,x2}U_{x1,x2}$ with
  \beq
  \label{eq:spin_spin_unitary}
  U_{x1,x2}= \ee^{-\frac{1}{2}P\!(x_1)J(x_1)\Delta_{m_0}\!(x_1-x_2)J(x_2)P\!(x_2)}.
  \eeq
  This unitary represents  a pairwise coupling between distant qubits, which must be mediated by the Klein-Gordon bosons acting as fundamental carriers of a spin-spin interaction. Since the Feynman propagator of the scalar field  appears  in Eq.~\eqref{eq:spin_spin_unitary}, there might be simpler measurement protocols to extract relevant properties of the QFT without resorting to the full generating-functional protocol just described.
  
  \begin{figure*}[t]
 \begin{centering}
  \includegraphics[width=2\columnwidth]{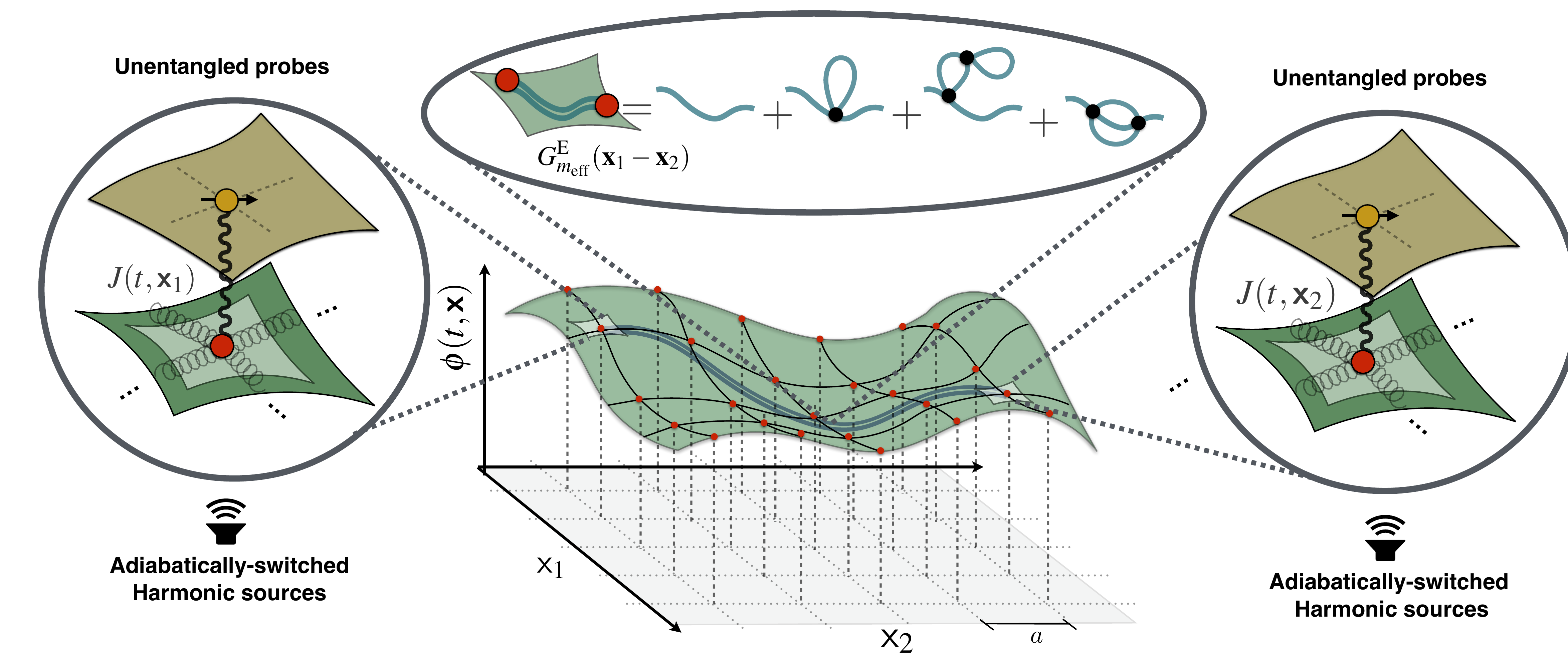}\\
  \caption{\label{Fig:scheme_harmonic_sensor} {\bf Scheme of the harmonic sensors for the generating functional:}  Two distant Ising spins at positions $\textsf{\textbf{x}}_1$ and $\textsf{\textbf{x}}_2$ are coupled to the real scalar field $\phi(t,\textsf{\textbf{x}})$  locally by harmonic Ising-Schwinger sources  $J(t,\textsf{\textbf{x}}_i)\propto\sin(\omega_Jt-\textsf{\textbf{k}}_J\cdot\textsf{\textbf{x}}_i)$ that is adiabatically switched during the whole probing lapse. As a consequence, the scalar bosons will mediate an effective Ising interaction between the  spins, which is represented by a double blue line. In the upper inset, we depict the strength of this spin-spin interactions, which is effectively controlled by a dimensionally-reduced Euclidean propagator of the scalar field. In the presence of quartic couplings, this propagator will include all possible scattering events that can be represented in terms of  Feynman diagrams, as displayed in the inset up to second order of the quartic interaction strength. }
\end{centering}
\end{figure*}

  Let us note, before moving on, that the crucial aspect of the above pairwise unitary is not restricted to the specific   form of the Ising-Schwinger sources, which involve  orthogonal projectors~\eqref{eq:Ising_source_coupling}. The key point is that the coupling must contain a qubit operator $P(x)\to O(x)$ that, in the Heisenberg picture with respect to the unsourced Hamiltonian~\eqref{eq:Ising_source_coupling}, commutes with itself at different instants of time $[O_H(x),O_H(y)]=0$. As will become clear when discussing the particular application for trapped-ion quantum computers~\cite{HAFFNER2008155,Ladd2010}, one can find various schemes where this operator is any of the Pauli matrices  $O(x)\in\{X(x),Y(x),Z(x)\}$. In the following, we will solely focus on the $Z$-type couplings  \beq
\label{eq:Ising_Z_source_coupling}
\tilde{\mathcal{V
}_J}=-J(x)\phi(x)Z(x),
\eeq
leading to Ising  $ZZ$  pairwise unitaries 
    \beq
  \label{eq:Ising_unitary}
  U_{x1,x2}= \ee^{-\frac{1}{2}Z\!(x_1)J(x_1)\Delta_{m_0}\!(x_1-x_2)J(x_2)Z\!(x_2)},
  \eeq
  but note that the results will be interchangeable to the other $XX$ or $YY$ Ising interactions.
  
   As shown below, for certain types of sources, these pairwise unitaries can be expressed in terms of an effective time-independent  Hamiltonian, which allow for a different sensing protocol. In particular, the probes evolve under an effective Ising model with long-range couplings $H_{\rm eff}$, which  controls completely the non-trivial part of the time evolution   \beq
  \label{eq:eff_unitary}
  U(t_{\rm f},t_0)\approx U_{\tilde{\mathcal{H}}_0}U_{\rm eff}=\ee^{-\ii (t_{\rm f}-t_0)\tilde{H}_0}\ee^{-\ii(t_{\rm f}-t_0)H_{\rm eff}}.
  \eeq
Instead of using non-adiabatic sources that are switched on/off in the impulsive regime, we consider a harmonic source 
  \beq
  \label{eq:sources_harmonic}
  J(x)=\mathsf{J}_0\sin(k_Jx)=\mathsf{J}_0\sin\big(\omega_Jt-\textsf{\textbf{k}}_J\cdot\textsf{\textbf{x}}\big),
  \eeq
  where $\mathsf{J}_0$ is the coupling strength density, and $k_J=(\omega_J,\textsf{\textbf{k}}_J)$ is the external $D$-momentum  determining the plane-wave harmonic source. We also consider that the source lies below  the resonance of the Klein-Gordon modes (i.e. $\omega_J\lesssim m_0$) and, moreover, its strength is constrained according to 
  \beq
  \label{eq:constraint_couplings}
  \mathsf{J}_0\ll(\omega_{\textsf{\textbf{k}}}-\omega_{J}){\rm d}^d\hspace{-0.1ex}\textsf{{k}}< (\omega_{\textsf{\textbf{k}}}+\omega_{J}){\rm d}^d\hspace{-0.1ex}\textsf{{k}} .
  \eeq

  In this regime, the time integrals  underlying the spacetime formulation of the evolution operator~\eqref{eq:int_U_spin} can be performed analytically using the forward and backward parts of the Feynman propagator~\eqref{eq:F_propagator}. We find that the normal-ordered part of Eq.~\eqref{eq:int_U_spin} becomes negligible, such that the evolution of the  scalar field and the Ising spins   gets effectively decoupled. Whereas the former evolves under the Klein-Gordon dynamics~\eqref{eq:KG_field}, the spins also experience
    the second-order term including the pairwise couplings~\eqref{eq:Ising_unitary}, which  leads to
   \beq
 U_{\rm eff}=\!\ee^{\ii(t_{\rm f}-t_0)\!\!\bigintssss\!\!\!{\rm d}^{d}\!\textsf{{x}}_1\!\!\bigintssss\!\!\!{\rm d}^{d}\!\textsf{{x}}_2\!\!\bigintssss_{\textsf{{k}}}\!\!\!\frac{  \mathsf{J}_0^2\cos\left((\textsf{\textbf{k}}-\textsf{\textbf{k}}_J)\cdot(\textsf{\textbf{x}}_1-\textsf{\textbf{x}}_2)\right)}{\omega_{\textsf{\textbf{k}}}^2-\omega_J^2}Z(\textsf{\textbf{x}}_1)Z(\textsf{\textbf{x}}_2)}\!\!,
 \eeq
 where $\int_{\textsf{{k}}}=\int\!{\rm d}^d\textsf{{k}}/(2\pi)^d$ stands for the integrals about the spatial momentum.
 Using Eq.~\eqref{eq:eff_unitary}, one readily identifies an  Ising Hamiltonian
 \beq
 \label{eq:eff_ham}
 H_{\rm eff}=\frac{1}{2}\!\int\!\!{\rm d}^{d}\!\textsf{{x}}_1\!\!\int\!\!{\rm d}^{d}\!\textsf{{x}}_2 \mathsf{J}(\textsf{\textbf{x}}_1-\textsf{\textbf{x}}_2)Z(\textsf{\textbf{x}}_1)Z(\textsf{\textbf{x}}_2), 
 \eeq
  where the factor of $1/2$ avoids double counting of pairs, and we have introduced   the long-range coupling-strength density
 \beq
 \label{coupling_density}
 \mathsf{J}(\textsf{\textbf{x}}_1-\textsf{\textbf{x}}_2)=-2\mathsf{J}_0^2\!\!\bigintssss_{\textsf{{k}}}\frac{  \cos\left((\textsf{\textbf{k}}-\textsf{\textbf{k}}_J)\cdot(\textsf{\textbf{x}}_1-\textsf{\textbf{x}}_2)\right)}{\textsf{\textbf{k}}^2+(m_0^2-\omega_J^2)}.
 \eeq
 It is worth pointing out that, despite starting with a local Lorentz-invariant field theory in Eqs.~\eqref{eq:KG_field} and~\eqref{eq:Ising_source_coupling}, it seems that we have arrived to an action at a distance between the Ising fields~\eqref{eq:eff_ham} that naively violates causality. Note, however,  that the relevant time-scale of the problem $t\sim1/\mathsf{J}(\textsf{\textbf{x}}_1-\textsf{\textbf{x}}_2){\rm d}^{2d}\!\textsf{{x}}\gg |\!|\textsf{\textbf{x}}_1-\textsf{\textbf{x}}_2|\!|/c=t_{\rm ret} $, where we have included again the speed of light to identify the corresponding retardation time $t_{\rm ret}$. Accordingly, the regime of validity of the effective Ising model~\eqref{eq:eff_ham}  assumes that the timescales are sufficiently large for the bosons to causally connect any pair of  qubits.  
 
 Let us  also point out that, in order to achieve this purely unitary dynamics,   keeping the Ising fields off-resonant is crucial. Otherwise, the scalar bosons could also act  as a source of dissipation, as   our field theory could be understood as a multi-qubit continuum limit of the  unbiased spin-boson model at zero temperature~\cite{RevModPhys.59.1,PhysRevA.51.992,doi:10.1098/rspa.1996.0029}. Here, the harmonic frequency  would play the role of the so-called quantum tunneling $\Delta_{\rm qt}\approx\omega_{J}$, provided that Eq.~\eqref{eq:constraint_couplings} is fulfilled. Neglecting possible collective dissipative effects, the main source of irreversible dynamics  would appear in the form of decoherence with a dephasing rate $\Gamma_{\rm d}$ that is proportional to the spectral density $\mathsf{S}(\omega)$ of the Klein-Gordon modes at a slightly-renormalised tunneling amplitude ${\omega}_{J,r}<\omega_J<m_0$~\cite{PhysRevA.93.043843}. Since there are no  modes with frequency $\omega_{\textsf{\textbf{k}}}< m_0$, the corresponding spectral density would be zero $\mathsf{S}({\omega}_{J,r})=0$, such that  $\Gamma_{\rm d}=0$, and we can simply focus on  the purely coherent dynamics of Eq.~\eqref{eq:eff_unitary}.
 
 Coming back to the   integral~\eqref{coupling_density}, we note that it can be expressed in terms of the dimensionally-reduced Euclidean  propagator  of a Gaussian field in $d$ dimensions~\cite{WILSON197475,zinn-justin_2012,zinn-justin_2013}, which is defined through the following Green's function
 \beq
 \left(-\boldsymbol{\nabla}^2+m_{\rm eff}^2\right) G_{m_{\rm eff}}^{\rm E}\!(\textsf{\textbf{x}})=\delta^d\!(\textsf{\textbf{x}}),
 \eeq
 where the effective mass  is shifted from the bare mass to
  \beq
  \label{eq:eff_mass}
  m_{\rm eff}^2=m_0^2-\omega_J^2.
  \eeq
  Since the Compton wavelength in natural units is simply the inverse of the bare mass $\xi_0=1/m_0$, we can define the following effective  wavelength 
  \beq
  \xi_{\rm eff}=m_{\rm eff}^{-1}=1/\left(m_0^2-\omega_J^2\right)^{1/2},
  \eeq 
  which shall control the range of the mediated interactions.
  
   This Euclidean propagator can be obtained by   lowering the dimension $D=d+1\to d$   of Feynman propagator of Eq.~\eqref{eq:feynman_propagator}, and  making a Wick rotation, yielding
    \beq
    \label{eq:Euclidean_propagator}
 G_{m_{\rm eff}}^{\rm E}\!(\textsf{\textbf{x}})\!=\!\!\bigintssss_{\textsf{{k}}}\frac{\ee^{\ii {\textsf{\textbf{k}}\cdot\textsf{\textbf{x}}}}}{\textsf{\textbf{k}}^2+m_{\rm eff}^2}=\left(\frac{m_{\rm eff}}{|\!|\textsf{\textbf{x}}|\!|}\right)^{\!\!\nu} \frac{\mathsf{K}_{\nu}(m_{\rm eff}|\!|\textsf{\textbf{x}}|\!|)}{(2\pi)^{\nu+1}},
 \eeq
 where $\nu=\frac{d}{2}-1$, and $\mathsf{K}_{\nu}\hspace{-0.1ex}(u)$ is the modified Bessel function of the second kind~\cite{fradkin_2021}, also called Basset function~\cite{000002668}. We have thus found that the effective spin-spin couplings mediated by a $D$-dimensional Klein-Gordon field subjected to harmonic Ising-Schwinger sources are controlled by the  dimensionally-reduced Euclidean propagator via
 \beq
 \label{eq:spin_spin_couplings}
  \mathsf{J}(\textsf{\textbf{x}}_1-\textsf{\textbf{x}}_2)=-2\mathsf{J}_0^2G_{m_{\rm eff}}^{\rm E}\!(\textsf{\textbf{x}}_1-\textsf{\textbf{x}}_2)\cos\big(\textsf{\textbf{k}}_J\cdot(\textsf{\textbf{x}}_1-\textsf{\textbf{x}}_2)\big).
 \eeq
 
 Let us pause for a moment and analyse this result. First of all, one can check that the above expressions are dimensionally correct in mass/energy units. According to Eq.~\eqref{eq:KG_field}, the natural  or scaling dimension of the scalar field is $d_\phi=(d-1)/2$, whereas the Ising field is dimensionless $d_\sigma=0$. Therefore, the Euclidean propagator~\eqref{eq:Euclidean_propagator}
 has natural dimension $d_{G_{m}^{\rm E}}=d-2$, whereas the coupling strength $\mathsf{J}_0$~\eqref{eq:sources_harmonic} has natural dimension $d_{\mathsf{J}_0}=(d+3)/2$,   and one finds that the effective Ising Hamiltonian~\eqref{eq:eff_ham} has  units of energy. We now discuss the implications of the form of the spin-spin couplings~\eqref{eq:spin_spin_couplings} in various dimensions. For the $D=3+1$ Klein-Gordon field, $\nu=1/2$, and the modified Bessel function is $\mathsf{K}_{1/2}(u)=\sqrt{\pi/2u}\ee^{-u}$, such that the Ising couplings become
 \beq
 \label{eq:yukawa_coupling}
d=3,\hspace{1.5ex} \mathsf{J}(\textsf{\textbf{x}})=-\frac{\mathsf{J}_0^2}{2\pi}\frac{\ee^{-\frac{|\!|\textsf{\textbf{x}}|\!|}{\xi_{\rm eff}}}}{|\!|\textsf{\textbf{x}}|\!|}\cos\big(\textsf{\textbf{k}}_J\cdot\textsf{\textbf{x}}\big).
 \eeq
Hence, we identify a Yukawa-type interaction between the qubits (i.e. screened Coulomb decay)~\cite{10.1143/PTPS.1.1}. In fact, the analogy becomes clearer when writing the  spins in terms of fermions via a Jordan-Wigner transformation~\cite{Jordan1928}, since the Ising-Schwinger source term~\eqref{eq:Ising_source_coupling} becomes a Yukawa coupling between the fermionic charge density and a propagating scalar field, although the former would be spinless and static rather than that described by a relativistic Dirac field.

 We note that the range of the   Yukawa interaction~\eqref{eq:yukawa_coupling}   is controlled by the inverse of the effective mass~\eqref{eq:eff_mass}, and can thus be  tuned by placing the frequency of the  harmonic source $\omega_{{J}}$ closer or further from the bare mass  $m_0$. If $\textsf{\textbf{k}}_J=\boldsymbol{0}$, the spin couplings  are always negative, and would thus describe  an Ising-Yukawa ferromagnet. Note that, as advanced above, the Ising field need not be densely distributed over the whole Minkowski spacetime, but can be arranged at 
  \beq
 \label{eq:energies_lattice_spins}
 \delta\epsilon(x)=\omega_0\sum_{i=1}^n\delta^d\!(\textsf{\textbf{x}}-\textsf{\textbf{x}}_i).
 \eeq
  In this  situation,  ferromagnetic couplings can  also be achieved for a harmonic source with non-zero momentum if the qubits are arranged in a plane/line orthogonal to the plane wave $\textsf{\textbf{k}}_J\perp (\textsf{\textbf{x}}_i-\textsf{\textbf{x}}_j)$. Likewise, if the qubits are contained in a region much smaller than the harmonic wavelength  $\textsf{\textbf{k}}_J\cdot(\textsf{\textbf{x}}_i-\textsf{\textbf{x}}_j)\approx 0$,  the interactions are approximately ferromagnetic. Away from these conditions, the Yukawa-decaying couplings will alternate between ferromagnetic and antiferromagnetic depending on the qubit distance, which can lead to magnetic  frustration.   

 For the $D=1+1$ Klein-Gordon field, $\nu=-1/2$, and the modified Bessel function is $\mathsf{K}_{-1/2}(u)=\mathsf{K}_{1/2}(u)$, such that
 \beq
 \label{eq:yukawa_coupling_1d}
 d=1,\hspace{1.5ex}\mathsf{J}(\textsf{{x}})=-\mathsf{J}_0^2\frac{\ee^{-\frac{|\textsf{{x}}|}{\xi_{\rm eff}}}}{m_{\rm eff}}\cos\big(\textsf{{k}}_{J,\textsf{x}}\textsf{{x}}\big).
 \eeq
 The situation is similar to the one discussed above, albeit with exponentially-decaying spin-spin interactions. Whereas for small distances $|\!| \textsf{\textbf{x}}_1-\textsf{\textbf{x}}_2|\!|\ll\xi_{\rm eff}$, we get a Coulomb-type decay in three-dimensions $\mathsf{J}(\textsf{\textbf{x}}_1-\textsf{\textbf{x}}_2)\propto1/|\!|\textsf{\textbf{x}}_1-\textsf{\textbf{x}}_2|\!|$, which gets exponentially screened  at larger spatial separations, the one-dimensional case~\eqref{eq:yukawa_coupling_1d} is always described by  exponentially-decaying interactions regardless of the distance.
 
 Finally, for the $D=2+1$ scalar field, $\nu=0$,  and the modified Bessel function cannot be expressed in terms of elementary functions. In this case,  one finds
  \beq
 d=2,\hspace{1.5ex} \mathsf{J}(\textsf{\textbf{x}})=-\frac{\mathsf{J}_0^2}{\pi}\mathsf{K}_0\!\left(\frac{|\!|\textsf{\textbf{x}}|\!|}{\xi_{\rm eff}}\right)\cos\big(\textsf{\textbf{k}}_J\cdot\textsf{\textbf{x}}\big)\!.
 \eeq
 For small spatial separations $|\!| \textsf{\textbf{x}}_1-\textsf{\textbf{x}}_2|\!|\ll\xi_{\rm eff}$,  there are  log-decaying couplings $\mathsf{J}_2(\textsf{\textbf{x}}_1-\textsf{\textbf{x}}_2)\approx-\gamma+\log(2\xi_{\rm eff}/|\!|\textsf{\textbf{x}}_1-\textsf{\textbf{x}}_2|\!|)$, where $\gamma\approx0.577$ is Euler's constant. Conversely, for long distances $|\!|\textsf{\textbf{x}}_1-\textsf{\textbf{x}}_2|\!|\gg\xi_{\rm eff}$,  we obtain a power-law coupling that gets exponentially screened.

  Before concluding this section, let us also note that one can include quantum fluctuations in the effective Ising models by modifying the perturbation in Eq.~\eqref{eq:Ising_source_coupling} to 
  \beq
\label{eq:Ising_source_coupling_new}
\tilde{V}\to\hat{V}=\int{\rm d}^dx\big( -J(x)\phi(x)Z(x)-H_{\textsf{t}}(x)X\!(x)\big),
\eeq
where $H_{\textsf{t}}(x)$ is a new coupling-strength density that controls the amount of quantum fluctuations in the Ising fields. We shall assume that the Ising fields are distributed in a certain spatial arrangement $\textsf{\textbf{x}}_i\in\Lambda_{\rm s}$, such that  
\beq
\label{eq:lattice_sources}
\begin{split}
H_{\textsf{t}}\!(x)&=2h_{\textsf{t}}(x)\cos(\omega_0t),\hspace{1ex}h_{\textsf{t}}(x)=h_{\textsf{t}}\sum_{i=1}^n\delta^d\!(\textsf{\textbf{x}}-\textsf{\textbf{x}}_i),\\
J(x)&=\sum_{i=1}^n{J}_0\sin\big(\omega_Jt-\textsf{\textbf{k}}_J\cdot\textsf{\textbf{x}}_i\big)\delta^d\!(\textsf{\textbf{x}}-\textsf{\textbf{x}}_i),
\end{split}
\eeq 
 where the couplings $J_0, h_{\textsf{t}}$ have now units of mass/energy. The non-trivial part of the time-evolution operator, considering that $h_{\textsf{t}}\ll2\omega_0$, can now be described as
  \beq
 \label{eq:unitaries_spin_Ising_model}
 U_{\hat{\mathcal{{V}}}_J}=\!\mathsf{T}\!\left\{\ee^{+\ii\int\!{\rm d}t\sum_i\big( J(t,\textsf{\textbf{x}}_i)\phi_H(t,\textsf{\textbf{x}}_i)Z(t,\textsf{\textbf{x}}_i)+h_{\textsf{t}}X(t,\textsf{\textbf{x}}_i)\big)}\!\right\},
 \eeq
 We can now repeat the previous procedure to  calculate the  evolution operator, but differences will arise as the Magnus expansion is no longer exact at second order~\eqref{eq:propagator}. Nonetheless, the additional terms can  be neglected if the coupling constraint~\eqref{eq:constraint_couplings} is changed so that it  encompasses both couplings  
\beq
\label{eq:ising_model_constraint}
|J_0|,|h_{\textsf{t}}|\ll(\omega_{\textsf{\textbf{k}}}-\omega_J),(\omega_{\textsf{\textbf{k}}}+\omega_J).
\eeq
We then arrive at a long-range version of the transverse-field quantum Ising model~\cite{PFEUTY197079,RevModPhys.51.659}, namely
\beq
\label{eq:eff_ising_lattice}
H_{\rm eff}=\frac{1}{2}\sum_{i=1}^n\sum_{j=1}^nJ_{ij}\hspace{0.25ex}Z(\textsf{\textbf{x}}_i)Z(\textsf{\textbf{x}}_j)-h_{\textsf{t}}\sum_{i=1}^n X(\textsf{\textbf{x}}_i),
\eeq
where the spin-spin couplings $J_{ij}=\mathsf{J}(\textsf{\textbf{x}}_i-\textsf{\textbf{x}}_j)J_0^2/\mathsf{J}_0^2$ have units of mass/energy, and are thus controlled by the $D$-dimensional Euclidean Green's function~\eqref{eq:spin_spin_couplings}. In analogy to the nearest-neighbor models~\cite{PFEUTY197079,RevModPhys.51.659}, there is a competition between magnetic phases that break spontaneously the $\mathbb{Z}_2$  symmetry  $P_1=\Pi_{\textsf{\textbf{x}}_i}X(\textsf{\textbf{x}}_i)$ that inverts $Z(\textsf{\textbf{x}}_i)\to P_1Z(\textsf{\textbf{x}}_i)P_1=-Z(\textsf{\textbf{x}}_i)$; and paramagnetic phases where all spins point in the direction of the external transverse field $h_{\textsf{t}}$. Neglecting the possible frustration due to the periodic alternation  between ferro- and antiferro-couplings, we expect that the exponential screening at large distances will lead to critical theories  in the  Ising universality class, which have the same scaling behaviour  as the corresponding nearest-neighbor models. 

Let us now switch to the discussion of the simplified quantum sensors of the massive Klein-Gordon field.
Regardless of the phase diagram of the Ising spins in the thermodynamic limit $n\to \infty$, if one is interested in the characterisation of the underlying scalar field,  a pair of Ising spins can suffice as sensor qubits $n=2$ (see Fig.~\ref{Fig:scheme_harmonic_sensor}), providing a much simpler scheme than the one presented above for the full generating functional. First of all, we do not require initialising the system in a multi-partite entangled state, but instead in $\ket{\Psi_0}=\ket{0}\otimes\ket{+_i}\otimes\ket{+_j}$, where we recall that $\ket{0}$ is the Klein-Gordon vacuum, and $\ket{+_i}=(\ket{\uparrow_{\textsf{\textbf{x}}_i}}+\ket{\downarrow_{\textsf{\textbf{x}}_i}})/\sqrt{2}$ is a coherent superposition for each of the Ising probes. Secondly, instead of  measuring the global parities~\eqref{eq:parity_generating_functional}, it will suffice to measure a single-qubit observable in the Pauli basis, e.g. $X(\textsf{\textbf{x}}_i)$, in the  time-evolved state $\ket{\Psi_{\rm f}}=U(t_{\rm f},(t_{\rm f}-t_0)/2)X(\textsf{\textbf{x}}_i)X(\textsf{\textbf{x}}_j)U((t_{\rm f}-t_0)/2,t_0)\ket{\Psi_0}$. Setting $h_{\textsf{t}}=0$,  this time evolution yields 
\beq
\bra{\Psi_{\rm f}}X(\textsf{\textbf{x}}_i)\ket{\Psi_{\rm f}}=\bra{\Psi_0}U_{\tilde{\mathcal{V}}_J}^\dagger X(\textsf{\textbf{x}}_i)U_{\tilde{\mathcal{V}}_J}^{\phantom{\dagger}}\!\ket{\Psi_0}=\cos\left(2J_{ij}(t_{\rm f}-t_0)\!\right)\!,
\eeq
such that one could extract the spin-spin coupling  strength $J_{ij}$ from the real-time evolution of the transverse magnetisation. Then,   the bare mass of the Klein-Gordon field $m_0$ and the bare coupling $J_0$ could be inferred  by repeating the same scheme for different frequencies of the harmonic source $\omega_J$, as this should change the  range of the interaction and lead to different oscillation frequencies of the observable. With this information, one can reconstruct  $\mathsf{Z}_0[J]$ in Eq.~\eqref{eq:free_Z}.

Note that, in addition to dispensing with the need of preparing large multi-partite entangled states, the previous evolution is of the spin-echo type~\cite{PhysRev.80.580}, and will thus refocus the dephasing caused by external fluctuations that affect the qubits on a   slower timescale than each of the single experimental runs. In contrast, spin echos or any other dynamical-decoupling sequence~\cite{RevModPhys.88.041001}, cannot be incorporated in the generating-functional sensing scheme of~\cite{PhysRevX.7.041012}, as it would also refocus some of the signals that are required to estimate the derivatives. We  conclude that the harmonic  scheme is not only simpler than the impulsive one, but also more robust to  noise.

\section{\bf Self-interacting scalar fields coupled to  $\mathbb{Z}_2$  fields}
  \label{sec:ren_Ising_models}
  In this section, we describe  the previous sensing protocols  in the presence of    self-interactions in the massive Klein-Gordon field, which leads to  the so-called $\lambda\phi^4$ QFT, and can be  addressed by the use of functional-integral methods.     In Appendix~\ref{sec:ren_gen_functional}, we review this functional approach for the full generating functional of the interacting scalar field $\mathsf{Z}[J]$ in absence of the Ising spins. This yields  a graphical representation of $\mathsf{Z}[J]$ in terms of Feynman diagrams, where the logarithm of the full generating functional can be expressed as a series in even powers of the renormalised source functions  weighted by the renormalised connected propagators at the corresponding order. To extend these results to the full problem with Ising-Schwinger sources, we describe in Sec.~\ref{sec:spin_path_integral} a spin-path integral representation of the amplitude of propagation between two arbitrary states of the Ising and scalar fields, which leads to an effective scalar-sigma QFT. In Sec.~\ref{sec:spin_scalalr_gen_functional},  we show that the $\lambda\phi^4$ bosons act as mediators of Ising-type interactions between $2n$ spins, each of which is controlled by the renormalised $2n$-point connected propagator of the self-interacting scalar bosons. For  harmonic sources in the specific regime of the previous section, the leading term leads again to  an effective quantum  Ising Hamiltonian with long-range couplings, the range of which now accounts for all the possible scattering events that the mediating boson can undergo while propagating between the corresponding pair of Ising spins. This shows that the sensing scheme of Sec.~\ref{sec:KG_ising}, when applied to the full interacting case, can gain information about the renormalisation effects, and thus probe this QFT in a novel manner.

  \subsection{Spin path integral and scalar-sigma field theory}
  \label{sec:spin_path_integral}
  
    As discussed in the previous section, the real Klein-Gordon field is the simplest model where one can introduce key concepts and  techniques to be  used later in more complicated QFTs. For instance,  including quartic self-interactions in the Klein-Gordon field~\eqref{eq:KG_field}  leads to the $\lambda\phi^4$ model
  \beq
  \label{eq:int_V}
  \mathcal{H}=\mathcal{H}_0+\mathcal{V}_{\rm int},\hspace{1.5ex}\mathcal{V}_{\rm int}=\frac{\lambda_0}{4!}\hspace{0.1ex}\phi^4(x),
  \eeq
  where $\lambda_0$ is the bare coupling strength. This model  has played a leading role in the development  of QFT. On the one hand, it is a cornerstone in our understanding of the spontaneous breakdown~\cite{PhysRevD.7.1888,PhysRevD.9.1686} and restoration~\cite{PhysRevD.9.3320,PhysRevD.9.3357} of symmetries in QFTs. This model has also been  a neat playground to understand    the conceptual implications of the renormalisation of QFTs~\cite{WILSON197475,PhysRevLett.28.240}, such as the role of  fixed points controlling the long-wavelength properties, which has found many applications in the theory of phase transitions and critical phenomena~\cite{RevModPhys.70.653}. Finally, we note that the $\lambda\phi^4$ model has also been a cornerstone  in the constructive approach to QFTs, allowing to prove rigorous results for various dimensions~\cite{PhysRev.176.1945,https://doi.org/10.1002/prop.19730210702,FELDMAN197680,PhysRevLett.33.440,LUSCHER198725,PhysRevLett.47.1,FROHLICH1982281}.

 In Appendix~\ref{sec:ren_gen_functional},  we  review the effects of the $\lambda\phi^4$ interactions using the path-integral formalism for the generating functional~\eqref{eq:ren_gen_function_quartic}. Paralleling our  approach for the free Klein-Gordon field, we can use this discussion as a guide to understand how the self-interactions affect the effective Ising models~\eqref{eq:eff_ising_lattice} when considering harmonic Ising-Schwinger sources.
As reviewed in the appendix, to get the path-integral formulation~\eqref{eq:path_integral_gen_functional} of the  full  generating functional $\mathsf{Z}[J]$~\eqref{eq:vpa_interction}, one uses the resolution of the identity in the field/momentum orthonormal basis of the scalar field. To include the Ising-Schwinger sources and the transverse field~\eqref{eq:Ising_source_coupling_new}, one may complement this basis with that of spin coherent states
  \cite{Radcliffe_1971,fradkin_2013}, which can be defined through the action of the spin ladder operators on a fiducial state
  \beq
  \ket{\boldsymbol{\Omega}(x)}=\ee^{\tan\left(\!\frac{\theta_s(x)}{2}\!\right)\ee^{\ii\phi_s(x)}S^+\!(x)}\ket{S,-S}_x.
  \eeq
  Here, $\ket{S,-S}_x=\ket{\downarrow}_x$, and $S^{+}\!(x)=\half(X(x)+\ii Y(x))$ for our original qubits $S=1/2$~\eqref{eq:Ising_source_coupling}, but can be readily generalised to any spin-$S$ representations of the $\mathfrak{su}(2)$ algebra. In addition, $\boldsymbol{\Omega}(x)$ is a unit vector field constrained to reside on the  $S_2$ sphere, which can thus  be characterised by the polar $\theta_s(x)$ and azimuthal $\phi_s(x)$ angles at each spacetime point.
  
   An important difference with respect to the aforementioned field/momentum basis is that the coherent states are not orthogonal, but satisfy instead
  \beq
  \label{eq:overlap_coherent_states}
  \langle\boldsymbol{\Omega}(x_1)|\boldsymbol{\Omega}(x_2)\!\rangle\!=\!\ee^{\ii S\Phi(\boldsymbol{\Omega}(x_1),\boldsymbol{\Omega}(x_2),{\bf e}_z)}\!\!\left(\!\!\frac{1+\boldsymbol{\Omega}(x_1)\!\cdot\boldsymbol{\Omega}(x_2)}{2}\!\right)^{\!\!\!\!S}\!.
  \eeq
  Here, $\Phi(\boldsymbol{\Omega}(x_1),\boldsymbol{\Omega}(x_2),{\bf e}_z)$ stands for the area of the spherical triangle with vertices determined by the tips of the coherent-state unit vectors and the north pole of the $S_2$ sphere. These scalar products appear after splitting in infinitesimal pieces the corresponding time-evolution operator~\eqref{eq:unitaries_spin_Ising_model}, where the fields now evolve in the Heisenberg picture with respect to the unsourced $\lambda\phi^4$ QFT (see App.~\ref{sec:ren_gen_functional}). Paralleling the path-integral construction for this scalar field theory, we  introduce the resolution of the identity of the field/momentum and coherent-state basis at nearby fixed instants of time, and obtain a functional integral  for the $\lambda\phi^4$ QFT coupled to Ising fields via  Ising-Schwinger sources.  
  Following~\cite{fradkin_2013}, for each spatial point, we find that the  amplitude of the above overlap  contributes to the  kinetic energy of a non-relativistic particle moving along the trajectory $\boldsymbol{\Omega}(t,\textsf{\textbf{x}})$ on the unit sphere. However, when recovering the continuum-time limit, the mass of this particle vanishes, and the trajectory  only depends on the accumulated phase~\eqref{eq:overlap_coherent_states} and the source couplings to $J(x),H_{\textsf{t}}(x)$. The latter can be obtained through the coherent-state expectation value of the Ising-Schwinger term~\eqref{eq:Ising_source_coupling_new}, 
$
{\rm d}t\langle\boldsymbol{\Omega}(t,\textsf{\textbf{x}}_i)|\hat{V}|\boldsymbol{\Omega}(t+{\rm d}t,\textsf{\textbf{x}}_i)\rangle\approx{\rm d}t\big( J(t,\textsf{\textbf{x}}_i
)\phi_H(t,\textsf{\textbf{x}}_i)2S{\Omega}_z(t,\textsf{\textbf{x}}_i)+h_{\textsf{t}}(t,\textsf{\textbf{x}}_i)2S{\Omega}_x(t,\textsf{\textbf{x}}_i)\big)
+\mathcal{O}({\rm d}t^2)$. In addition, the  accumulated phase~\eqref{eq:overlap_coherent_states} for the whole time evolution can  be expressed as the integral of a Berry connection  along the trajectory $\boldsymbol{\Omega}(t,\textsf{\textbf{x}})$, which is equivalent to the Aharonov-Bohm phase acquired by a test particle moving  in the background gauge field  generated by  a monopole of charge $S$ situated at the origin of $S_2$. Remarkably, for the Euclidean path integral that arises from the finite-temperature $T$ partition function of a spin model, these trajectories must be closed $\boldsymbol{\Omega}(0,\textsf{\textbf{x}}_i)=\boldsymbol{\Omega}(1/T,\textsf{\textbf{x}}_i)$, and the integral of the Berry connection can then be expressed as a  topological Berry phase or, equivalently,  as a Wess-Zumino  $\theta$ term. This term  plays an important role for both Heisenberg ferromagnets~\cite{fradkin_2013} and anti-ferromagnets~\cite{HALDANE1983464,PhysRevB.38.7215}. 

In our  case, given the form of the coupling~\eqref{eq:Ising_source_coupling_new}, it suffices to parametrise the dynamics of the spins  using coherent states along a great circle of $S_2$, i.e. setting $\phi_s(x)=0$ in Eq.~\eqref{eq:overlap_coherent_states}. Accordingly, the Berry connection vanishes, and there  is no accumulated phase.  The   amplitude of propagation of the full field theory  between an arbitrary pair of states in this basis finds the following path-integral representation 
\beq
\label{eq:path_integral_spins}
\langle \{\!\varphi(x),\boldsymbol{\Omega}(x)\} | U_{ \hat{\mathcal{V}}_J}| \{\varphi(y),\boldsymbol{\Omega}(y)\} \rangle\!=\!\!\int_{\rm b.c.}\!\!\!\!\!\!\!\!{\rm D}[{\varphi,\boldsymbol{\Omega}}]\delta(\boldsymbol{\Omega}^2-1)\ee^{\ii S}.
\eeq
Here, we have introduced the action
\beq
\label{eq:sclalar_spin_action}
S=\int{\rm d}^Dx\big(\mathcal{L}_0-\mathcal{V}_{\rm int}(\varphi)-\hat{\mathcal{V}}_J(\varphi,\boldsymbol{\Omega})\big),
\eeq
where we get the Klein-Gordon Lagrangian~\eqref{eq:KG_lag} with the $\lambda\phi^4$ interactions~\eqref{eq:int_V}, and an additional spin-scalar coupling  
\beq
\hat{\mathcal{V}}_J= -J(x)2S\varphi(x)\Omega_z(x)-h_{\textsf{t}}(x)2S\Omega_x(x)
\eeq
Note that the spin-scalar path integral~\eqref{eq:path_integral_spins} has an integration measure with a constraint that enforces the vector fields to lie on $S_2$, and one integrates over all possible fields consistent with  (b.c.), namely the boundary conditions associated to  the initial/final spin configurations of the asymptotic states.
  
  In the following, we  implement the constraint on the vector fields explicitly by making the following substitution on the symmetry-breaking and transverse components
  \beq
  \sigma(x)={\Omega}_z(x), \hspace{1.2ex}  \pi(x)=\sqrt{1-{\Omega}_z^2(x)}=\!\sum_{\ell=0}^{\infty}g_\ell \sigma^{2\ell}\!(x),
  \eeq
  where $g_\ell=(-1)^\ell\binom{\half}{\ell}$, and $\binom{\alpha}{\ell}=\alpha(\alpha-1)\cdots(\alpha-\ell+1)/\ell!$ is the generalised binomial coefficient. Upon substitution, the action~\eqref{eq:sclalar_spin_action} can be rewritten as a scalar-sigma model
\beq
\label{eq:sclalar_spin_action_final}
S_{}=\int{\rm d}^Dx\big(\mathcal{L}_0-\mathcal{V}_{\rm int}(\varphi,\sigma)+J_s(x)\sigma(x)\varphi(x)\big),
\eeq
with a non-linear part that  includes  the self-interactions of  the $\varphi$ field, but also those  of the $\sigma$ field
\beq
\label{eq:int_V_sigma_scalar}
\mathcal{V}_{\rm int}(\varphi,\sigma)=\frac{\lambda_0}{4!}\varphi^4(x)+\!\sum_{\ell}h_\ell(x)  \sigma^{2\ell}\!(x).
\eeq
Here, we have defined the couplings
\beq
J_s(x)=2SJ(x), \hspace{2ex} h_\ell(x)=2Sg_\ell h_{\textsf{t}}(x).
\eeq
  This  QFT partially resembles a non-linear sigma model~\cite{Gell-Mann1960}, but note that, in this case,  the $\sigma$ fields  are non-propagating. Non-trivial effects  come from the coupling to the  interacting scalar bosons, and hence the name scalar-sigma model.

This coupled QFT has a global $\mathbb{Z}_2\times\mathbb{Z}_2$ symmetry where any of the fields gets inverted. In presence of the Ising-Schwinger source term, this symmetry is reduced to $\mathbb{Z}_2$ acting simultaneously on both fields $(\varphi(x),\sigma(x))\to(-\varphi(x),-\sigma(x))$. Note that  spontaneous symmetry breaking (SSB) can take place in any dimension $d$
by the appearance of a vacuum expectation value in any of the fields.  This is the field-theoretic interpretation of the SSB in the effective quantum Ising model~\eqref{eq:eff_ising_lattice}, which would lead to the ferromagnetic ordered mentioned in the previous sections. In that case, SSB could only occur in the $\sigma$ sector, as the scalar sector is described by free Klein-Gordon fields that do not support a non-zero vacuum expectation value. Even if the generic $\lambda_0\phi^4$ field theory can support this scalar SSB channel, we shall not explore it in this paper for the reasons outlined in Sec.~\ref{sec:anh_ions}, where we  connect to trapped-ion experiments.

  \begin{figure*}[t]
 \begin{centering}
  \includegraphics[width=2\columnwidth]{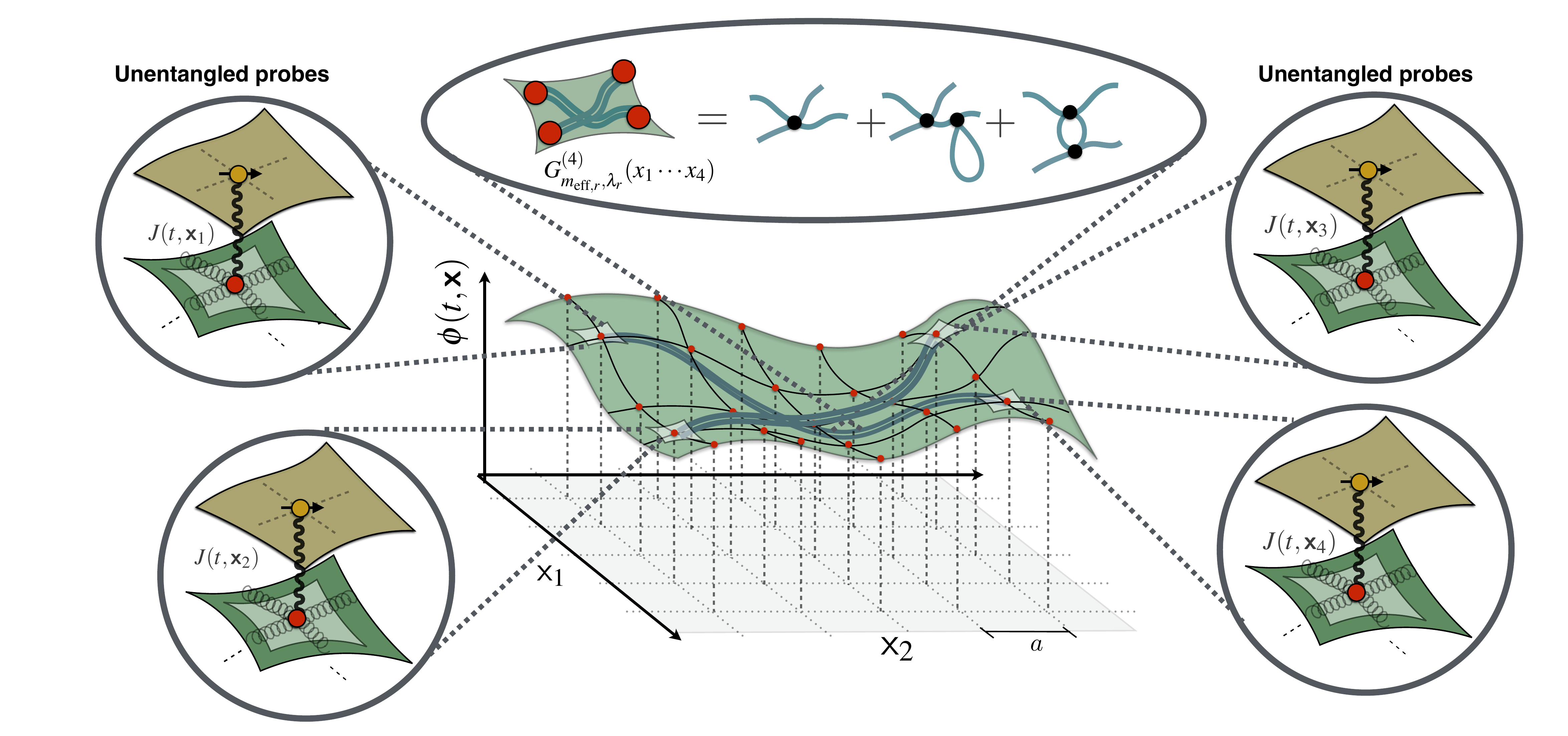}\\
  \caption{\label{Fig:scheme_harmonic_sensor_4_spin} {\bf Scheme of the 4-spin interactions for the harmonic sensors:}  Four distant Ising spins at positions $\textsf{\textbf{x}}_1,\cdot, \textsf{\textbf{x}}_4$  are coupled to the real scalar field $\phi(t,\textsf{\textbf{x}})$  locally by harmonic Ising-Schwinger sources  $J(t,\textsf{\textbf{x}}_i)\propto\sin(\omega_Jt-\textsf{\textbf{k}}_J\cdot\textsf{\textbf{x}}_i)$ . As a result, there are 4-spin Ising interactions mediated by the scalar bosons, the strength of which will be related to a connected 4-point propagator. In the upper inset, we show that  in the presence of quartic couplings, these interactions will include all possible scattering events that involve two incoming and two outgoing virtual bosons, together with internal loops due to the self-interaction vertices. }
\end{centering}
\end{figure*}

 \subsection{Renormalised long-range quantum Ising models}
 \label{sec:spin_scalalr_gen_functional}
 
 In this subsection, we start from the scalar-sigma field theory~\eqref{eq:sclalar_spin_action_final}, and exploit  functional methods to find an effective description of the unitary dynamics that governs the model of scalar $\lambda\phi^4$ fields coupled to the Ising spins.

We note that 
 the action of the scalar-sigma model~\eqref{eq:sclalar_spin_action_final} is formally equivalent to the standard functional description of the self-interacting Klein-Gordon field in Eq.~\eqref{eq:path_integral_gen_functional} of  Appendix~\ref{sec:ren_gen_functional}, provided that one substitutes  the sources $J(x)\to J_s(x)\sigma(x)$. If we are interested in a situation where, as customarily, the scalar field evolves in time from the vacuum into the vacuum,  we can then follow a similar approach as described in Appendix~\ref{sec:ren_gen_functional} to express the amplitude of propagation $\mathcal{T}_{\boldsymbol{\Omega}(y)\to\boldsymbol{\Omega}(x)}=\langle 0,\{\!\boldsymbol{\Omega}(x)\} | U_{ \hat{\mathcal{V}}_J}| 0,\{\boldsymbol{\Omega}(y)\} \rangle\!$ as follows
  \beq
\label{eq:path_integral_spins}
\mathcal{T}_{\boldsymbol{\Omega}(y)\to\boldsymbol{\Omega}(x)}=\!\!\int_{\rm b.c.}\!\!\!\!\!\!\!\!{\rm D}{\sigma}\frac{\ee^{-\ii\!\!\bigintssss\!\!{\rm d}^Dx \mathcal{V}_{\rm int}\left(\!-\ii\delta_{J_s(x)\sigma(x)},\sigma(x)\!\right)}{\mathsf{Z}}_0[J_s\sigma]}{\ee^{-\ii\!\!\bigintssss\!\!{\rm d}^Dx \mathcal{V}_{\rm int}\left(\!-\ii\delta_{J_s(x)\sigma(x)},\sigma(x)\!\right)}{\mathsf{Z}}_0[J_s\sigma]\big|_{0}\!\!\!\!},
\eeq
 where   we use a short-hand notation for the functional derivatives $\delta_{J_s(x)\sigma(x)}=\delta/\delta J_s(x)\sigma(x)$, and (b.c.) now only refers to the initial and final configurations of the $\sigma$ field, which are determined by the asymptotic spin states of the transition amplitude. Note that, in this expression,   the free generating functional ${\mathsf{Z}}_0[J]\to{\mathsf{Z}}_0[J_s\sigma]$ has the same form as  Eq.~\eqref{eq:free_Z}, but now describes the mediated couplings between $\sigma$ fields. In addition, the self-interaction potential must include the $\sigma$-field non-linearities~\eqref{eq:int_V_sigma_scalar}. 
 
 This expression can  be treated using perturbation theory, and admits a description in terms of Feynman diagrams. Note that, in principle, the new interaction term~\eqref{eq:int_V_sigma_scalar} not only includes the scalar vertex with four legs, but also all possible $\sigma$ vertices involving  $2n$ legs. These additional vertices could in principle combine into a wider landscape of Feynman diagrams as that of the pure $\lambda\phi^4$ theory discussed in  Appendix~\ref{sec:ren_gen_functional}, including additional insertions of the  $\sigma$ vertices. However, the constraints in Eq.~\eqref{eq:ising_model_constraint}, which allowed us to neglect the terms of the Magnus expansion beyond second order~\eqref{eq:propagator} stemming from higher-order nested commutators, are equivalent to considering only Feynman diagrams with the  $\sigma$ vertices at tree level, i.e. neglecting the combination of the scalar and sigma scattering processes in the diagrams. Accordingly, the diagrammatic description of this transition amplitude~\eqref{eq:path_integral_spins} is
   \begin{widetext}
  \beq
  \label{eq:spin_feynman}
\int_{\rm b.c.}\!\!\!\!\!\!{\rm D}{\sigma}\bigg( 1\hspace{-8ex}
 \setlength{\unitlength}{1cm}
\thicklines
\begin{picture}(18,0)
\put(1.25,0.0){$+\frac{\ii}{4}$}
\put(1.75,0.05){$\color{gray}{\boldsymbol{\Uparrow}}$}
\put(1.9,.1){\line(1,0){0.8}}
\put(2.3,0.3){\circle{0.4}}
\put(2.21,0.01){$\color{myred}\bullet$}
\put(2.64,0.0){$\color{gray}{\boldsymbol{\Downarrow}}$}
\put(2.84,0.){$+\frac{1}{8}$}
\put(3.45,0.05){$\color{gray}{\boldsymbol{\Uparrow}}$}
\put(3.6,0.1){\line(1,0){0.8}}
\put(4,0.3){\circle{0.4}}
\put(4,0.69){\circle{0.4}}
\put(3.91,0.01){$\color{myred}\bullet$}
\put(3.91,0.41){$\color{myred}\bullet$}
\put(4.34,0.0){$\color{gray}{\boldsymbol{\Downarrow}}$}
\put(4.65,0.){$+\frac{1}{8}$}
\put(5.17,0.05){$\color{gray}{\boldsymbol{\Uparrow}}$}
\put(5.3,0.1){\line(1,0){1.1}}
\put(5.6,0.30){\circle{0.4}}
\put(6.1,0.30){\circle{0.4}}
\put(5.51,0.01){$\color{myred}\bullet$}
\put(6.02,0.01){$\color{myred}\bullet$}
\put(6.34,0.0){$\color{gray}{\boldsymbol{\Downarrow}}$}
\put(6.6,0.){$+\frac{1}{12}$}
\put(7.25,0.05){$\color{gray}{\boldsymbol{\Uparrow}}$}
\put(7.4,0.1){\line(1,0){1}}
\put(7.9,0.1){\circle{0.55}}
\put(8.08,0.01){$\color{myred}\bullet$}
\put(7.54,0.01){$\color{myred}\bullet$}
\put(8.34,0.0){$\color{gray}{\boldsymbol{\Downarrow}}$}
\put(8.5,0.0){$-\frac{\ii}{4!}$}
\put(9.22,0.1){\line(1,0){0.8}}
\put(9.61,-0.3){\line(0,1){0.8}}
\put(9.52,0.01){$\color{myred}\bullet$}
\put(9.08,0.05){$\color{gray}{\boldsymbol{\Uparrow}}$}
\put(9.94,0.0){$\color{gray}{\boldsymbol{\Downarrow}}$}
\put(9.47,0.44){$\color{gray}{\Rightarrow}$}
\put(9.4,-0.43){$\color{gray}{\Leftarrow}$}
\put(10.2,0.0){$-\frac{1}{12}$}
\put(10.95,0.1){\line(1,0){1.05}}
\put(11.61,-0.3){\line(0,1){0.8}}
\put(11.3,0.29){\circle{0.4}}
\put(11.52,0.01){$\color{myred}\bullet$}
\put(11.2,0.01){$\color{myred}\bullet$}
\put(10.8,0.05){$\color{gray}{\boldsymbol{\Uparrow}}$}
\put(11.94,0.0){$\color{gray}{\boldsymbol{\Downarrow}}$}
\put(11.47,0.44){$\color{gray}{\Rightarrow}$}
\put(11.37,-0.43){$\color{gray}{\Leftarrow}$}
\put(12.15,0.0){$-\frac{1}{32}$}
\put(12.75,0.18){$\color{gray}{\boldsymbol{\Uparrow}}$}
\put(12.9,.25){\line(1,0){0.8}}
\put(13.3,0.45){\circle{0.4}}
\put(13.21,0.16){$\color{myred}\bullet$}
\put(13.64,0.18){$\color{gray}{\boldsymbol{\Uparrow}}$}
\put(12.75,-0.18){$\color{gray}{\boldsymbol{\Downarrow}}$}
\put(12.9,-.08){\line(1,0){0.8}}
\put(13.3,-0.26){\circle{0.4}}
\put(13.21,-0.16){$\color{myred}\bullet$}
\put(13.64,-0.18){$\color{gray}{\boldsymbol{\Downarrow}}$}
\put(14.05,0.0){$-\frac{1}{16}$}
\put(14.75,0.28){$\color{gray}{\boldsymbol{\Uparrow}}$}
\put(14.9,.35){\line(1,0){0.8}}
\put(15.3,0.06){\circle{0.6}}
\put(15.64,0.28){$\color{gray}{\boldsymbol{\Uparrow}}$}
\put(14.75,-0.33){$\color{gray}{\boldsymbol{\Downarrow}}$}
\put(14.9,-.24){\line(1,0){0.8}}
\put(15.21,0.26){$\color{myred}\bullet$}
\put(15.21,-0.35){$\color{myred}\bullet$}
\put(15.64,-0.35){$\color{gray}{\boldsymbol{\Downarrow}}$}
\put(15.9,0.01){$+\cdots\bigg) \tilde{\mathsf{Z}}_0.$}
\end{picture}
\eeq
  \end{widetext}
This equation should be read as follows:   the Ising-Schwinger source terms,  proportional to the $\sigma$ fields, are depicted by arrows  ${\color{gray}{\boldsymbol{\Uparrow}}}=J_s(x)\sigma(x)$ to represent the underlying spins at different spacetime locations $x$. The blobs ${\color{myred}\boldsymbol{\bullet}}=\lambda_0$ stand for interaction vertices with the bare quartic coupling. Solid lines that join  an arrow and a blob should be translated into ${\color{gray}{\boldsymbol{\times}}}\boldsymbol{\!\!\!\!\!-\!\!\!\!-\!\!\!\!{\color{myred}{\bullet}}}=\Delta_{m_0}(x-z)$, and thus involve the free Feynman propagator of the scalar field~\eqref{eq:feynman_propagator} from the  point $x$ of the $\sigma$ field  to  the  self-interaction vertex at $z$. Likewise, solid lines  connected to the same blob stand for interaction loops that should be translated for  $\bigcirc\hspace{-2.5ex}{\color{myred}\bullet}\hspace{2ex}=\Delta_{m_0}(0)$, while those connecting two distant blobs must be substituted by the scalar-field  propagator between the  corresponding spacetime points  $\boldsymbol{{\color{myred}\bullet}\!\!\!-\!\!\!\!-\!\!\!\!{\color{myred}\bullet}}=\Delta_{m_0}(z_1-z_2)$. For each of the above diagrams, we should integrate over all possible spacetime locations  of the sources $\int\!{\rm d}^Dx_i$, and those of the intermediate interaction vertices $\int\!{\rm d}^Dz_i$.

 Note that  the arrows in Eq.~\eqref{eq:spin_feynman} are drawn in different directions to emphasise that, contrary to the standard Schwinger sources, the dynamics of which is fixed externally, the new Ising-Schwinger sources involve  $\sigma$ fields with their own dynamics. In particular, from the perspective of  Sec.~\ref{sec:KG_Ising_spins} on the free Klein-Gordon field coupled to Ising spins, we know that there are boson-mediated processes where the spins, when initialised in a particular basis,  can be flipped dynamically. Additionally, if the transverse field is non-zero, there will be quantum fluctuations that will also flip the spins in a different basis. Our choice of alternating directions of the arrows ${\color{gray}{\boldsymbol{\Uparrow}}},{\color{gray}{{\Rightarrow}}},{\color{gray}{\boldsymbol{\Downarrow}}},{\color{gray}{{\Leftarrow}}},$ thus tries to depict this inherent spin  dynamics. Additionally, in Eq.~\eqref{eq:spin_feynman}, the free generating functional should be substituted for  $\mathsf{Z}_0[J_s\sigma]\to\tilde{Z}_0[J_s\sigma,h_\ell]$, which now includes the Ising-Schwinger sources and the tree-level $\sigma$ vertices
\beq
\label{eq:tilde_Z_0}
\tilde{Z}_0\!=\ee^{-\frac{1}{2}\!\!\bigintssss\!\!\!{\rm d}^Dx_1\!\!\bigintssss\!\!\!{\rm d}^Dx_2{J}_s(x_1\!)\sigma(x_1\!)\Delta_{{m}_{ 0}}\!(x_1-x_2)\sigma(x_2\!){J}_s(x_2\!)+\ii\!\!\bigintssss\!\!\!\!{\rm d}^Dx\sum_\ell\! h_\ell\sigma^{2\ell}\!(x)}\!\!.
\eeq

The graphical representation in terms of Feynman diagrams and $\tilde{Z}_0$ in Eq.~\eqref{eq:spin_feynman} then has a clear interpretation. Distant Ising spins can virtually excite the bosonic field via the source coupling, and effectively interact by a long-range coupling mediated by the exchange of a virtual scalar boson~\eqref{eq:tilde_Z_0}. Additionally, these spins are also subjected to  quantum fluctuations captured by the non-linearities of the $\sigma$ vertices~\eqref{eq:tilde_Z_0}. The Feynman diagrams~\eqref{eq:spin_feynman} show that this virtual  exchange process is not entirely captured by the free terms~\eqref{eq:tilde_Z_0}, as the scalar field also has self-interactions, and the mediating boson may get scattered through various processes as it propagates between the  spins, as depicted in the upper inset of Fig.~\ref{Fig:scheme_harmonic_sensor}.

In addition to modifying the  interactions between pairs of spins~\eqref{eq:tilde_Z_0}, self-interactions introduce further possibilities where $2n$-spins also get effectively coupled.   In the fifth diagram of Eq.~\eqref{eq:spin_feynman}, we show that a 4-spin interaction can occur by the action of two Ising-Schwinger sources  that virtually excite a pair of bosons, which then scatter into another pair of bosons that carry the interactions to a different pair of spins. Note that these two-boson mediated processes also allow for further scattering  of the carriers due to the self-interaction of the scalar field, as  described by the following diagrams of Eq.~\eqref{eq:spin_feynman} and  depicted in the upper inset of Fig.~\ref{Fig:scheme_harmonic_sensor_4_spin}.

We can turn this qualitative explanation of the mediated interactions into a quantitative analysis by handling the    perturbative series in  analogy to our exposition of  Appendix~\ref{sec:ren_gen_functional}. From now onwards, for our spin $S=1/2$ case, we will make no distinction between the bare and spin sources since $J_s(x)=J(x)$. The expressions below can be, in any case, readily generalised to arbitrary $S$. The amplitude of propagation between two different $\sigma$-field  configurations can be written in a condensed form that takes into account all the aforementioned scattering processes of the bosonic carriers. In summary, these scattering events lead to  additive and multiplicative  renormalisations of the bare mass $m_0\to {m}_r$, and the bare coupling strength  $\lambda_0\to\lambda_r$. As discussed in  Appendix~\ref{sec:ren_gen_functional}, using the power series of the self-energy $\mathsf{\Sigma}(k)=\mathsf{\Sigma}(0)+k^2\partial_{k^2}\mathsf{\Sigma}(k)|_{k^2=0}+\cdots$ to second order in the coupling strength, one can obtain the so-called  tadpole and sunrise  contributions to $\Sigma_{m_0,\lambda_0}(0)=\mathsf{\Sigma}^{(1,{\rm td})}_{m_0,\lambda_0}+\mathsf{\Sigma}^{(2,{\rm td})}_{m_0,\lambda_0}+ \mathsf{\Sigma}^{(2,{\rm sr})}_{m_0,\lambda_0}$, namely
 \beq
 \label{eq:mass_reb_terms}
 \begin{split}
\mathsf{\Sigma}^{(1,{\rm td})}_{m_0,\lambda_0}&= \frac{\lambda_0}{2}\!\!\bigintssss_{k_1}\!\!\tilde{\Delta}_{m_{ 0}}\!(k_1),\\
\mathsf{\Sigma}^{(2,{\rm td})}_{m_0,\lambda_0}&= -\ii\frac{\lambda_0^2}{4}\!\!\bigintssss_{k_1}\!\!\bigintssss_{k_2}\!\!\tilde{\Delta}_{m_{ 0}}^2\!(k_1)\tilde{\Delta}_{m_{ 0}}\!(k_2),\\
 \mathsf{\Sigma}^{(2,{\rm sr})}_{m_0,\lambda_0}&=-\ii\frac{\lambda_0^2}{6}\!\!\bigintssss_{k_1}\!\!\bigintssss_{k_2}\!\!\tilde{\Delta}_{m_{ 0}}\!(k_1)\tilde{\Delta}_{m_{ 0}}\!(k_2)\tilde{\Delta}_{m_{ 0}}\!(k_1+k_2),
\end{split}
 \eeq
 as well as the sunrise contribution to the wavefunction renormalisation $z_{m_0,\lambda_0}^{\!-1}=1-\partial_{k^2}\mathsf{\Sigma}(k)|_{k^2=0}$, namely
 \beq
\left.\frac{\partial\mathsf{\Sigma}^{(2,{\rm sr)}}}{\partial k^2}\right|_{0}=\frac{\lambda_0^2}{6}\!\!\bigintssss_{k_1}\!\!\bigintssss_{k_2}\!\!\tilde{\Delta}_{m_{ 0}}\!(k_1)\tilde{\Delta}_{m_{ 0}}\!(k_2)\tilde{\Delta}^2_{m_{ 0}}\!(k_1+k_2).
 \eeq
Accordingly, to leading order,   the renormalised mass reads
\beq
\label{eq:mass_renormalisation}
m_r^2=\left(m_0^2\hspace{0.2ex}+\mathsf{\Sigma}_{m_0,\lambda_0}(0)\right)z_{m_0,\lambda_0},
\eeq
whereas the source functions become
 \beq
 \label{eq:source_ren}
  {J}_{ r}(x)=J(x)\sqrt{z_{m_0,\lambda_0}}.
  \eeq
  Finally, the interaction strength also gets renormalised due to the last diagram of Eq.~\eqref{eq:spin_feynman}, leading to
   \beq
   \label{eq:int_renormalisation}
 {\lambda}_r=\left(\lambda_0-\ii\frac{3{\lambda}_0^2}{2}\!\!\bigintssss_{k_1}\!\!\tilde{\Delta}^2_{m_{ 0}}\!(k_1)\right)z_{m_0,\lambda_0}^2.
 \eeq

With these renormalisations, the   amplitude of propagation in Eq.~\eqref{eq:spin_feynman} can thus be compactly  rewritten as
  \begin{widetext}  
      \beq
        \label{eq:renormalised_transition_matrix}
\mathcal{T}_{\boldsymbol{\Omega}(y)\to\boldsymbol{\Omega}(x)}\!\!=\!\!\!\int_{\rm b.c.}\!\!\!\!\!\!\!\!{\rm D}{\sigma}\ee^{-\frac{1}{2}\!\!\bigintssss\!\!{\rm d}^{2D}x{J}_{r}(x_1)\sigma(x_1)\Delta_{{m}_{ r}}\!(x_1-x_2)\sigma(x_2){J}_{r}(x_2)-\frac{\ii}{4!}\!\!\bigintssss\!\!{\rm d}^{4D}x{J}_{r}(x_1){J}_{r}(x_2)\sigma(x_1)\sigma(x_2){G}^{(4,{\rm c})}_{{m}_r,{\lambda}_r}\!\!(x_1\cdots x_4)\sigma(x_3)\sigma(x_4){J}_{r}(x_3){J}_{r}(x_4)+\ii\!\!\bigintssss\!\!\!\!{\rm d}^{D}x\sum_\ell\! h_\ell\sigma^{2\ell}\!(x)}\!\!,
\eeq
\end{widetext}
where we readily identify the quadratic and quartic long-range interactions of the $\sigma$ fields mediated by the scalar bosons. Note that, as discussed in  Appendix~\ref{sec:ren_gen_functional}, one could consider higher-order terms in the external sources, which would introduce  $2n$-spin  interactions with $n\geq3$.

We now argue that, among all these possible mediated interactions for harmonic Ising-Schwinger sources~\eqref{eq:sources_harmonic},  all but the 2-spin couplings will be negligible. The source amplitude  gets renormalised ${\mathsf{J}}_0\to {\mathsf{J}}_{0,r}$ via Eq.~\eqref{eq:source_ren} yielding
 \beq
  \label{eq:sources_harmonic_renormalised}
  {J}_r(x)={\mathsf{J}}_{0,r}\sin(k_Jx)={\mathsf{J}}_{0,r}\sin\big(\omega_Jt-\textsf{\textbf{k}}_J\cdot\textsf{\textbf{x}}\big),
  \eeq
 To estimate  the strength of the $2n$-spin  interactions, we  recall our  discussion of Sec.~\ref{sec:KG_Ising_spins}, and the importance of tuning the sources below resonance $\omega_J\lesssim{m}_r\lesssim{\omega}_{\textsf{\textbf{k},r}}=(\textsf{\textbf{k}}^2+{m}_r^2)^{1/2}$ to avoid dissipative processes where propagating bosons instead of virtual ones get excited. As discussed there, the source couplings  should also  be constrained by Eq.~\eqref{eq:constraint_couplings} to effectively decouple the dynamics of the Ising and scalar fields. This allowed us to get an effective unitary evolution where the only non-trivial part was encoded in the  Ising Hamiltonian~\eqref{eq:eff_ham}. The 
 renormalised version of such a constraint~\eqref{eq:constraint_couplings} reads
\beq
\label{eq:constraint_couplings}
  {\mathsf{J}}_{0,r}\ll({\omega}_{\textsf{\textbf{k}},r}-\omega_{J}){\rm d}^d\hspace{-0.1ex}\textsf{{k}}< ({\omega}_{\textsf{\textbf{k}},r}+\omega_{J}){\rm d}^d\hspace{-0.1ex}\textsf{{k}},
  \eeq
  which guarantees that the back-action of the Ising spins onto the  $\lambda\phi^4$ field is negligible and that, if initialised  in the vacuum (or any other state), the scalar field will remain in such a vacuum during the whole evolution. The discussion of Sec.~\ref{sec:KG_Ising_spins} 
  then proceeds by performing explicitly the time-integrals involving the harmonic sources in the forward and backward directions of the Feynman propagator. As a result of these integrations and the  coupling constraint~\eqref{eq:constraint_couplings}, one can neglect rapidly oscillating terms, and describe the evolution unitary by a time-independent effective Hamiltonian~\eqref{eq:eff_ham}. We see in Eq.~\eqref{coupling_density} that the long-range spin-spin coupling scales with $\mathsf{J}_0\cdot \mathsf{J}_0/(\omega_{\textsf{\textbf{k}}}-\omega_J)(\omega_{\textsf{\textbf{k}}}+\omega_J)$, which is  already small according to the  constraint~\eqref{eq:constraint_couplings}. hence, the  dynamical effects of these interactions   can only be observed by letting the system evolve for sufficiently long times. For the current renormalised equations~\eqref{eq:renormalised_transition_matrix}, we get a similar behaviour, and one can check that the additional 4-$\sigma$ term in Eq.~\eqref{eq:renormalised_transition_matrix}  also contributes  to the time-independent  Hamiltonian, but this time with 4-spin interactions that scale with   terms like ${\mathsf{J}}_{0,r}\cdot {\mathsf{J}}_{0,r}\cdot {\mathsf{J}}_{0,r}\cdot {\mathsf{J}}_{0,r}/({\omega}_{\textsf{\textbf{k}}_1,r}\pm\omega_J)({\omega}_{\textsf{\textbf{k}}_2,r}\pm\omega_J)({\omega}_{\textsf{\textbf{k}}_3,r}\pm\omega_J)({\omega}_{\textsf{\textbf{k}}_4,r}\pm\omega_J)$. These terms should be integrated over in momentum space with a total momentum conservation, and will lead to spin interactions between 4 distant spins with some specific long-range coupling. In any case, Eq.~\eqref{eq:constraint_couplings} shows that these 4-spin interactions will be negligible in comparison to the 2-spin terms, and more so if one considers higher-order $2n$-spin  interactions.
  
  Once we have arrived at this point, the last step is  to convert the leading contribution to the amplitude of propagation from the path-integral formulation~\eqref{eq:renormalised_transition_matrix} onto a canonical time-evolution operator, reversing the steps of the path-integral construction. Altogether, when considering that the Ising spins only occupy a certain spatial arrangement~\eqref{eq:lattice_sources}, we arrive at the final result: the full time evolution can be represented by the two concatenated unitaries in Eq.~\eqref{eq:eff_unitary}, where the non-trivial dynamics of the Ising spins is controlled by a renormalised long-range quantum Ising Hamiltonian
  \beq
\label{eq:ren_ising_lattice}
{H}_{{\rm eff},r}=\frac{1}{2}\sum_{i=1}^n\sum_{j=1}^n{J}^r_{ij}\hspace{0.25ex}Z(\textsf{\textbf{x}}_i)Z(\textsf{\textbf{x}}_j)-h_{\textsf{t}}\sum_{i=1}^n X(\textsf{\textbf{x}}_i).
\eeq
Here, the  spin-spin couplings have a range that is controlled by the dimensionally-reduced Euclidean propagator 
  \beq
 \label{eq:spin_spin_couplings_lattice}
{J}^r_{ij}=-2{J}_{0,r}G_{{m}_{{\rm eff},r}}^{\rm E}\!(\textsf{\textbf{x}}_i-\textsf{\textbf{x}}_j)\cos\big(\textsf{\textbf{k}}_J\cdot(\textsf{\textbf{x}}_i-\textsf{\textbf{x}}_j)\big){J}_{0,r},
 \eeq
 with an effective renormalised mass given by
 \beq
 \label{eq:ren_eff_mass}
 {m}_{{\rm eff},r}^2={m}_r^2-\omega_J^2.
 \eeq
 Therefore,  tuning the frequency $\omega_J$ of the harmonic source~\eqref{eq:sources_harmonic_renormalised}  closer to, or further from, the renormalised mass $m_r$ of the scalar bosons,   one will be able to control the range of the Ising interactions, and extract the information about the renormalisation of the scalar field using the sensing method of Sec.~\ref{sec:KG_Ising_spins}. Let us recall again that this  mass gets additive and multiplicative renormalisations~\eqref{eq:mass_renormalisation} from all the intermediate scattering events of the  bosons  mediating the Ising interaction. Let us finally note that  the spin-spin coupling strength scales with the renormalised source amplitudes ${J}_{0,r}$, and that these terms also get a multiplicative renormalisation described in Eq.~\eqref{eq:source_ren}. Typically, for the decoupled $\lambda\phi^4$ QFT, such source renormalisations do not have any physical consequence, as one is ultimately interested in the renormalised propagators. The generating functional in these cases is a mathematical tool, and it suffices to take functional derivatives with respect to these new sources, or directly assume that the sources couple to the renormalised fields. On the contrary, for the current spin-scalar model, the renormalisation of the sources has an impact on the strength of the spin-spin couplings, which is relevant as these control the time-scale of the spin dynamics.

Considering Eq.~\eqref{eq:ren_ising_lattice} and our previous discussions, the new sensing protocol that monitors the real-time oscillations of a pair of Ising spins coupled to the $\lambda\phi^4$ QFT,  will allow us to recover the renormalised parameters of the field theory. Once we have them, one can reconstruct the generating functional in Eq.~\eqref{eq:ren_gen_function_quartic}, in this case to second-order in the sources
\beq
\mathsf{Z}[{J}_r]=\ee^{-\frac{1}{2}\!\!\bigintssss\!\!{\rm d}^Dx_1\!\!\bigintssss\!\!{\rm d}^Dx_2{J}_r(x_1)\Delta_{{m}_{ r}}\!(x_1-x_2){J}_r(x_2)}.
\eeq
 We note that, if the timescales of the ever-present experimental noise on the qubits is sufficiently low, it may also be possible to devise schemes that sense the effect of the 4-spin (generally 2$n$-spin) interactions, gaining access to higher-order propagators and the renormalised quartic coupling.

We close this section by noting that we have so far not mentioned any ultra-violet (UV) cutoff of the QFT, although it is implicit in the discretised drawings of Figs.~\ref{Fig:scheme_impulsive_sensor}-\ref{Fig:scheme_harmonic_sensor_4_spin}. Such a cutoff is not required for the results presented in Sec.~\ref{sec:KG_ising} for the effective Ising models mediated by free Klein-Gordon bosons. In this case, the only UV diverging quantity is the zero-point energies of the  fields.   However, as soon as $\lambda\phi^4$ interactions are included, a different kind of UV divergences appear, as we see that the  renormalisations involve certain Feynman diagrams, such as the tadpole term in Eq.~\eqref{eq:mass_reb_terms}, which includes the propagator at infinitesimally short distances
 ${\Delta}_{m_0}(0)$, which  displays such UV divergences. The QFT thus needs to be regularised by the introduction of a  cutoff 
\beq
\label{eq:UV_cutoff}
|k^\mu|\leq\Lambda_{\rm c}.
\eeq
 As a consequence, the  renormalised parameters~\eqref{eq:mass_renormalisation}-\eqref{eq:int_renormalisation} will depend on the cutoff scale. The central result of the renormalisation group is that the bare coupling constants of the QFT, $m_0,\lambda_0,J_0$ in this case, will flow with the cutoff scale and give rise to physical renormalised quantities ${m_r},{\lambda}_r,{J}_{0,r}$  that no longer depend on the arbitrary cutoff. As we will discuss below, in certain situations, this UV cutoff is not arbitrary but fixed  by the physical system at hand (e.g. physical lattice). In this case, the renormalisation group allows us to extract the universal long-wavelength properties, and find a predictive field theory valid at low energies~\cite{RevModPhys.66.129}. We will discuss this in more detail in the following section.

\section{\bf Renormalisation of sound in trapped-ion crystals}

In this section, we present a detailed discussion on how the previous results can be applied to a long-wavelength description of  crystals of atomic ions confined  in radio-frequency traps. This description can be formalised within the  framework of the  theory of  elasticity for the quantised sound waves of  crystals, which is reviewed in Appendix~\ref{sec:elastodynamics_solids}, where we also comment on  the main obstacles for a solid-state implementation of the Ising-Schwinger sensing scheme introduced in this work. We  show in Sec.~\ref{sec:elastodyn_ions} that the transverse sound waves of  harmonic trapped-ion  crystals subjected to state-dependent forces offer a neat realisation of the massive Klein-Gordon field coupled to  Ising spins. This allows us to apply the results of Sec.~\ref{sec:KG_Ising_spins},  connecting the predicted long-range Ising models with previous works on  phonon-mediated spin-spin interactions between trapped ions. This  clarifies the distance dependence  of the latter, and shows that our theory is a valid long-wavelength description of trapped-ion crystals under state-dependent forces. In Sec.~\ref{sec:anh_ions}, we abandon the harmonic limit by  approaching a structural phase transition of the ion crystal, where  non-linear effects lead to scattering of  the quanta related to the sound waves, namely the phonons,  which requires us to use  the renormalisation predictions of Sec.~\ref{sec:spin_scalalr_gen_functional}. In particular, in Sec.~\ref{sec:RG_predictions}, we use an explicit integration of the renormalisation-group flow equations in the limit of small quantum fluctuations to predict the renormalisation of the range of the Ising spin-spin interactions. We conclude in Sec.~\ref{sec:drivings} by showing that additional parametric modulations of the trap frequencies can give a knob to control the amount of quantum fluctuations, exploring  regimes that go beyond the previous perturbative predictions, where the results of the proposed QS would be very interesting.

\subsection{ Harmonic trapped-ion crystals:  rigidity, massive Klein-Gordon fields and  effective Ising models.}
\label{sec:elastodyn_ions}

In this section, we discuss the realisation of the previous ideas using crystals of trapped atomic ions~\cite{PhysRevX.7.041012}. Clouds of singly-ionised atoms can be confined in finite regions of space for hours/days  using storage rings or Penning and Paul traps~\cite{ghosh_1995}. These traps are held inside ultra-high vacuum chambers, and additionally shielded from external fluctuating fields, such that one can explore  quantum many-body properties of the ion cloud in a pristine and controlled environment.
In this way, one can observe the plasma-like behaviour of  large   thermal clouds~\cite{RevModPhys.71.87}, or focus on ions in medium- to small-sized clouds. In the latter context,  crucial progress in our understanding of laser cooling  during the past decades~\cite{PhysRevA.20.1521,PhysRevA.25.35,PhysRevA.36.2220, PhysRevA.46.2668,PhysRevLett.85.4458} has allowed the development of various techniques~\cite{Eschner:03} to reach ever-lower   temperatures. These ultra-cold atomic ions crystalize as a result of the competition between trapping and Coulomb forces~\cite{PhysRevLett.59.2931,PhysRevLett.59.2935,PhysRevA.45.6493,PhysRevLett.60.2022,Birkl1992}, leading to  the so-called Coulomb clusters/crystals~\cite{RevModPhys.71.87}.

In the following, we   shall focus on linear Paul traps,  a type of  radio-frequency traps that   confine  the ions in stable linear configurations  along the trap symmetry  axis~\cite{PhysRevA.45.6493}. These kind of traps have been exploited, among other things,  for   the  manipulation of individual quantum systems~\cite{RevModPhys.75.281}, the development of frequency standards based on optical clocks~\cite{RevModPhys.87.637}, and the demonstration of various quantum computing algorithms~\cite{primer_qc_ions,Schindler_2013,doi:10.1063/1.5088164}. In this configuration, the microscopic description of a single trapped ion~\cite{doi:10.1063/1.367318}, or its extension to a collection of them~\cite{Bermudez_2017}, shows how   slow secular vibrations  with a typical timescale set by the inverse of the trap frequencies $\{\omega_{\alpha}\}_{\alpha=\textsf{x,y,z}}$ get decoupled from a fast driven motion  synchronous with the external radio-frequency field $\Omega_{\rm rf}\gg\omega_\alpha$.

 For the ion crystal, the long-wavelength description of the collective secular dynamics~\cite{James1998}  resembles the elasto-dynamical theory of  phonons in a 1D solid-state crystal, as reviewed in Appendix~\ref{sec:elastodynamics_solids}, albeit with  important differences that we now stress. On the one hand,  while  strict 1D crystals are thermodynamically unstable,   the ion chains are dynamically-driven inhomogeneous  clusters of ions, such that there is no  conflict with the spontaneous breakdown of translational symmetry in static situations~\cite{PhysRev.158.383,PhysRevLett.17.1133,Coleman1973}. However, due to the separation of timescales mentioned above, the much slower secular motion can be discussed using static equilibrium tools. Secondly, whereas  1D crystals can only support compressional longitudinal waves~\eqref{eq:long_waves}, ions in a Coulomb chain can vibrate in both longitudinal and transverse directions with respect to the trap axis. One of the key advantages of trapped-ion crystals with respect to the solid state is that these different  vibrations can be selectively cooled or excited by controlling both the frequency and  the propagation direction of additional laser beams. This contrasts the typical situation in solids,  where a  macroscopic thermal strain   excites all modes simultaneously. The last key difference with respect to the elasto-dynamical description of sound waves in solids of Appendix~\ref{sec:elastodynamics_solids} is that, contrary to the screened inter-atomic potentials of solids, which are in accordance with the form of the coarse-grained  stress forces, trapped-ion crystals are subjected to long-range Coulomb potentials. These long-range forces will lead to important differences in the context of the present work. Additionally, trapped ions allow one to tune the analogue of the bulk and shear moduli $K_e,\mu_r$, which characterise the stiffness and rigidity of the crystal to external strain and, in turn, determine the effective speeds of sound as discussed in the appendix in  Eqs.~\eqref{eq:long_waves} and~\eqref{eq:transv_waves}. More importantly, trapped-ion crystals allow to control microscopically the importance of non-elastic corrections,  providing a neat route to explore  renormalised Ising interactions mediated by $\lambda\phi^4$ fields, as discussed in the following subsection.

  \begin{figure}[t]
 \begin{centering}
  \includegraphics[width=1\columnwidth]{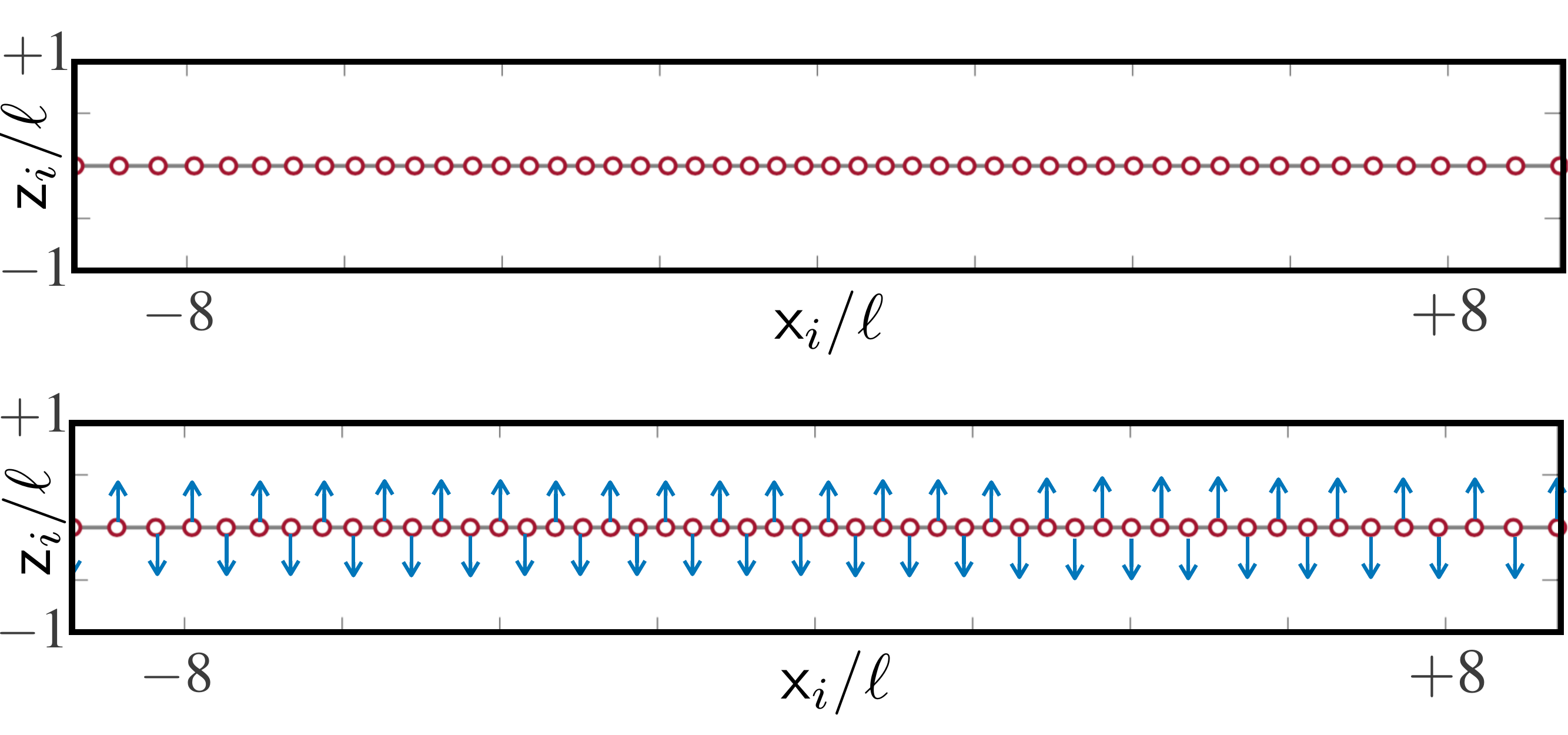}\\
  \caption{\label{Fig:eq_poitions} {\bf Trapped-ion chain and effective Ising couplings:} {\bf (a)} Equilibrium positions for a chain of  $N=50$ $^{171}$Yb$^+$ ions confined in a Paul trap with secular frequencies $\omega_{\mathsf{x}}/2\pi=0.1$MHz, and $\omega_{\mathsf{z}}/2\pi= 3.75$MHz. The microscopic length-scale corresponds to $\ell=12.7\mu$m in this case, while the minimum distance at the bulk of the ion chain is $a=4.4\mu$m.  {\bf (b)} The blue alternating arrows depict the lowest vibrational mode in the transverse branch, the zigzag mode. }
\end{centering}
\end{figure}

Let us start by describing in detail the first differences, which can already be understood  by inspecting the trapped-ion chain in the secular harmonic approximation~\cite{PhysRevLett.93.170602,PhysRevE.70.066141}, and the realisation of phonon-mediated Ising-type interactions thereof~\cite{PhysRevLett.92.207901,PhysRevA.72.063407}, making contact with Sec.~\ref{sec:KG_ising}. 
In the secular approximation, there is a competition between the  harmonic trapping potential and the Coulomb repulsion such that, for $\omega_{\textsf{x}}\ll\omega_{\textsf{y}},\omega_{\textsf{z}}$, the $N$ trapped ions arrange in equilibrium positions $\textsf{\textbf{x}}_i=\textsf{x}_i{\bf  e}_{\textsf{x}}$ along the trap symmetry axis  (see Fig.~\ref{Fig:eq_poitions}~{\bf (a)}). The position of each ion can thus be expressed as $\boldsymbol{r}_{i}(t)={\textsf{x}}_{i}{\bf e}_x+\sum_\alpha u_{i,\alpha}(t){\bf e}_\alpha$, where $u_{i,\alpha}(t)$ represents the small local vibrations around equilibrium of the $i$-th ion  along the $\alpha$-th axis, which can be considered as a discretised version of the displacement field $u_\alpha(t,\textsf{\textbf{x}})$ introduced in the coarse-grained elasticity theory~\eqref{eq:long_waves}. In a harmonic approximation, one expands the overall potential to second order, and finds that the  dynamics of the ion chain can be described by a Hamiltonian of local harmonic oscillators with long-range couplings 
\beq
\label{eq:coulomb_crystal_phonons}
H_h\!=\!\sum_{i,\alpha}\!\!\left(\!\frac{1}{2}m_a(\partial_tu_{i,\alpha})^2+\frac{1}{2}\kappa^\alpha u^2_{i,\alpha}+\frac{1}{4}\!\sum_{j\neq i }\kappa_{i,j}^\alpha\big(u_{i,\alpha}-u_{j,\alpha}\big)^{\!2}\!\right)\!.
\eeq
Here, $m_a$ is the mass of the atomic ions, and
\beq
\label{eq:harmonic_couplings_ions}
\kappa_{i,j}^{\mathsf{z}}=-\frac{e^2}{4\pi\epsilon_0}\frac{1}{|\textsf{x}_i-\textsf{x}_j|^3}=\kappa_{i,j}^{\mathsf{y}}=-\half \kappa_{i,j}^{\mathsf{x}},
\eeq
 can be understood as interatomic spring constants  with a dipolar decay due to the net charge of the ions. In addition, 
 \beq
 \kappa^\alpha=m_a\omega_{\alpha}^2
 \eeq
  is an effective spring constant leading to an elastic local force  that aims at restoring the  equilibrium, i.e. $ \boldsymbol{u}_{i}=\boldsymbol{0}$.

Let us note that the elasticity theory of Appendix~\ref{sec:elastodynamics_solids} relies on the short-range character of the interatomic forces. In  solid state,  this is typically modelled by  restricting the interatomic   springs to nearest neighbours  along the chain $\kappa_{i,j}^{\textsf{x}}\to \kappa_{i,i+1}^{\textsf{x}}$, and setting $\kappa^{\textsf{x}}=0$ as there is no additional trapping potential~\cite{goodstein_2017,ashcroft_2016}. Note that,  in order to describe a stable crystalline configuration, the nearest-neighbour couplings must be positive in this case $\kappa_{i,i+1}^{\textsf{x}}>0$. For the Coulomb chain~\eqref{eq:coulomb_crystal_phonons}, the situation is richer, as the non-zero  local spring constants $\kappa^\alpha>0$  allow for stable configurations with both attractive and repulsive elastic couplings  $\kappa^\alpha_{i,j}\gtrless0$. The attractive/repulsive character can be  understood by considering that the ion excursions from equilibrium define distant electric dipoles $\boldsymbol{P}_i=e\boldsymbol{u}_i, \boldsymbol{P}_{\!j}=e\boldsymbol{u}_j$, and that these interact  by repelling (attracting) each other when aligned orthogonal (parallel) to the axis of separation, here the $\textsf{x}$-axis, yielding $\kappa_{i,j}^{\textsf{y
}},\kappa_{i,j}^{\textsf{z}}>0$ ($\kappa_{i,j}^{\textsf{x}}<0$). This has a direct  consequence for the collective vibrational branches, as the lowest longitudinal vibration will correspond to all dipoles lying in parallel, which requires the ions to vibrate in phase and leads to the so-called center-of-mass mode~\cite{James1998}. On the other hand,  the lowest transverse vibrations correspond to alternating dipoles along the $\textsf{y,z}$ axis, such that the ions  vibrate out of phase and lead to the so-called zigzag mode (see Fig.~\ref{Fig:eq_poitions}~{\bf (b)}). For reasons that will become clear below, we will restrict to the transverse vibrations along the $\textsf{z}$ axis in a situation where the corresponding trapping potential has been partially relaxed $\omega_{\textsf{x}}<\omega_{\textsf{z}}\ll\omega_{\textsf{y}}$.

We can now address the connection of this microscopic theory~\eqref{eq:coulomb_crystal_phonons} to a Klein-Gordon field theory~\eqref{eq:KG_field} at long wavelengths or, equivalently, low energies. As noted above, the lowest-energy transverse vibration will correspond to out-of-phase  displacement of the ions, which can be associated to a  quasi-momentum $\textsf{k}_{s}=\pi/a$. Here, $a$ is  a characteristic microscopic length-scale of  the Coulomb chain~\cite{James1998} that will be proportional to the combination of microscopic parameters 
\beq
\label{eq:lengthsclae}
\ell=(e^2/4\pi\epsilon_0 m_a\omega_{\textsf{x}}^2)^{1/3}.
\eeq
 Although we could carry on in the most generic situation~\cite{PhysRevX.7.041012}, in order to make a more direct contact with the  scalar QFT of the previous sections, we shall assume that the ions are homogeneously distributed in the chain $\textsf{x}_i=ia$ with inter-ion  distance, i.e. lattice spacing, $a$ set by the minimum distance between the bulk ions (see Fig.~\ref{Fig:eq_poitions}~{\bf (a)}). This sets  directly the  aforementioned microscopic length scale, which  turns out to be  $a\lesssim\ell$. The homogeneous  approximation is valid for the bulk of the ion crystal,  neglecting the differences that arise due to inhomogeneities as one approaches the chain edges~\cite{RevModPhys.71.87}, which  is consistent with the spirit of our long-wavelength theory. We will thus apply periodic boundary conditions, and  discuss numerically  the inhomogeneous  chain at the end.

 As  is customary  in  other long-wavelength descriptions of  condensed-matter problems~\cite{Affleck1990,Shankar}, one proceeds by separating fast and slow variations of the degrees of freedom of interest~\cite{PhysRevX.7.041012,PhysRevLett.106.010401,PhysRevLett.109.010501}, here the transverse displacement
\beq
\label{eq:scalar_fields_ions}
u_{j,\textsf{z}}(t)=a\ee^{-\ii \textsf{k}_s \textsf{x}_j}\tilde{\phi}\!(t,\mathsf{x}),\hspace{2ex} \partial_t u_{j,\textsf{z}}(t)=\frac{1}{m_a}\ee^{-\ii \textsf{k}_s \textsf{x}_j}\tilde{\pi}(t,\mathsf{x}),
\eeq
where $\tilde{\phi}(t,{\textsf{x}})$ is the slowly-varying component  that will play the role of the scalar field in the continuum ${ \textsf{x}_j}\to {\textsf{x}}$, and $\tilde{\pi}(t,\mathsf{x})$ is its conjugate momentum. These operators are defined such that one recovers the correct equal-time commutation relations $[\tilde{\phi}\!(t,\mathsf{x}_i),\tilde{\pi}(t,\mathsf{x}_j)]=\ii\hbar\delta_{i,j}/a\to\ii\hbar\delta(\textsf{{x}}-\textsf{{x}}')$ in the continuum  limit, corresponding to the canonical commutator of the scalar field defined below Eq.~\eqref{eq:KG_field} for natural units. Note, however, that we need to work in  SI units as we have not yet identified the quantity that will play the role of the speed of light. Accordingly, we must rescale the fields through a canonical transformation to achieve the correct scaling dimensions of the Klein-Gordon model, which for $D=1+1$ dimensions is
\beq
\label{eq:rescaling_SI}
{\phi}(t,\mathsf{x})=\frac{1}{\sqrt{m_aa}}\tilde{\phi}(t,\mathsf{x}), \hspace{2ex} {\pi}(t,\mathsf{x})=\sqrt{m_aa}\tilde{\pi}(t,\mathsf{x}).
\eeq

Since the   field varies slowly, one proceeds by performing a gradient expansion to capture the  long-wavelength physics
\beq
\label{eq:gradient_expansion}
{\phi}(t,{\textsf{x}_j})\approx{\phi}(t,{\textsf{x}_i})+({\mathsf{x}_j-\mathsf{x}_i})\hat{\nabla}_{\!\!\mathsf{x}_i}{\phi}(t,{\textsf{x}_i})+\cdots,
\eeq
where we have introduced the discretized  derivative 
\beq
\label{eq:discrete_differences}
\hat{\nabla}_{\!\mathsf{x}_i}{\phi}(t,{\textsf{x}_i})=\frac{{\phi}(t,{\textsf{x}_{i+1}})-{\phi}(t,{\textsf{x}_i})}{a}.
\eeq
As a consequence of this gradient expansion, the long-range terms~\eqref{eq:coulomb_crystal_phonons}  effectively reduce to nearest-neighbour couplings, leading to a clear similarity with the  solid-state harmonic crystal. As will become clear below, the long-range couplings should decay sufficiently fast such that the corresponding series obtained by summing over all neighbours converge, and the truncated gradient expansion is thus meaningful. This procedure improves upon the so-called phonon-like approximation~\cite{PhysRevE.70.066141}, where one  truncates the dipolar tail to  nearest neighbours  without a previous  gradient expansion, and thus yields an incorrect  speed of sound. Although there exist other methods to obtain a more  accurate  dispersion relation at long wavelengths~\cite{PhysRevLett.93.170602,PhysRevE.70.066141}, the current gradient expansion is likely the simplest and most-economic one. Substituting the  expressions~\eqref{eq:scalar_fields_ions}-\eqref{eq:gradient_expansion} in the Coulomb-chain Hamiltonian~\eqref{eq:coulomb_crystal_phonons}, we find that the  vibrations along the transverse $\mathsf{z}$-axis can be described by 
\beq
\label{eq:scalar_qft_crystal}
H_{ h}\!=\!a\!\!\sum_{\mathsf{x}\in\Lambda_s}\!\!\frac{1}{2}\!\!\left(\frac{c_{\mathsf{t}}^2{\pi}^2\!(t,\mathsf{x})}{K^2\hbar^2}+K^2\hbar^2\!({\hat{\nabla}_{\!\!\mathsf{x}}{\phi}\!(t,\mathsf{x}))^2}+{m}_0^2c_{\mathsf{t}}^2K^2\!{\phi}^2\!(t,\mathsf{x})\!\!\right)\!\!,
\eeq 
where $\Lambda_s$ stands for the positions of the ions in the chain, and we have introduced the parameters
\beq
\label{eq:mic_parameters}
 c_{\textsf{t}}=\omega_{\textsf{x}}a\sqrt{\frac{\ell^3}{a^3}\log 2 },\hspace{1.5ex}m_{0}=\frac{\hbar\omega_{\textsf{zz}}}{c_{\textsf{t}}^2},\hspace{1.5ex} K=\frac{m_aac_{\textsf{t}}}{\hbar},
 \eeq
 where  the zigzag mode frequency reads
 \beq
 \label{eq:zz_freq}
 \omega_{\textsf{zz}}^2=\omega_{\textsf{z}}^2-\frac{7}{2}\zeta(3)\omega_{\textsf{x}}^2\frac{\ell^3}{a^3},
 \eeq
  and  we have made use of the Riemann zeta function
 \beq
  \zeta\!(z)=\sum_{r=1}^\infty\frac{1}{r^z}.
 \eeq 
 
 The physical interpretation  of these parameters  is now  given. In the continuum limit, one gets $a\sum_{\mathsf{x}\in\Lambda_s}\to\int\!{\rm d}{\mathsf{x}}$, such that the term inside the parenthesis of Eq.~\eqref{eq:scalar_qft_crystal} can be identified with  a Hamiltonian density. After a   canonical rescaling  $\phi(x)\to\phi(x)/{K}$,  $\pi(x)\to{K}\pi(x)$, this Hamiltonian  density can be derived from   $\mathcal{L}_{\rm KG}=\half(\hbar^{2}\partial_\mu\phi\partial^\mu\phi-m_0^2c_{\textsf{t}}^2\phi^2)$, which is the Klein-Gordon Lagrangian density in SI units. Hence, one clearly sees  that $c_{\textsf{t}}$ plays the role of an effective speed of light, and $m_0$ of an effective bare mass. To understand the role of the dimensionless parameter  $K$, we need to use a different normalisation of the fields~\eqref{eq:rescaling_SI}, such that the canonical algebra is $[{\phi}\!(t,\mathsf{x}_i),{\pi}(t,\mathsf{x}_j)]=\ii\delta_{i,j}/a\to\ii\delta(\textsf{{x}}-\textsf{{x}}')$,  the scalar field is dimensionless, and its conjugate momentum has units of inverse length. This is achieved by the  transformation  $\phi(x)\to\phi(x)({\frac{{c_{\textsf{t}}}}{K\hbar}})^{1/2}$,  $\pi(x)\to\hbar\pi(x)({\frac{K\hbar}{{c_{\textsf{t}}}}})^{1/2}$ which, in the massless limit $m_0=0$, leads to the  Hamiltonian density $\mathcal{H}_{\rm cb}=\frac{\hbar c_{\textsf{t}}}{2}\big(\frac{1}{K}\pi^2+K(\nabla\phi)^2\big)$. Note that, in this case,  this Hamiltonian density corresponds to the field theory of a conformal boson with  Tomogana-Luttinger parameter $K$~\cite{fradkin_2013}.  This parameter   appears in the treatment  of  strongly-correlated 1D models~\cite{PhysRevLett.47.1840,giamarchi_2010},  where  various gapless phases of matter arising from very different microscopic models~\cite{PhysRevB.12.3908,PhysRevB.14.2989,PhysRevLett.45.1358}  share fundamental properties  with the groundstate of an interacting-fermion model~\cite{Tomonaga,Luttinger}. This leads to  the concept of the Tomonaga-Luttinger liquid~\cite{Haldane_1981} where, regardless of the nature of the fundamental constituents, the relevant low-energy excitations are  bosonic particle-hole pairs with a long-wavelength description in terms of this conformal boson QFT with additional local interactions: the so-called sine-Gordon model~\cite{PhysRevD.11.2088}. Due to the very nature of these interactions,  the  boson field can be compactified on a circle with a  radius that depends on the Tomonaga-Luttinger parameter $R=1/\sqrt{4\pi K}$ which, in turn, controls the power-law decay of two-point functions of the original microscopic degrees of freedom~\cite{senechal}.  In the present situation, the additional local interactions in the trapped-ion crystal shall not be those of the sine-Gordon model~\cite{PhysRevD.11.2088}, such that there is no reason to compactify the scalar field. Moreover, as the original degrees of freedom are already a discretised version of the bosonic field, the Luttinger parameter will not appear in the two-point functions. In fact, one can readily see that this parameter does not modify the Heisenberg equations of motion of the scalar field, and will thus not appear in the dispersion relation. In the following subsection, we shall see how it plays an important role in the presence of $\lambda\phi^4$ interactions, controlling the amount of quantum fluctuations.
However, prior to that, let us  give some additional insight on the derivation and physical origin of the  microscopic expressions for $c_{\mathsf{t}},{m}_0$, and $K$.


 The effective speed  of light  $c_{\mathsf{t}}$ can also be derived from a coarse-grained elasto-dynamical theory of the ion crystal. As discussed around  Eq.~\eqref{eq:transv_waves} of the Appendix, deformations with shear strain lead to  sound waves that do not alter the length of  chain, but describe shape deformations. The speed of such    transverse    sound waves  can be expressed in terms of  the crystal density $\rho$ and the shear modulus $\mu_r$ through
 \beq
 c_{\mathsf{t}}=\sqrt{\frac{\mu_r}{\rho}}.
 \eeq
 In this case, the shear modulus measures  the rigidity of the  ion chain  to  changes to its linear shape, as opposed to the   compression/expansion  that would be described by the  bulk modulus and lead to longitudinal sound waves. Within this gradient expansion, the shear modulus reads
\beq
\label{eq:shear_modulus}
\mu_r=\frac{1}{2a}\sum_{j\neq i}\kappa^{\textsf{z}}_{i,j}|\textsf{x}_j-\textsf{x}_i|^2\cos(\mathsf{k}_s|\textsf{x}_j-\textsf{x}_i|)=m_a\omega_{\textsf{x}}^2a\frac{\ell^3}{a^3}\log2
\eeq
 where, in the last equality,  we have taken the thermodynamic limit $N\to\infty$, substituted $k_s|\textsf{x}_j-\textsf{x}_i|= \pi|j-i|$, and summed the corresponding series $\sum_{r=1}^\infty(-1)^r/r=-\log2$ . Within this homogeneous approximation, the mass density of the ion chain is $\rho=m_a/a$, and the transverse  speed of sound 
 \beq
 \label{eq:transevrse_speed_sound}
 c_{\textsf{t}}=\sqrt{\frac{e^2}{4\pi\epsilon_0 }\frac{\log2}{m_aa }},
 \eeq
 which leads directly to the corresponding expression in Eq.~\eqref{eq:mic_parameters}, upon using the length scale in Eq.~\eqref{eq:lengthsclae}.

 
Let us now derive the expression for the bare mass or, equivalently, the  Compton wavelength $\xi_0$  associated to these massive scalar bosons. By  multiplying the first and third terms in Eq.~\eqref{eq:scalar_qft_crystal} and, likewise, in the trapped-ion  Hamiltonian~\eqref{eq:coulomb_crystal_phonons}, we find the following expression
\beq
\xi_{0}^{\!-2}=\frac{m_0^2c_{\textsf{t}}^2}{\hbar^2}\!=\!\frac{1}{c_{\textsf{t}}^2}\!\!\left(\!{\!\omega_{\textsf{z}}^2-\!\sum_{j\neq i}\!\frac{2\kappa_{j,i}^{\mathsf{z}}(-1)^{j-i}}{m_a}\sin^2\!\!\left(\half \mathsf{k}_s|\textsf{x}_j-\textsf{x}_i|\right)}\!\!\right)\!.
\eeq
Once again, taking the thermodynamic limit in the homogeneous bulk for the ion chain, we find
\beq
\label{eq:compton_wavelength}
\xi_{0}=a\frac{\omega_{\textsf{x}}}{\omega_{\textsf{zz}}}\sqrt{\frac{\ell^3}{a^3}\log 2},
\eeq
where the zigzag mode frequency~\eqref{eq:zz_freq} is obtained by summing the series $\sum_{r=1}^\infty(1-(-1)^r)/r^3=7\zeta(3)/4$.
 This Compton wavelength~\eqref{eq:compton_wavelength} will be a key quantity below, and  leads directly to the bare mass of Eq.~\eqref{eq:mic_parameters}.

 Let us now give a physical interpretation of the Tomonaga-Luttinger parameter. First of all,  the conformal boson is a self-dual QFT under  the transformation $\pi(x)\leftrightarrow\phi(x)$, $K\leftrightarrow1/K$, such that the definition of the Tomonaga-Luttinger parameter as $K$ or  $K^{-1}$  depends on  convention.  We follow the choice in~\cite{PhysRevX.7.041012,fradkin_2013}, although we note that it is also customary to take the inverse convention in condensed matter~\cite{giamarchi_2010,Shankar}. Also note that the normalization of the fields may differ, and one often finds definitions $K\to K/2\pi$~\cite{bosonization}. With these different choices, one can  relate the Tomonaga-Luttinger parameter to   the sound speed and the thermal compressibility of a 1D harmonic chain~\cite{doi:10.1142/1882,gonzalez_1995}. With our current convention, and considering the transverse nature of the trapped-ion displacements, this relation   would involve the rigidity modulus instead 
 \beq
 \label{eq:therm_K}
 c_{\textsf{t}}K=\frac{\mu_ra^2}{\hbar},
 \eeq
  which  leads directly to the corresponding parameter of Eq.~\eqref{eq:mic_parameters}. Being proportional to the shear modulus, the Tomonaga-Luttinger parameter is thus a measure of the rigidity of the ion chain to shear deformations. As will be discussed in more detail in the following section, given the current microscopic interpretation in terms of shear  rigidity, it is not surprising that the Tomonaga-Luttinger parameter gets renormalised  by additional anharmonicities. Indeed, prior to the canonical rescaling mentioned below Eq.~\eqref{eq:scalar_qft_crystal}, the coarse-grained Lagrangian of the ion chain would be $\mathcal{L}_{\rm KG}=\frac{K^2}{2}(\hbar^{2}\partial_\mu\phi\partial^\mu\phi-m_0^2c_{\textsf{t}}^2\phi^2)$, which can be readily identified with our discussion of the wavefunction renormalisation for the $\lambda\phi^4$ theory in the Appendix below Eq.~\eqref{eq:wavefunction_renormalisation}. The difference here is that, for the trapped-ion chain, the Tomonaga-Luttinger parameter can be interpreted as a wavefunction renormalisation $K^2\leftrightarrow z_\phi^{-1}$ that occurs already in the free theory, and is a consequence of the underlying rigidity of the ion chain. As noted above, from this perspective, it is not surprising that this rigidity gets renormalised by anharmonicities, which gives a physical interpretation of the field-theoretic multiplicative renormalisations~\eqref{eq:source_ren} due to the sunrise Feynman diagram. Indeed, such 'non-interacting multiplicative renormalisations' are directly taken care of by the parameter definition of the coarse-grained theory~\eqref{eq:scalar_qft_crystal}.

Let us now present an alternative account of these results that will be useful later. Since the harmonic trapped-ion Hamiltonian~\eqref{eq:coulomb_crystal_phonons} is quadratic, one can derive an exact description of the collective vibrational branches in the homogeneous limit by a simple Fourier transform~\cite{ashcroft_2016}. The  collective-mode frequencies for transverse $\mathsf{z}$ vibrations  are 
\beq
\label{eq:full_dispersion}
\omega^2({\textsf{k}})={\omega_{\mathsf{z}}^2+\sum_{j\neq i}\frac{2}{m_a}\kappa_{j,i}^{\mathsf{z}}\sin^{2}\left(\half\mathsf{k}|\textsf{x}_j-\textsf{x}_i| \right)},
\eeq
where the momentum belongs to the  first Brillouin zone ${\rm BZ}=\{\textsf{k}=\frac{2\pi}{Na}n\hspace{0.2ex}:\hspace{0.2ex} n\in\mathbb{Z}_N\}$. In the thermodynamic limit,  the exact dispersion relation is expressed in terms of the Riemann zeta function and the series representation of the polylogarithm
\beq
\label{eq:series_representation_polylog}
 {\rm Li}_n(z)=\sum_{r=1}^{\infty}\frac{z^r}{r^n}, 
 \eeq
 where $z\in\mathbb{C}$ and the radius of convergence is $|z|\leq1$~\cite{000002668}. In the present case,  the polylogarithms are evaluated at the $N$ roots of unity to yiled the dispersion relation
\beq
\label{eq:full_dispersion_relation}
\omega^2({\textsf{k}})=\omega_{\mathsf{z}}^2+\omega_{\textsf{x}}^2\frac{\ell^3}{a^3}\!\left(2\zeta\!(3)-{\rm Li}_3\!\big(\ee^{\ii \textsf{k} a}\big)-{\rm Li}_3\!\big(\ee^{-\ii \textsf{k} a}\big)\right).
\eeq
On can now Taylor expand the polylogarithms around  the lowest-energy mode for the transverse vibrations $\textsf{k}a\approx\pi+\delta\textsf{k}a$ for $|\delta\textsf{k}|\ll\pi/a$, leading to
\beq
\label{eq:dispersion_ions}
\omega^2({\delta\textsf{k}})\!\approx\omega_{\mathsf{z}}^2-\omega_{\textsf{x}}^2\frac{\ell^3}{a^3}\!\!\!\left(\!\frac{7}{2}\zeta(3)-(\delta\textsf{k}a)^2\log 2\!\!\right)\!\!=\!\frac{{m}_0^2c_{\textsf{t}}^4}{\hbar^2}+c_{\textsf{t}}^2\delta{\textsf{k}}^2\!,
\eeq
where one obtains the same expressions for the effective speed of light~\eqref{eq:transevrse_speed_sound} and Compton wavelength~\eqref{eq:compton_wavelength}. As remarked above, the dispersion relation gives us no information about the Tomonaga-Luttinger parameter, which can  be either extracted from the gradient expansion, or using the above relationship $c_{\textsf{t}}K=\mu_ra^2/\hbar=c_{\textsf{t}}^2m_aa/\hbar$. It is  interesting to note that, while the longitudinal vibrational modes for a harmonic crystal with dipolar couplings do not lead to a well-defined sound speed $\omega_{\delta\textsf{k}}^2\approx|\textsf{k}|^2\log|\textsf{k}|$~\cite{ashcroft_2016}, the transverse sound speed of the ion crystal is perfectly valid~\eqref{eq:dispersion_ions} due to the repulsive nature of the harmonic couplings~\eqref{eq:coulomb_crystal_phonons} and the out-of-phase oscillations of neighbouring ions in the lowest-energy  mode. This is the underlying reason for our choice to focus on the transverse vibrations of the ions along the $\textsf{z}$ axis.

 We have seen that the transverse displacement of each ion can be described by  a  scalar quantum field discretised in a lattice formed by the positions of the ions in the Coulomb crystal. In addition, at each of these positions, 
we have all the electronic degrees of freedom of each ion, among which we shall select only a pair of energy levels $\{\ket{0_{\textsf{{x}}}}=\ket{\uparrow_{\textsf{{x}}}}, \ket{1_{\textsf{{x}}}}=\ket{\downarrow_{\textsf{{x}}}}\}$. These will typically correspond to  long-lived levels in the ground- and/or metastable-state manifolds. The specific choices lead to the so-called hyperfine, Zeeman, optical, or fine-structure qubits~\cite{doi:10.1063/1.5088164}, where we remind that $\omega_0$ is the transition frequency between the two corresponding  levels. Using the discretised version of the orthogonal  projectors introduced around~\eqref{eq:Ising_source_coupling}, $Q(t,\mathsf{x})=\ket{\uparrow_{\textsf{x}}}\bra{{\uparrow_{\textsf{x}}}}=1-P(t,\mathsf{x})$, the dynamics of the internal and motional degrees of freedom is described by
\beq
H_0=H_{h}+a\!\sum_{\mathsf{x}\in\Lambda_s}\delta\epsilon(\mathsf{x})Q(t,\mathsf{x}), \hspace{1.25ex} \delta\epsilon(\mathsf{x})=\frac{\hbar\omega_0}{a}.
\eeq 
where the vibrational Hamiltonian in the long-wavelength harmonic approximation   corresponds to Eq.~\eqref{eq:scalar_qft_crystal}. According to our long-wavelength description based on spin path integrals, this system will then correspond to $S=\half$ scalar-sigma model.

The coupling of the qubits to the vibrational degrees of freedom is realised by shining  electromagnetic radiation into the ion crystal, which stems  from additional laser or microwave sources, again depending on the chosen qubit type~\cite{doi:10.1063/1.5088164}. In order to achieve the form of the Ising-Schwinger sources, either  in the $Z(x)$ basis~\eqref{eq:Ising_Z_source_coupling} or any other Pauli basis, one typically works in the  Lamb-Dicke regime of resolved sidebands~\cite{RevModPhys.75.281}. Here,  one selects the  propagation  direction and frequency of the radiation to couple  selectively  to the desired vibrational branch by creating/annihilating a single phonon. The desired Ising-Schwinger sources can be achieved by using a M\o lmer-S\o rensen~\cite{PhysRevLett.82.1971} or Leibfried~\cite{Leibfried2003} scheme, both of which generate a so-called state-dependent  force that underlies the realization of multi-qubit entangling gates for quantum computation~\cite{Soderberg_2010}. For concreteness, following the discussion in the first part of this manuscript, we focus on $Z$-dependent  forces and hyperfine/Zeeman qubits, but remark that similar expressions can be found for other Pauli operators  depending on the qubit choice. Following the notation introduced in~\cite{PhysRevX.7.041012}, but sticking to the use of fields in  SI units, the corresponding spin-phonon coupling can be rewritten as
\beq
\hat{V}=a\!\sum_{\mathsf{x}\in\Lambda_s}\cos(\textsf{k}_s\mathsf{x})\frac{g}{a}\sin(\Delta\omega_L t-\Delta\textsf{\textbf{k}}_L\cdot\textsf{\textbf{x}})Z(t,\textsf{x})\phi(t,\textsf{x}).
\eeq
Here, $\Delta \textsf{\textbf{k}}_L$ ($\Delta \omega_L$) is the wavevector (frequency) of the beat note arising from a pair of  interfering laser beams, and $g=\hbar\Omega_L\Delta \textsf{\textbf{k}}_L\cdot {\bf e}_{\textsf{z}}\sqrt{m_aa^3}$ is the coupling strength between the qubits and the scalar field, which is controlled by the two-photon Rabi frequency $\Omega_L$~\cite{Leibfried2003}.
By a direct comparison with the harmonic Ising-Schwinger sources introduced in Eq.~\eqref{eq:lattice_sources}, we directly find that the harmonic dependence of the source is fully tunable via the laser's beat note
\beq
\label{eq:freq_source}
\omega_J=\Delta \omega_L,\hspace{1ex} \textsf{\textbf{k}}_J=\Delta \textsf{\textbf{k}}_L.
\eeq

Let us now discuss the role of the Tomonaga-Luttinger parameter. As already discussed above,  in order to arrive at the  corresponding expression of the Klein-Gordon QFT, we should use a canonical transformation of the fields    $\phi(x)\to\phi(x)/{K}$,  $\pi(x)\to{K}\pi(x)$. For a pure scalar theory, this rescaling has no effect on the dynamics, such that the Tomonaga-Luttinger parameter does not have any dynamical manifestation. However, as soon as we couple the scalar field to the Ising spins, the situation changes. Since the  Tomonaga-Luttinger parameter is a measure of the  harmonic internal forces that quantify the shear stress of the crystal, it does not appear directly in the bare coupling strength of the Ising-Schwinger sources, which can be understood as external sources that lead to shear strain. Note that, under the aforementioned canonical transformation  of the fields,   the Ising-Schwinger sources will get effectively rescaled by the Tomonaga-Luttinger parameter, which can be interpreted as Hook's law in disguise (i.e. strain proportional to stress)~\cite{landau_elasticity,thorne_blandford_2017}. Considering   the additional alternation due to the oscillating nature of the lowest-energy  $\textsf{k}_s=\pi/a$ vibrational mode, one finds that the amplitude of Ising-Schwinger sources is
\beq
\label{eq:IS_couplings_ions}
J_0 \to J_0(\mathsf{x})=-J_0\cos(\textsf{k}_s\mathsf{x}),\hspace{1ex} J_0=\frac{g}{K}.
\eeq
Hence, the Tomonaga-Luttinger parameter will appear in the strength of the boson-mediated Ising interactions~\eqref{eq:spin_spin_couplings}. This will be particularly important in  presence of non-linearities for the sound waves. As commented in the paragraph below~\eqref{eq:therm_K}, the Tomonaga-Luttinger parameter can be understood as a non-interacting wavefunction renormalisation, and one will get further  renormalisations when including interactions,  modifying the strength of the spin-spin couplings. 

  
Finally, to conclude with the discussion about the trapped-ion realisation, the additional transverse field $H_{\textsf{t}}(x)$ in Eq.~\eqref{eq:lattice_sources} can be  obtained by adding another source of radiation, either from additional laser beams, or from an external microwave source, tuned in resonance with the  carrier transition  instead of the vibrational sidebands~\cite{RevModPhys.75.281}. Altogether, the time-evolution unitary~\eqref{eq:unitaries_spin_Ising_model} would correspond to
\beq
\label{eq:dipole_force}
\hat{V}=\!\sum_{\mathsf{x}\in\Lambda_s}{J_0(\mathsf{x})}\sin(\Delta\omega_L t-\Delta\textsf{\textbf{k}}_L\cdot\textsf{\textbf{x}})K\phi(t,\textsf{x})Z(t,\textsf{x})-{h}_{\textsf{t}}X(t,\textsf{x}),
\eeq
where the transverse field $h_{\textsf{t}}=\hbar\Omega_0/2\ll\hbar\omega_0$ depends on the carrier Rabi frequency $\Omega_0$.

According to the generic results of Sec.~\ref{sec:KG_ising}, the source term~\eqref{eq:dipole_force} will lead to Ising  interactions mediated by the effective massive Klein-Gordon fields, which can be written as   
\beq
\label{eq:Ising_trapped_ions}
H_{\rm eff}=\frac{1}{2}\sum_{i=1}^n\sum_{j=1}^nJ_{ij}\hspace{0.25ex}Z(\textsf{{x}}_i)Z(\textsf{{x}}_j)-h_{\textsf{t}}\sum_{i=1}^n X(\textsf{{x}}_i).
\eeq
Here, the Ising couplings~\eqref{eq:spin_spin_couplings} must be updated  to account for the spatial modulation of the source  couplings~\eqref{eq:IS_couplings_ions}, yielding
\beq
\label{eq:spin_spin_couplings_1d}
J_{ij}=-2J_0(\mathsf{x}_i)G_{m_{\rm eff}}^{\rm E}\!(\textsf{{x}}_i-\textsf{{x}}_j)\cos\big(\textsf{{k}}_{J,\textsf{x}}(\textsf{{x}}_i-\textsf{{x}}_j)\big)J_0(\mathsf{x}_j).
\eeq
We recall that the  decay of the interactions with the distance is controlled by the dimensionally-reduced Euclidean propagator, which corresponds to  $D=1$ dimensions~\eqref{eq:Euclidean_propagator} in this case, with an effective mass~\eqref{eq:eff_mass}
 that depends on the frequency of the harmonic sources. Note, however, that all the expressions in Secs.~\ref{sec:KG_ising}-\ref{sec:ren_Ising_models} use natural units. In the context of trapped ions,   one can translate them into SI units by exchanging  the mass parameter in favour of  an  effective  Compton wavelength $m_{\rm eff}^2\to\xi_{\rm eff}^{-2}=\xi_{0}^{-2}-\Delta\omega_L^2/c_{\textsf{t}}^2$, such that
 \beq
 \label{eq:range_length}
 \xi_{\rm eff}= \frac{c_{\textsf{t}} }{\sqrt{{\omega_{\textsf{zz}}^2-\Delta\omega_{L}^2}}},
 \eeq
 where we  recall that the beat-note is always red-detuned with respect to the transverse vibrational branch $\Delta\omega_L\lesssim\omega_{\textsf{zz}}=\min\{\omega(\textsf{k}), \textsf{k}\in[0,2\pi/a)\}$.
In addition, to be consistent with the SI units, the Euclidean propagator in Eq.~\eqref{eq:spin_spin_couplings} should be rescaled as $G_{m_{\rm eff}}^{\rm E}\!(\textsf{{x}}_1-\textsf{{x}}_2)\to G_{m_{\rm eff}}^{\rm E}\!(\textsf{{x}}_1-\textsf{{x}}_2)/\hbar^2$.

We thus see that, as announced in the previous sections,  the trapped-ion setup offers an ideal scenario where to apply our results for the harmonic Ising-Schwinger sources in  Sec.~\ref{sec:KG_ising}, and the effective Ising interactions mediated by the scalar Klein-Gordon field. In fact, following the seminal works~\cite{PhysRevLett.92.207901,PhysRevA.72.063407}, there have been numerous trapped-ion experiments~\cite{Friedenauer2008,Islam2011,Britton2012,Islam583,Richerme2014,Jurcevic2014,Smith2016, Lanyon2017,PhysRevX.8.021012,monroe_review_2015} that have  explored the physics of phonon-mediated Ising-type interactions with an increasing number of qubits. Our results indicate that the long-wavelength QFT describing these experiments in the harmonic regime is indeed the previous scalar-sigma model~\eqref{eq:sclalar_spin_action_final} in the absence of self-interactions, unveiling the key role played by the generating functional of the massive Klein-Gordon QFT for a particular type of sources: the harmonic Ising-Schwinger sources~\eqref{eq:lattice_sources}. This  not only allows for a neat connection of recent trapped-ion experiments to the physics of relativistic QFTs, but shall also be useful in the two following aspects. On the one hand, it   sheds light on the specific decay law of the spin-spin couplings for sufficiently-large ion crystals, complementing the results presented in~\cite{PhysRevA.93.013625,PhysRevB.95.024431} as discussed in the following paragraph. On the other hand, the results presented in Sec.~\ref{sec:ren_Ising_models} show a very interesting path for the characterisation of  renormalisation effects in such mediated Ising interactions, which take place when the scalar bosons are not described by free Klein-Gordon fields, but instead through  a self-interacting $\lambda\phi^4$ QFT. The  following two sections will present a detailed account of that situation.

Let us then finish this section by  discussing  the effective range of the phonon-mediated spin-spin interactions in the harmonic approximation of the trapped-ion crystal. To simplify the discussion, we shall orient the laser beams such that $\Delta\textsf{\textbf{k}}_L\cdot\textsf{\textbf{x}}_i=0$, neglecting in this way the additional  frustration of the spin-spin couplings~\cite{PhysRevLett.107.207209,Bermudez_2012}. Using the explicit form of the  Euclidean propagator~\eqref{eq:yukawa_coupling_1d}, and substituting in Eq.~\eqref{eq:spin_spin_couplings} the trapped-ion expressions of the effective speed of light and the Compton wavelength, we find 
\beq
\label{eq:exp_tail}
J_{ij}=	-(-1)^{i-j}J_{\rm eff}\frac{\xi_{\rm eff}a^2}{\ell^3}\ee^{-\frac{|\textsf{x}_i-\textsf{x}_j|}{\xi_{\rm eff}}}.
\eeq
Here, we have defined a coupling strength
\beq
J_{\rm eff}=\frac{\hbar\Omega_L^2\eta_{\mathsf{x}}^2}{\omega_{\mathsf{x}}}\frac{2}{\log 2},
\eeq
where $\eta_{\mathsf{x}}=\Delta \textsf{k}_L\sqrt{\hbar/2m_a\omega_{\textsf{x}}}$ is the so-called Lamb-Dicke parameter. The crucial point is that, according to the expression of the effective Compton wavelength~\eqref{eq:range_length}, by controlling the detuning  of the laser beat note with respect to the resonance with the lowest-energy zigzag mode $\Delta\omega_L^2\lesssim\omega_{\textsf{zz}}^2$, one can control the exponential decay of the Ising interactions.

Interestingly,  expression~\eqref{eq:exp_tail} shows the same distance dependence as the exponential tail of the spin-spin couplings derived in~\cite{PhysRevA.93.013625}, where no reference  is made  to an effective QFT for the ion chain, nor to the role of the Feynman propagators or the generating functional. This agreement serves as a validation of our approach,  and gives a concrete physical interpretation of the decay length $\xi_{\rm eff}$ as the effective Compton wavelength of the bosons. It also clarifies that the alternation in~\eqref{eq:exp_tail}  is the remnant of the separation of fast and slow variations in the coarse graining~\eqref{eq:scalar_fields_ions}. We also note that  Eq.~\eqref{eq:exp_tail}  gives a more-accurate estimate of the  exponential decay~\eqref{eq:range_length} with respect to the approach of~\cite{PhysRevA.93.013625}, since the latter  uses additional approximations approximation for both the dispersion relation and the spin-spin coupling strengths.

On the other hand, the authors of~\cite{PhysRevA.93.013625} identify a key aspect of systems with long-range interactions: there can be additional contributions to the spin-spin interactions due to the  power-law couplings between the oscillators~\eqref{eq:harmonic_couplings_ions}, as also  corroborated by alternative perturbative studies~\cite{PhysRevB.95.024431}. In order to account for these terms,  we should work with the full lattice propagator in  continuous time  

  \beq
  \label{eq:feynman_propagator_latt}
 \Delta_{m_0}^{\rm lat}\!(x)=\int_{\hat{k}} \tilde{\Delta}_{m_0}^{\rm lat}\!(\hat{k})\ee^{-\ii kx},\hspace{2ex} \tilde{\Delta}_{m_0}^{\rm lat}\!(\hat{k})=\frac{\ii}{\hat{k}^2-\omega_{\textsf{z}}^2+\ii\epsilon},
 \eeq
where the lattice version of the 2-momentum is
\beq
\hat{k}=\left(\omega,\sum_{j\neq i}\frac{2}{m_a }\kappa_{j,i}^{\mathsf{z}}\sin^{2}\left(\half\mathsf{k}|\textsf{x}_j-\textsf{x}_i| \right)\right),
\eeq
and we use  the short-hand notation $\int_{\hat{k}}=\int_{\mathbb{R}}\frac{{\rm d}\omega}{2\pi}\frac{1}{N}\sum_{\textsf{k}\in{\rm BZ}}$. In the thermodynamic limit $N\to\infty$, the dimensionally-reduced Euclidean  lattice propagator reads
  \beq
    \label{eq:Euclidean_propagator_ions}
G_{m_{\rm eff}}^{\rm lat}\!(\textsf{{x}})\!=a\!\!\!\bigintssss_{\hspace{0.1ex}0}^{\!\!\frac{2\pi}{a}}\!\!\!\frac{{\rm d}\textsf{{k}}}{2\pi}\hspace{0.2ex}\frac{\ee^{\ii {\textsf{{k}}\textsf{{x}}}}}{\omega^2\!({\textsf{{k}}})-\Delta\omega_L^2},
 \eeq
 where $\omega({\textsf{{k}}})$ stands for  the   trapped-ion dispersion relation~\eqref{eq:full_dispersion} with the full contribution of the long-range couplings expressed in terms of  polylogarithms~\eqref{eq:full_dispersion_relation}, and we integrate over the Brillouin zone $\textsf{{k}}\in[0,2\pi/a)$, such that the low-energy zigzag mode $\textsf{{k}}_s=\pi/a$ lies exactly in the middle. The spin-spin couplings~\eqref{eq:spin_spin_couplings} for $\Delta\textsf{\textbf{k}}_L\cdot\textsf{\textbf{x}}_i=0$ read
 \beq
 \label{eq:spin_spin_ions}
J_{ij}=-\hbar\Omega_L^2\eta_{\textsf{x}}^22\omega_{\textsf{x}}G_{m_{\rm eff}}^{\rm lat}\!(\textsf{{x}}_i-\textsf{{x}}_j).
\eeq

  \begin{figure}[t]
 \begin{centering}
  \includegraphics[width=1\columnwidth]{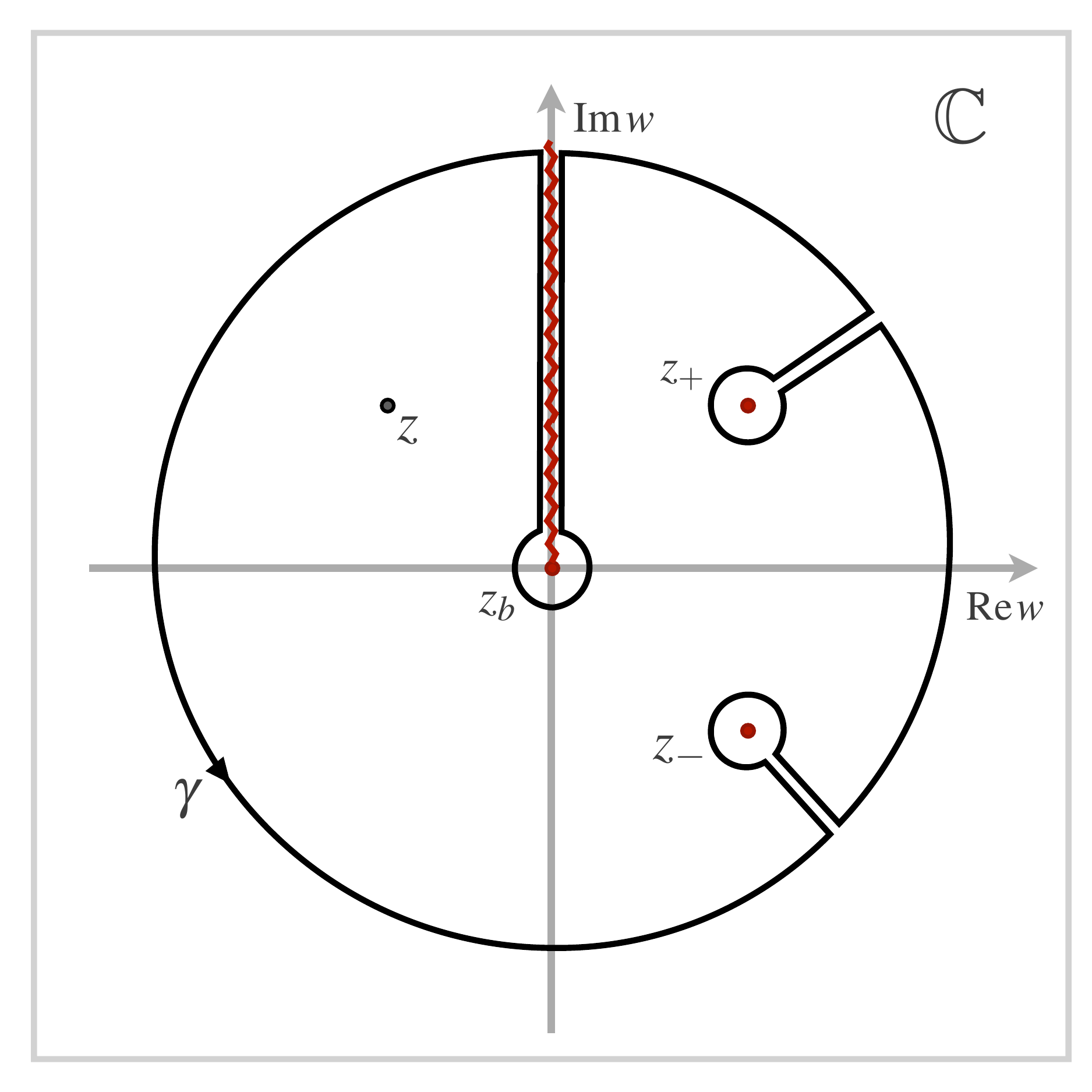}\\
  \caption{\label{Fig:contour} {\bf Cauchy's theorem and branch-cut structure:} We use Cauchy's theorem $f(z)=\frac{1}{2\pi i}\oint_\gamma{\rm d}w\frac{f(w)}{w-z}$ for the multiple keyhole contour $\gamma$, which avoids the isolated poles and the branch cut such that the function is analytic inside. The simple poles lie at $z_\pm=\pi\pm\ii/\xi_{\rm eff}$, where the effective Compton wavelength is given in Eq.~\eqref{eq:range_length}. This describes the broadening/screening of the zigzag mode $k_{s}a=\pi$. The residues of the poles are $\textsf{r}=\textsf{r}^*=\pi \xi_{\rm eff}a/c_{\textsf{t}}^2$, where the effective speed of light is given by Eq.~\eqref{eq:transevrse_speed_sound}. The branch point at $ee^{\ii z_b}=1$ is connected to the point at infinity by a branch cut along the positive imaginary axis, displaying a discontinuity that is calculated using the properties of the polylogarithms, leading to Eq.~\eqref{eq:dispersion_ions}.   }
\end{centering}
\end{figure}

The integral of Eq.~\eqref{eq:Euclidean_propagator_ions} can be evaluated   by extending the momentum $\textsf{k}a\to z$ to the complex plane $z=x+\ii y\in\mathbb{C}$. As discussed below, the properties of the lattice propagator can be understood via the analytic structure of $f(z)=1/(\omega^2\!(z)-\Delta\omega_L^2)$, where the complex-valued dispersion relation
\beq
\label{eq:complex_dispersion_relation}
\omega^2(z)=\omega_{\mathsf{z}}^2+\omega_{\textsf{x}}^2\frac{\ell^3}{a^3}\!\left(2\zeta\!(3)-{\rm Li}_3\!\big(\ee^{\ii z}\big)-{\rm Li}_3\!\big(\ee^{-\ii z}\big)\right)
\eeq 
requires analytical continuation of the polylogarithms  as one leaves the real axis, since $|\ee^{\pm\ii z}|=\ee^{\mp y}>1$ lies beyond the radius of convergence of the power series~\eqref{eq:series_representation_polylog}. Inspired by the analytical techniques underlying the K\"{a}ll\'en-Lehman spectral representation of the  2-point Feynman propagator~\cite{Kallen1952,Lehmann1954,Peskin:1995ev,zwicky2016brief}, we now provide a derivation of the  spin-spin couplings alternative to~\cite{PhysRevA.93.013625}. In this spectral representation, the full propagator of an interacting QFT can be described as a sum of free propagators, which  arise from either isolated single-particle  poles at the renormalised mass, or from a branch-cut discontinuity that appears above a certain threshold due to multi-particle excitations. This situation finds a clear parallelism in our case due to the power-law couplings, as $f(z)$ contains both a pair of complex-conjugate simple poles $z_-=(z_+)^*$, and a branch cut  connecting the branch point $z_b$ of the polylogarithms   at $\ee^{\pm\ii z_b}=1$  with the point at infinity along the imaginary axis (see Fig.~\ref{Fig:contour}). To understand why the corresponding contributions to the spin-spin couplings are simply added, one can apply Cauchy's theorem~\cite{stein_shakarchi_2005} along the multiple keyhole contour $\gamma$ displayed in this figure, which allows us to rewrite the complex function as follows
\beq
\label{eq:dispersion_ions}
f(z)=\frac{\textsf{r}}{z-z_+}+\frac{\textsf{r}^*}{z-z_-}+\int_0^\infty\!\!\!\frac{{\rm d}s}{2\pi}\frac{{\rm disc}(f(z))}{\ii s-z}.
\eeq
Here, $\textsf{r},\textsf{r}^*$ are the corresponding residues, and ${\rm disc}(f(z))={\rm lim}_{\epsilon\to 0^+}(f(\ii s+\epsilon)-f(\ii s-\epsilon))$ is the branch-cut discontinuity.  For any short-range discretization of the scalar Klein-Gordon field, which substitutes the derivatives by finite differences that only involve a finite number of neighbours like   Eq.~\eqref{eq:discrete_differences}, this discontinuity is absent and one only gets the contribution of the simple poles. On the other hand, for power-law couplings such as those in Eq.~\eqref{eq:harmonic_couplings_ions}, the complex dispersion relation~\eqref{eq:complex_dispersion_relation} contains polylogarithms  with the branch-cut discontinuity ${\rm Li}_n(\ee^{s}+\ii\epsilon)-{\rm Li}_n(\ee^{s}-\ii\epsilon)=2\pi\ii s^{n-1}/\Gamma(n)$ for $\epsilon\to0^+$ and $s>0$, where $\Gamma(n)$ is the Gamma function. Considering the simplified dependence on $z\in\mathbb{C}$ of the function~\eqref{eq:dispersion_ions},  calculating the Fourier transform in Eq.~\eqref{eq:Euclidean_propagator_ions} that yields the spin-spin couplings~\eqref{eq:spin_spin_ions} becomes direct.

 In the context of the  K\"{a}ll\'en-Lehman representation of the 2-point function of an interacting  QFT at zero temperature~\cite{Peskin:1995ev}, the single-particle pole  would be aligned along the real axis $z_+=z_-\in\mathbb{R}$,  leading  to a renormalised propagator corresponding to the term $\textsf{r}/(z-z_+)\to \ii\textsf{r}/( k^2-m_r^2+\ii\epsilon)$. The  branch cut, which would  also be aligned along the real axis, would start at a certain threshold above the single-particle pole where multi-particle excitations can be created by the interactions. The associated discontinuity, which is related to the continuous part of the so-called spectral function $\rho_c(s)$,  weights the contribution of each of the free Klein-Gordon propagators, which would correspond to the term $1/(z-\ii s)\to\ii/(k^2-M^2+\ii\epsilon)$, and be characterised by a mass $M$ above the multi-particle threshold of the interacting theory~\cite{Peskin:1995ev}. 
 
\begin{figure}[t]
 \begin{centering}
  \includegraphics[width=1\columnwidth]{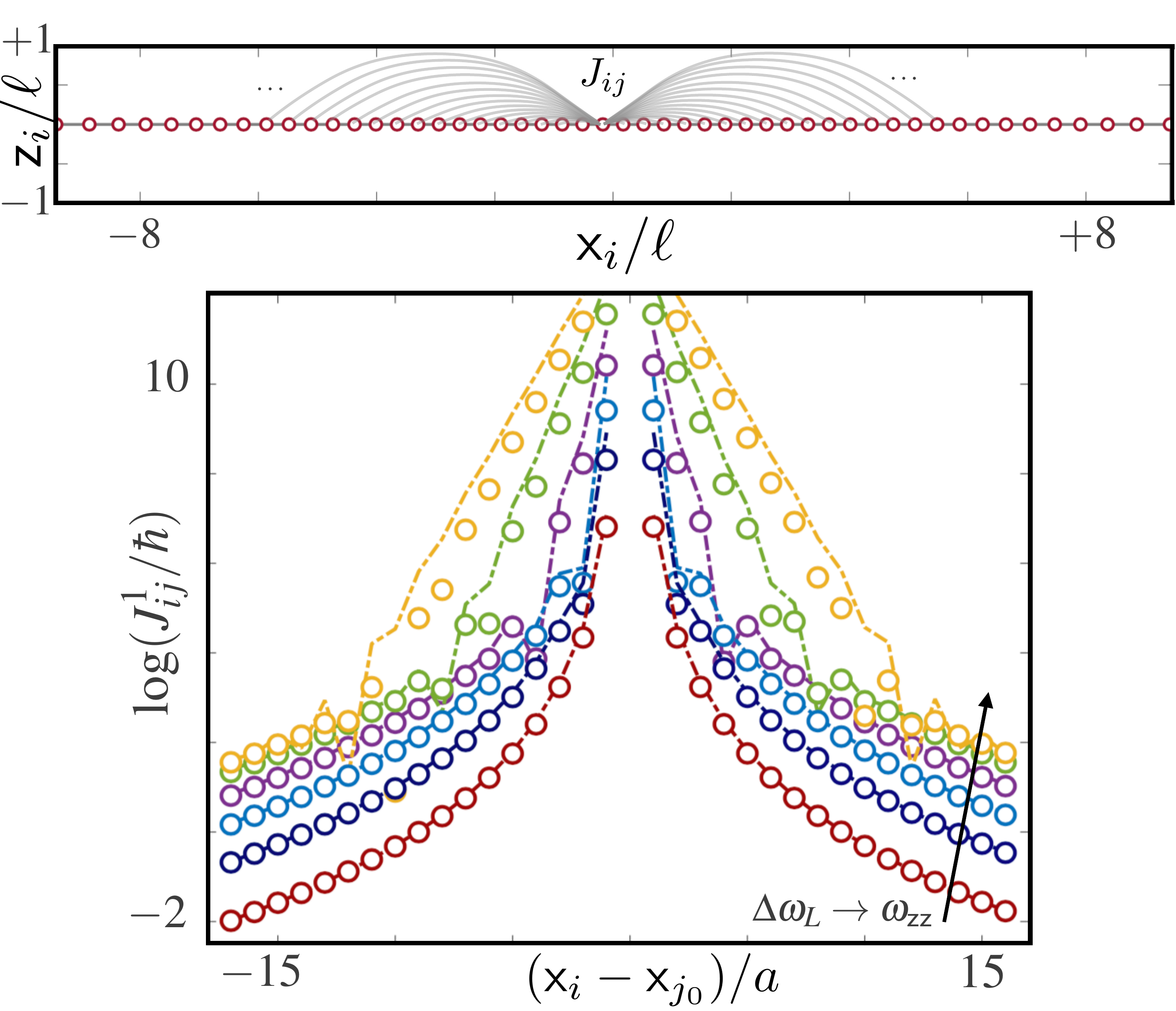}\\
  \caption{\label{Fig:spin_couplings} {\bf Trapped-ion chain and effective Ising couplings:} {\bf (a)} Equilibrium positions for a chain of  $N=50$ $^{171}$Yb$^+$ ions. The grey lines depict the long-range spin-spin couplings $J^1_{ij}$ between the centre-most ion $j_0=N/2$ and the rest of the chain. {\bf (b)} The exact spin-spin couplings, represented by circles, are calculated from $J^1_{ij}=|\Omega_L|^2E_r\sum_{n}\frac{\mathcal{M}_{i,n}\mathcal{M}_{jn}}{\Delta\omega_{L}^2-\omega_{\mathsf{z},n}^2}$, where  $\{\omega_{\mathsf{z},n},\mathcal{M}_{i,n}\}_{n=1}^N$ are the eigenvalues and eigenvectors obtained from the diagonalisation of Eq.~\eqref{eq:coulomb_crystal_phonons}, and $E_r=\mathsf{\mathbf{p}}_L^2/2m_a$ is the recoil energy of the ions due to the beat-note momentum $\mathsf{\mathbf{p}}_L=\hbar\Delta\textsf{\textbf{k}}_L$~\cite{monroe_review_2015}. We note that for the  $^{171}$Yb$^+$ ions, the Ising interactions typically use a M\o lmer-S\o rensen scheme, such that the roles of the $X, Z$ operators must be reversed. The dashed-dotted lines correspond to the coarse-grained predictions of Eq.~\eqref{eq:dip_exp_tail}  with no fitting parameter, but directly using the microscopic expressions of all the parameters derived in the main text, and substituting the homogeneous approximation $\mathsf{x}_i^0=ia$ for the inhomogeneous positions of {\bf (a)}. The different curves/data correspond to different detunings of the laser beat-notes with respect to the zigzag mode $(\omega_{\textsf{zz}}-\Delta\omega_L)/2\pi\in\{18.75,37,5, 93.75,187.5,937.5 \}$kHz. }
\end{centering}
\end{figure}

 Following this analogy, the additive structure for the spin-spin couplings has the following interpretation. The isolated poles give rise to an exponentially-decaying contribution, which coincides exactly with the long-wavelength result~\eqref{eq:exp_tail}. In addition,   due to the  branch-cut discontinuity, we get an additive contribution that can be understood as the superposition of each of the Euclidean propagators above threshold, which in this case corresponds to  the imaginary semi-axis. Thanks to the use of Cauchy's theorem~\eqref{eq:dispersion_ions}, there is no need to identify steepest-descent directions, nor to perform any further approximations as discussed in~\cite{PhysRevA.93.013625}, and the corresponding integral in $s$  leads to
 \beq
\label{eq:dip_exp_tail}
J_{ij}=	J_{\rm eff}\!\!\left(\!\!\!(-1)^{i-j+1}\frac{\xi_{\rm eff}a^2}{\ell^3}\ee^{-\frac{|\textsf{x}_i-\textsf{x}_j|}{\xi_{\rm eff}}}+\frac{\omega_{\textsf{x}}^4\log2}{(\omega_{\textsf{z}}^2-\Delta\omega_L^2)^2}\frac{\ell^3}{|\mathsf{x}_i-\mathsf{x}_j|^3}\!\!\right)\!\!\!.
\eeq
This expression shows that  for distances much larger than  the effective Compton wavelength~\eqref{eq:range_length}, $|\textsf{x}_i-\textsf{x}_j|\gg\xi_{\rm eff}$,  the decay of the spin-spin couplings is dominated by an antiferromagnetic 
 dipolar tail. As one approaches the resonance with the zigzag mode from below $\Delta\omega_L\to\omega_{\rm zz}$, the contribution of the alternating exponential becomes more pronounced, and there are deviations from the dipolar decay. In   Fig.~\ref{Fig:spin_couplings}~{\bf (b)}, we compare these analytical predictions to the exact expression of the Ising couplings for an inhomogeneous crystal of $^{171}$Yb$^+$ ions. As shown in this figure, there is a clear agreement of the spin-spin couplings even for an inhomogeneous lattice spacing.  
 
  We  note that using  the exact expressions for the dispersion relation and the boson-mediated Ising couplings, as we do in this work,  should give more accurate estimates of the spin-spin couplings, especially in the vicinity of the structural phase transition. Regardless of these quantitative details, our conclusion is the same as in~\cite{PhysRevA.93.013625}: although  phonon-mediated Ising couplings in small trapped-ion crystals can be approximated by a power-law decay  with a tunable exponent~\cite{Britton2012,Islam583,Richerme2014,Jurcevic2014,Smith2016, Lanyon2017,PhysRevX.8.021012},  theoretical predictions about static or dynamical effects that aim to describe  the thermodynamic limit should consider that Eq.~\eqref{eq:dip_exp_tail} is not a power-law decay with a tunable exponent~\cite{PhysRevB.95.024431}. Otherwise, some of the predicted  phenomena,  such as phase diagrams or quantum phase transitions that depend on the power-law exponent  would  not represent what can be  explored with trapped-ion experiments. 
 
\subsection{ Anharmonic trapped-ion crystals:   $\lambda\phi^4$ fields, sound renormalisation and  Ising models.}
  \label{sec:anh_ions}
  
  The real advantage of the present approach is that it  allows us to explore situations beyond the harmonic approximation of the trapped-ion crystal. Anharmonic corrections to the vibrational Hamiltonian~\eqref{eq:coulomb_crystal_phonons} become particularly relevant in the vicinity of  structural  transitions for    anisotropically-confined ion crystals~\cite{PhysRevLett.70.818}. By increasing  the ratio of the secular trap frequencies $\omega_{\textsf{x}}/\omega_{\textsf{z}}$ in the regime $\omega_{\textsf{y}}\gg\omega_{\textsf{z}}\gtrsim\omega_{\textsf{x}}$, the linear configuration becomes unstable towards  ladder-like structures with an increasing number of legs, which has been characterised in several experiments~\cite{Bermudez_2012,Ulm2013,Pyka2013,PhysRevLett.110.133004,PhysRevA.87.051401,PhysRevLett.119.153602}. The first structural change between the ion chain and a so-called zigzag ladder  corresponds to a second-order  phase transition taking place at some critical $\omega_{\textsf{x}}/\omega_{\textsf{z}}\big|_c$, which can be described by an effective Landau model~\cite{PhysRevB.77.064111}. In our work, we shall work in the regime $\omega_{\textsf{x}}/\omega_{\textsf{z}}\lesssim\omega_{\textsf{x}}/\omega_{\textsf{z}}\big|_c$, such that the linear configuration is stable, but the anharmonic corrections become  important.  In the context of  elasticity  theory, this is the so-called yield limit, where a small 
  increase in stress causes a large strain and one expects large deviations from Hook's law~\cite{thorne_blandford_2017}. Note that we are focusing on  the vicinity of the structural  phase transition in the linear configuration, as the zigzag structure will be accompanied by additional excess micromotion synchronous with the rf potential of the linear Paul trap~\cite{doi:10.1063/1.367318,Bermudez_2017}, and combining this driven motion with a coarse-grained QFT model lies beyond the scope of this work.
  
   In the vicinity of this phase transition, the most relevant  corrections $H=H_h+H_{a}$ to Eq.~\eqref{eq:coulomb_crystal_phonons},  coming from higher orders in the Taylor expansion of the Coulomb potential, are 
  \beq
\label{eq:coulomb_crystal_quartic}
H_{a}\!=\frac{1}{2}\!\sum_{i}\sum_{j\neq i }\frac{\beta_{i,j}^{\textsf{z}}}{4!}\big(u_{i,\textsf{z}}-u_{j,\textsf{z}}\big)^{\!4},
\eeq
where we have introduced the anharmonic coupling strengths
\beq
\label{eq:anharmonic_couplings_ions}
\beta_{i,j}^{\mathsf{z}}=\frac{e^2}{4\pi\epsilon_0}\frac{9}{|\textsf{x}_i-\textsf{x}_j|^5}.
\eeq
One now proceeds with the substitution of the gradient expansion in SI units~\eqref{eq:scalar_fields_ions}-\eqref{eq:rescaling_SI}, to find that the long-wavelength anharmonic correction to the massive Klein-Gordon QFT~\eqref{eq:scalar_qft_crystal} corresponds to a quartic self-interaction
\beq
H_{ a}\!\approx\!a\!\sum_{\mathsf{x}\in\Lambda_s}\frac{1}{4!}{\lambda_0K^4}{\phi}^4\!(t,\mathsf{x}).
\eeq 
Let us note that  the canonical rescaling $\phi(x)\to\phi(x)/K$, $\pi(x)\to K\pi(x)$ yields the standard  $\lambda\phi^4$ QFT in SI units
\beq
\label{eq:rescaled_QFT_ions}
H=a\!\!\sum_{\mathsf{x}\in\Lambda_s}\!\!\frac{1}{2}\!\!\left(\!\!\left(\frac{c_{\mathsf{t}}^2}{\hbar^2}\pi^2\!(x)+\hbar^2\!({\hat{\nabla}_{\!\!\mathsf{x}}{\phi}\!(x)^2}+{m}_0^2c_{\mathsf{t}}^2\!{\phi}^2\!(x)\!\!\right)\!\!+\frac{\lambda_0}{4!}{\phi}^4\!(x)\right),
\eeq
where we have introduced the quartic coupling
\beq
\label{eq:lambda_0_SI}
\lambda_0=8\frac{m_a^2}{K^4}\sum_{j\neq i}\!{\beta_{j,i}^{\mathsf{z}}}\sin^4\!\!\left(\half \mathsf{k}_s|\textsf{x}_j-\textsf{x}_i|\right),
\eeq
which reads as follows in the thermodynamic limit
\beq
\label{eq:coupling_strength_ions}
\lambda_0=\frac{279\zeta(5)}{2K^4}m_a^3\omega_{\textsf{x}}^2\ell^3.
\eeq

Once we have derived the complete coarse-grained theory, and found the microscopic trapped-ion expressions of the bare parameters in Eqs.~\eqref{eq:mic_parameters},~\eqref{eq:freq_source},~\eqref{eq:IS_couplings_ions}, and~\eqref{eq:lambda_0_SI}, we can directly apply the results of Sec.~\ref{sec:ren_Ising_models} to understand how the dynamics of the Ising spins changes due to the leading anharmonicities of the ion crystal in the vicinity of the structural phase transition. As concluded around Eq.~\eqref{eq:ren_ising_lattice}, the main contribution to the dynamics can still be described by the long-range quantum Ising model~\eqref{eq:Ising_trapped_ions}, but one must consider renormalised spin-spin couplings
\beq
\label{eq:spin_spin_couplings_ions}
{J}^r_{ij}=-2{J}_r(\mathsf{x}_i)G_{{m}_{{\rm eff},r}}^{\rm E}\!(\textsf{{x}}_i-\textsf{{x}}_j){J}_r(\mathsf{x}_j)
\eeq
 with renormalised sources~\eqref{eq:IS_couplings_ions}, and an effective renormalised mass that flows with the quartic coupling~\eqref{eq:ren_eff_mass}. As discussed in Sec.~\ref{sec:ren_Ising_models}, all these renormalisations account for the various scattering processes of the phonon that mediates the interactions, as it propagates between the  spins~\eqref{eq:spin_feynman}.

In order to connect to the perturbative calculations of Sec.~\ref{sec:ren_Ising_models}, which use  natural units, we recall that the effective mass term $m_{\rm  eff}$  should be substituted by an inverse length scale related to the Compton wavelength $1/\xi_{\rm eff}=m_{\rm eff} c_{\textsf{t}}/\hbar$ in SI units~\eqref{eq:range_length}. Likewise, in natural units, the renormalisation equations~\eqref{eq:int_renormalisation} indicate that the coupling constant $\lambda_0$ has natural dimension $d_{\lambda_0}=4-D=2$. Hence, it should be substituted by a combination of SI parameters $\lambda_0,\hbar,c_{\textsf{t}}$ that carries units of inverse length squared. In the present case, this leads to
\beq
\label{eq:lambda_0_ions}
\frac{\lambda_0c_{\textsf{t}}}{\hbar^3}=\frac{a^{-2}}{K}\frac{279\zeta(5)}{2\log 2}.
\eeq 
Accordingly, we can use  the perturbative  equations~\eqref{eq:mass_renormalisation}-\eqref{eq:int_renormalisation}
by substituting $\lambda_0\to{\lambda_0c_{\textsf{t}}}/{\hbar^3}$, as well as $m_{\rm eff}^2\to {m_{\rm eff}^2c_{\textsf{t}}^2}/{\hbar^2}$. However, we face the problem that was already alluded to, and that appears in  interacting QFTs: these perturbative corrections introduce UV divergences which must be dealt with by means of the renormalisation group (RG). In the present trapped-ion context, these divergences are an artefact of our coarse-grained description, and will always be regularised by the  physical UV  cutoff due to the underlying lattice
\beq
\label{eq:UV_cutoff_ions}
|\textsf{k}|\leq\Lambda_{\rm c}=\frac{\pi}{a},\hspace{1.5ex} |k^0|\leq c_{\textsf{t} }\frac{\pi}{a}.
\eeq
Note that this equation coincides exactly with  Debye's cutoff in solid-state crystals discussed in Eq.~\eqref{eq:Debye_cutoff} of the appendix, since the length of the 1D crystal would be $L=Na$ when $a$ is the lattice spacing.
The goal of the renormalisation group is then not to deal with infinities, but to extract the universal features of the  renormalised theory~\eqref{eq:spin_spin_couplings} that will not depend on short wavelengths on the order of the lattice spacing  $a$, but describe instead length scales $\xi\gg a$. We will deal with these RG predictions in the following section.


\subsection{Rigidity and renormalisation-group predictions}
\label{sec:RG_predictions}

The typical starting point of the  Wilsonian RG~\cite{WILSON197475, RevModPhys.66.129} is the normalization factor $\mathsf{N}$ in Eq.~\eqref{eq:path_integral_gen_functional}. After a Wick rotation  $\tau =\ii t$, this factor can be interpreted as the  partition function $\mathsf{N}\to Z$ of a model of statistical mechanics in a higher spatial dimension $d=D+1$, where the free energy corresponds to the  QFT action~\eqref{eq:path_integral_gen_functional} in imaginary time, the so-called Euclidean action. This partition function displays the same   renormalisation of the microscopic couplings $g_i$, together with the aforementioned UV divergences, as those   discussed in Sec.~\ref{sec:ren_gen_functional} via the normalised generating functional. The key idea of the RG is that, after regularising these divergences by introducing a UV cutoff $\Lambda_{\rm c}$,  the microscopic couplings will flow in the larger  parameter space   of all possible couplings of the theory $\{g_i(\Lambda_{\rm c})\}$, the so-called theory space. This flow occurs as one lowers the  cutoff  in order to focus on the low-energy or infra-red (IR) behaviour~\cite{WILSON197475,PhysicsPhysiqueFizika.2.263}. A quantitative description of this RG flow can be obtained by first performing a coarse-graining step on the partition function, where one integrates out the high-energy modes at the cutoff scale $k\in[\Delta_{\rm c}/s,\Delta_{\rm c}]$ for some parameter $s>1$. After such a  coarse graining,  a second RG step consists of rescaling the momenta and fields  in order to  compare the original  partition function $Z$ with  the coarse-grained one, extracting how the microscopic parameters run with the cutoff, and finding  differential equations for the RG flow~\cite{WILSON197475, RevModPhys.66.129}. Given our focus on the normalised generating functional in this work, as it is the key quantity to understand the effective Ising dynamics for the full model of $\lambda\phi^4$ fields coupled to  spins~\eqref{eq:renormalised_transition_matrix}, we will revisit this RG procedure using the results of App.~\ref{sec:ren_gen_functional} and Sec.~\ref{sec:spin_scalalr_gen_functional}. 

Let us start by discussing the first RG step. From now on, we set natural units $\hbar=c_{\textsf{t}}=1$, keeping in mind our previous discussion to connect the results to the SI trapped-ion units. After the Wick rotation, we can define the  spacetime points and wavevectors as follows $\boldsymbol{x}=(\ii t,\textsf{x})$ and $\boldsymbol{k}=(-\ii k^0,\textsf{x})$, which are $D=2$ Euclidean vectors.  In the absence of the Ising spins, one can  calculate the perturbative expansion of the full generating functional~\eqref{eq:int_gen_functional} by separating the slow $\varphi_<(x)$ and fast  $\varphi_>(x)$ field components 
\beq
\begin{split}
\varphi_<(x)&=\bigintssss_{0}^{\frac{\Lambda_{\rm c}}{s}}\!\!\frac{{\rm d}^2{k}}{(2\pi)^2}\ee^{-\ii \boldsymbol{k}\cdot\boldsymbol{x}}\varphi(\boldsymbol{k}),\\
\varphi_>(x)&=\bigintssss_{\frac{\Lambda_{\rm c}}{s}}^{\Lambda_{\rm c}}\!\!\frac{{\rm d}^2{k}}{(2\pi)^2}\ee^{-\ii \boldsymbol{k}\cdot\boldsymbol{x}}\varphi(\boldsymbol{k}),
\end{split}
\eeq
where the scalar product is now Euclidean.
The diagrammatic expansion in the absence of Ising spins in Eq.~\eqref{eq:Z_interacting} can be rewritten in Euclidean time as follows
  \begin{widetext}
  \beq
   \label{eq:Z_interacting_RG} 
 \mathsf{Z}^{ E}[J]=\bigg( 1\hspace{-8ex}
 \setlength{\unitlength}{1cm}
\thicklines
\begin{picture}(19,0)
\put(1.25,0.0){$-\frac{1}{4}$}
\put(1.9,.1){\line(1,0){0.8}}
\put(1.75,0.01){$\color{gray}{\boldsymbol{\times}}$}
\put(2.08,0.1){\begin{tikzpicture}
\begin{scope}[ thick,dashed]
\draw (0,0) circle (.2cm);
\end{scope}
\end{tikzpicture}}
\put(2.21,0.01){$\color{myred}\bullet$}
\put(2.84,0.){$+\frac{1}{8}$}
\put(2.55,0.01){$\color{gray}{\boldsymbol{\times}}$}
\put(3.45,0.01){$\color{gray}{\boldsymbol{\times}}$}
\put(3.6,0.1){\line(1,0){0.8}}
\put(3.78,0.1){\begin{tikzpicture}
\begin{scope}[ thick,dashed]
\draw (0,0) circle (.2cm);
\end{scope}
\end{tikzpicture}}
\put(3.78,0.49){\begin{tikzpicture}
\begin{scope}[ thick,dashed]
\draw (0,0) circle (.2cm);
\end{scope}
\end{tikzpicture}}
\put(3.91,0.01){$\color{myred}\bullet$}
\put(3.91,0.41){$\color{myred}{\bullet}$}
\put(4.25,0.01){$\color{gray}{\boldsymbol{\times}}$}
\put(4.65,0.){$+\frac{1}{8}$}
\put(5.3,0.1){\line(1,0){0.35}}
\put(5.67,0.09){\begin{tikzpicture}
\begin{scope}[ thick,dashed]
\draw [dashed] (10,1) -- (10.5,1);
\end{scope}
\end{tikzpicture}}
\put(6.05,0.1){\line(1,0){0.35}}
\put(5.17,0.01){$\color{gray}{\boldsymbol{\times}}$}
\put(5.38,0.1){\begin{tikzpicture}
\begin{scope}[ thick,dashed]
\draw (0,0) circle (.2cm);
\end{scope}
\end{tikzpicture}}
\put(5.88,0.1){\begin{tikzpicture}
\begin{scope}[ thick,dashed]
\draw (0,0) circle (.2cm);
\end{scope}
\end{tikzpicture}}
\put(5.51,0.01){$\color{myred}\bullet$}
\put(6.02,0.01){$\color{myred}\bullet$}
\put(6.25,0.01){$\color{gray}{\boldsymbol{\times}}$}
\put(6.6,0.){$+\frac{1}{12}$}
\put(7.4,0.1){\line(1,0){0.3}}
\put(8.1,0.1){\line(1,0){0.3}}
\put(7.72,0.09){\begin{tikzpicture}
\begin{scope}[ thick,dashed]
\draw [dashed] (10,1) -- (10.5,1);
\end{scope}
\end{tikzpicture}}
\put(7.25,0.01){$\color{gray}{\boldsymbol{\times}}$}
\put(7.6,-0.2){\begin{tikzpicture}
\begin{scope}[ thick,dashed]
\draw (0,0) circle (.28cm);
\end{scope}
\end{tikzpicture}}
\put(8.08,0.01){$\color{myred}\bullet$}
\put(7.54,0.01){$\color{myred}\bullet$}
    \put(8.25,0.01){$\color{gray}{\boldsymbol{\times}}$}
\put(8.5,0.0){$-\frac{1}{4!}$}
\put(9.22,0.1){\line(1,0){0.8}}
\put(9.61,-0.3){\line(0,1){0.8}}
\put(9.52,0.01){$\color{myred}\bullet$}
\put(9.1,0.01){$\color{gray}{\boldsymbol{\times}}$}
\put(9.86,0.01){$\color{gray}{\boldsymbol{\times}}$}
\put(9.465,0.41){$\color{gray}{\boldsymbol{\times}}$}
\put(9.465,-0.41){$\color{gray}{\boldsymbol{\times}}$}
\put(10.2,0.0){$+\frac{1}{12}$}
\put(10.95,0.1){\line(1,0){0.35}}
\put(11.4,0.09){\begin{tikzpicture}
\begin{scope}[ thick,dashed]
\draw [dashed] (10,1) -- (10.35,1);
\end{scope}
\end{tikzpicture}}
\put(11.55,0.1){\line(1,0){0.45}}
\put(11.61,-0.3){\line(0,1){0.8}}
\put(11.08,0.1){\begin{tikzpicture}
\begin{scope}[ thick,dashed]
\draw (0,0) circle (.2cm);
\end{scope}
\end{tikzpicture}}
\put(11.52,0.01){$\color{myred}\bullet$}
\put(11.2,0.01){$\color{myred}\bullet$}
\put(10.8,0.01){$\color{gray}{\boldsymbol{\times}}$}
\put(11.85,0.01){$\color{gray}{\boldsymbol{\times}}$}
\put(11.465,0.41){$\color{gray}{\boldsymbol{\times}}$}
\put(11.465,-0.41){$\color{gray}{\boldsymbol{\times}}$}
\put(12.15,0.0){$+\frac{1}{32}$}
\put(12.9,.25){\line(1,0){0.8}}
\put(13.075,0.25){\begin{tikzpicture}
\begin{scope}[ thick,dashed]
\draw (0,0) circle (.2cm);
\end{scope}
\end{tikzpicture}}
\put(13.21,0.16){$\color{myred}{\bullet}$}
\put(13.55,0.16){$\color{gray}{\boldsymbol{\times}}$}
\put(12.9,-.08){\line(1,0){0.8}}
\put(12.75,0.16){$\color{gray}{\boldsymbol{\times}}$}
\put(12.75,-0.165){$\color{gray}{\boldsymbol{\times}}$}
\put(13.075,-0.48){\begin{tikzpicture}
\begin{scope}[ thick,dashed]
\draw (0,0) circle (.2cm);
\end{scope}
\end{tikzpicture}}
\put(13.21,-0.165){$\color{myred}{\bullet}$}
\put(13.55,-0.165){$\color{gray}{\boldsymbol{\times}}$}
\put(14.05,0.0){$+\frac{1}{16}$}
\put(14.9,.35){\line(1,0){0.8}}
\put(14.9,-.24){\line(1,0){0.8}}
\put(14.75,0.26){$\color{gray}{\boldsymbol{\times}}$}
\put(14.98,-0.26){\begin{tikzpicture}
\begin{scope}[ thick,dashed]
\draw (0,0) circle (.3cm);
\end{scope}
\end{tikzpicture}}
\put(15.21,0.26){$\color{myred}{\bullet}$}
\put(15.55,0.26){$\color{gray}{\boldsymbol{\times}}$}
\put(14.75,-0.33){$\color{gray}{\boldsymbol{\times}}$}
\put(15.21,-0.33){$\color{myred}{\bullet}$}
\put(15.55,-0.33){$\color{gray}{\boldsymbol{\times}}$}
\put(15.9,0.01){$+\cdots\bigg) \mathsf{Z}_{0}^{ E}[J],$}
\end{picture}
\eeq
  \end{widetext}
  where we see that some of the factors preceedding the diagrams  change sign, and all become real. Additionally, we note that  the  lines now stand for the free Euclidean propagator
  \beq
  \Delta_{m_{0}}^E\!(\boldsymbol{x})\!=\!\!\bigintssss_{\boldsymbol{k}}\tilde{ \Delta}^E_{m_{0}}(\boldsymbol{k})\ee^{-\ii {\boldsymbol{k}\cdot\boldsymbol{x}}},\hspace{1ex}\tilde{ \Delta}^E_{m_{0}}(\boldsymbol{k})=\frac{1}{\boldsymbol{k}^2+m_{0}^2}=:\tilde{ \Delta}^E_{m_{0},\boldsymbol{k}},
  \eeq
  such that, for  slow components (solid lines), the momentum integrals extend to $|\boldsymbol{k}|\in[0,\Lambda_{\rm c}/s]$, whereas for  fast components (dashed lines), they extend to  extend to an small annulus around the UV cutoff  $|\boldsymbol{k}|\in[\Lambda_{\rm c}/s,\Lambda_{\rm c}]$. Note that the ellipsis in Eq.~\eqref{eq:Z_interacting_RG} now contains, in addition to higher-order terms in the interaction strength, also Feynman diagrams that involve other combinations of slow and fast components, e.g. tree level diagrams with only fast-mode propagators. However, the current combinations in Eq.~\eqref{eq:Z_interacting_RG}  suffice  to understand the RG flow, as they capture the effect that the coarse-graining over fast modes within the annulus around the cutoff  will have on the slow modes, and thus on the low-energy IR theory.
  
    As discussed around Eq.~\eqref{eq:ren_gen_function}, the additional scattering terms~\eqref{eq:Z_interacting_RG}  lead to an effective generating functional in which the bare  parameters get renormalised. In this case, such a renormalisation is a result of the coarse graining over fast modes, such that the dressed parameters become cutoff dependent. For instance, the bare mass $m_0^2\to m_0^2+\delta m^2_0$ changes due to the tadpoles and sunrise mass renormalisation of the fast modes similarly to Eq.~\eqref{eq:mass_renormalisation}, namely
     \beq
\delta m^2_0= \frac{\lambda_0}{2}\!\!\bigintssss_{\boldsymbol{k}_1}^{\rm f}\!\!\!\!\!\!\tilde{\Delta}^{ E}_{m_{ 0},\boldsymbol{k}_1}\!\!\!\left(\!\!1-\lambda_0\!\!\bigintssss_{\boldsymbol{k}_2}^{\rm f}\!\!\!\!\!\left(\half\tilde{\Delta}^{ E}_{m_{ 0},\boldsymbol{k}_1}+\third\tilde{\Delta}^{ E}_{m_{ 0},\boldsymbol{k}_1+\boldsymbol{k}_2}\right)\tilde{\Delta}^{ E}_{m_{ 0},\boldsymbol{k}_2}\!\!\right)\!\!,
 \eeq
  where the integrals restricted  to  momenta of the  fast modes, as indicated by $\int_{\boldsymbol{k}}^{\rm f}= \int_{\frac{\Lambda_{\rm c}}{s}<|\boldsymbol{k}|<\Lambda_{\rm c}}|\boldsymbol{k}|\frac{{\rm d}|\boldsymbol{k}|}{2\pi}$. Likewise, the interaction strength changes according to $\lambda_0\to\lambda_0+\delta\lambda_0$, where 
   \beq
\delta\lambda_0=-\frac{3\lambda_0^2}{2}\!\!\bigintssss_{\boldsymbol{k}_1}^{\rm f}\!\!\left(\tilde{\Delta}^{ E}_{m_{ 0},\boldsymbol{k}_1}\right)^{\!\!2}\!\!.
 \eeq

 The next RG step is to rescale the variables and fields, recovering the original cutoff such that the resulting partition function or, in this case, the generating functional can be compared to the original  one, and the flow in theory space can be extracted. At this step, the sunrise diagram and the  wavefunction renormalisation~\eqref{eq:wavefunction_renormalisation} leading to the multiplicative mass and source renormalisations~\eqref{eq:source_ren} come into play
 \beq
 \boldsymbol{k}\to \boldsymbol{k}'=s\boldsymbol{k},\hspace{1.5ex}\varphi(\boldsymbol{k})\to \varphi\big(\boldsymbol{k}'s^{-1}\big)=s^{2}z^{-1}_{m_0,\lambda_0}\varphi'(\boldsymbol{k}'),
 \eeq
 where the integrals  are again restricted to the fast modes
 \beq
\label{eq:wavefunction_renormalisation}
z_{m_0,\lambda_0}^{\!-1}=1+\frac{\lambda_0^2}{6}\!\!\bigintssss_{\boldsymbol{k}_1}^{\rm f}\!\!\tilde{\Delta}^{ E}_{m_{ 0},\boldsymbol{k}_1}\!\!\bigintssss_{\boldsymbol{k}_2}^{\rm f}\!\!\!\left(\tilde{\Delta}^{ E}_{m_{ 0},\boldsymbol{k}_1+\boldsymbol{k}_2}\!\right)^{\!\!2}\!\tilde{\Delta}^{ E}_{m_{ 0},\boldsymbol{k}_2}.
\eeq
After this rescaling, one can obtain the total running of the microscopic couplings  in theory space. The only step left is to extract the system of differential equations for infinitesimal  coarse-graining $s=\ee^{\delta\ell}\approx 1+\delta\ell$, which allows to approximate the above Euclidean integrals, and leads to 
\beq
\label{eq:RG_phi^4}
\begin{split}
\frac{{\rm d} m_0^2}{{\rm d}\delta\ell}&=2m_0^2+\frac{\lambda_0}{4\pi}\frac{\Lambda_{\rm c}^2}{\Lambda_{\rm c}^2+m_0^2},\\
\frac{{\rm d} \lambda_0}{{\rm d}\delta\ell}&=2\lambda_0-\frac{3\lambda_0^2}{4\pi}\frac{\Lambda_{\rm c}^2}{(\Lambda_{\rm c}^2+m_0^2)^2}.\\
\end{split}
\eeq
Up to this point, the discussion parallels the standard RG treatment of the $\lambda\phi^4$ QFT based on the partition function~\cite{WILSON197475, RevModPhys.66.129}. For the $d=1+1$ case, these RG equations are not very informative, as both couplings are relevant already at tree level (i.e. they grow exponentially  as one approaches long wavelengths). Indeed, any self-interacting term $\varphi^{2n}$ with $n\geq 3$ will also be relevant, bringing us away from the Gaussian fixed point $(m_0,\lambda_0)=(0,0)$. 

We note  that  the Tomonaga-Luttinger parameter of the original low-energy description~\eqref{eq:scalar_qft_crystal} does not seem to play any role in the RG treatment, as we have rescaled the fields to get the standard QFT~\eqref{eq:rescaled_QFT_ions}. Note, however, that this parameter is proportional to the shear modulus of the ion crystal, and thus quantifies how rigid the system is.    Accordingly, the effect of the quantum fluctuations introduced by the canonically-conjugate momentum operators in Eq.~\eqref{eq:scalar_qft_crystal} will be inversely proportional to the Tomonaga-Luttinger parameter. As neatly discussed in~\cite{PhysRevB.89.214408}, even if the above RG equations~\eqref{eq:RG_phi^4} do not contain information about the non-perturbative fixed point that controls the scaling of the phase transition (i.e. the D=2 analogue of the Wilson-Fisher fixed point for $D=4-\epsilon$); they can detrmine how the classical critical point $\left.m_0^2\right|_c=0$  changes with the strength  of the quartic interactions  towards $\left.m_0^2(\lambda_0)\right|_c\neq0$ when  quantum fluctuations are very small. Following~\cite{PhysRevB.89.214408}, quantum fluctuations are controlled by  a dimensionless effective Planck's constant 
\beq
\label{eq:eff_hbar}
\tilde{\hbar}=\frac{\hbar}{m_a\omega_{\mathsf{x}}a^2}=\frac{\sqrt{\frac{\ell^3}{a^3}\log 2 }}{K},
\eeq
which we find to be proportional  to  the inverse of the Tomonaga-Luttinger parameter. For the crystal of $^{171}$Yb$^+$ ions  (see Fig.~\ref{Fig:eq_poitions}), $K=1.3\cdot 10^5$ and quantum fluctuations are thus highly suppressed $\tilde{\hbar}=3.1\cdot 10^{-5}$. Since the quartic-interaction parameter in SI units~\eqref{eq:lambda_0_ions} is  proportional to $K^{-1}\propto\tilde{\hbar}\ll1$,  the  flow equations~\eqref{eq:RG_phi^4} can be rewritten in a way that the dependence on  quantum fluctuations becomes apparent. In particular, linearising them  around  $m_0^2=0$ allows to predict   how this classical critical point changes  as a function of the quantum fluctuations. To make contact with our previous exposition of the trapped-ion problem, we  rewrite this result in terms of the Luttinger parameter and use SI units for the interaction strength~\eqref{eq:lambda_0_ions}, such that
\beq
\label{eq:flow_class_crictial}
\left.\frac{m_0^2c_{\textsf{t}}^2}{\hbar^2}\right|_c=\frac{279\zeta(5)}{16\pi Ka^2 \log2}\big(\log K-\log K^*\big).
\eeq
Here, $K^*$ is a non-universal parameter that can be extracted from the numerical results of~\cite{PhysRevB.89.214408} via Eq.~\eqref{eq:eff_hbar}, such that $K^*\approx 58.98$ for the $^{171}$Yb$^+$ crystal.
 We note that this is a non-universal property, unlike the scaling behaviour of the structural phase transition, which is controlled by a non-perturbative fixed point that, due to symmetry considerations, must be in the universality class of the 2D Ising model~\cite{silvi_zigzag_numerics}. 

With this perturbative RG prediction, we can analyse how  the boson-mediated Ising interactions will change due to the quartic interactions. In particular,  their decay~\eqref{eq:dip_exp_tail} is  controlled by the effective Compton wavelength of the bosons in the coarse-grained theory before the  dipolar tail takes on.  
Since the classical critical point moves according to Eq.~\eqref{eq:flow_class_crictial}, the effective Compton wavelength can still be described by Eq.~\eqref{eq:range_length} if we substitute  $\omega_{\mathsf{zz}}^2\to \omega_{\mathsf{zz}}^2+\delta \omega_{\mathsf{zz}}^2$, where
\beq
\label{eq:shift}
  \delta\omega_{\mathsf{zz}}^2=\omega_{\mathsf{x}}^2\frac{\ell^3}{a^3}\frac{279\zeta(5)}{16\pi Ka^2}\big(\log K-\log K^*\big).
\eeq
includes the renormalization of the bare parameters due to the quartic interactions.
We now make the following two remarks. On the one hand, the combination of this shift with the scaling of Eq.~\eqref{eq:range_length} is consistent with the scaling properties of the Gaussian fixed point. We know, on the other hand, that the correlation length should be controlled by the Ising universality class, in which case the scaling will differ. In any case, since the correction in Eq.~\eqref{eq:shift} turns out to be rather small (i.e. $\sqrt{\delta\omega^2_{\mathsf{zz}}}/\omega_{\mathsf{zz}}=2.5\cdot 10^{-3}$ for the $^{171}$Yb$^+$ crystal of  Fig.~\ref{Fig:spin_couplings}), it will be very challenging to probe these differences experimentally, which  is a result of the very large rigidity of the ion crystal and the suppressed quantum fluctuations.  In the last section of this article, we discuss how this rigidity could be controlled by introducing additional parametric drivings of the  crystal that allow one to tune the value of the shear modulus and, in turn, lower the Tomonaga-Luttinger parameter.

\subsection{Parametric drivings to control  the shear modulus}
\label{sec:drivings}

 In this section, we discuss how the shear rigidity and, in turn, the Tomonaga-Luttinger parameter can be controlled by introducing a fast parametric modulation of the transverse trap frequency $\omega_{\textsf{z}}$. As discussed in~\cite{PhysRevLett.93.263602}, the transverse modes of the harmonic ion crystal are rather special in the sense that  the total number of local vibrational excitations of the ions  is a conserved quantity  for the time-scales of interest. This is a consequence of the microscopic parameters in Eq.~\eqref{eq:coulomb_crystal_phonons}, which fulfil $\hbar|\kappa_{ij}^{\textsf{z}}|/4m_a\omega_{\textsf{z}}\ll2\hbar\omega_{\textsf{z}}$. Accordingly, one can treat the local vibrations as  particles tunnelling between neighbouring ions, and explore experimentally  their propagation along the Coulomb crystal, as realised in various recent experiments~\cite{Brown2011,Harlander2011,PhysRevA.85.031401,Ramm_2014,PhysRevLett.120.073001,PhysRevLett.111.160501,PhysRevLett.120.073001,Abdelrahman2017,PhysRevLett.123.100504}. Moreover, the corresponding tunnelling amplitude can be controlled by Floquet engineering via periodic modulations~\cite{RevModPhys.89.011004}, mimicking situations considered for electrons in solids~\cite{PhysRevB.34.3625} and beyond~\cite{PhysRevLett.107.150501,PhysRevLett.123.213605}. In the present context, we consider a site-dependent parametric modulation of the  trap frequency
\beq
\label{eq:modulation}
\omega_{\textsf{z}}\to\omega_{\textsf{z}}+{\omega_{d}}\eta_i\cos(\omega_{d} t),
\eeq
where $\omega_{d}$ is the parametric driving frequency, and $\eta_i=i\Delta\eta$ is a relative modulation amplitude that increases linearly along the ion chain. We shall consider 
the fast-modulation regime, such that  the aforementioned tunnelling  gets dressed in analogy to what occurs for electrons in solids~\cite{PhysRevB.34.3625}. For the transverse vibrations, this effect can in turn be recast as a 
dressing of the shear modulus~\eqref{eq:shear_modulus} of the ion crystal
\beq
\label{eq:shear_modulus_dressed}
\tilde{\mu}_r\approx\frac{1}{2a}\sum_{j\neq i}\kappa^{\textsf{z}}_{i,j}|\textsf{x}_j-\textsf{x}_i|^2\cos(\mathsf{k}_s|\textsf{x}_j-\textsf{x}_i|)\mathfrak{J}_0\big(\Delta\eta(j-i)\big)
\eeq
where we have introduced the zero-th Bessel function of the first kind $\mathfrak{J}_0(x)\approx \sqrt{\frac{2}{\pi x}}\cos\left(\frac{\pi}{4}-x\right)$ for $x\gg 1$. This approximation requires $\hbar|\kappa_{i,j}^z|/4m_a\omega_{\textsf{z}}\ll\hbar\omega_{d},\hbar\omega_{\textsf{z}}$, where $\omega_d\neq\omega_{\textsf{z}}$, where corrections due to rapidly-rotating terms will be vanishingly small, and the main contribution of the parametric drivings is the dressing of the shear modulus~\eqref{eq:shear_modulus_dressed}.

\begin{figure}[t]
 \begin{centering}
  \includegraphics[width=0.8\columnwidth]{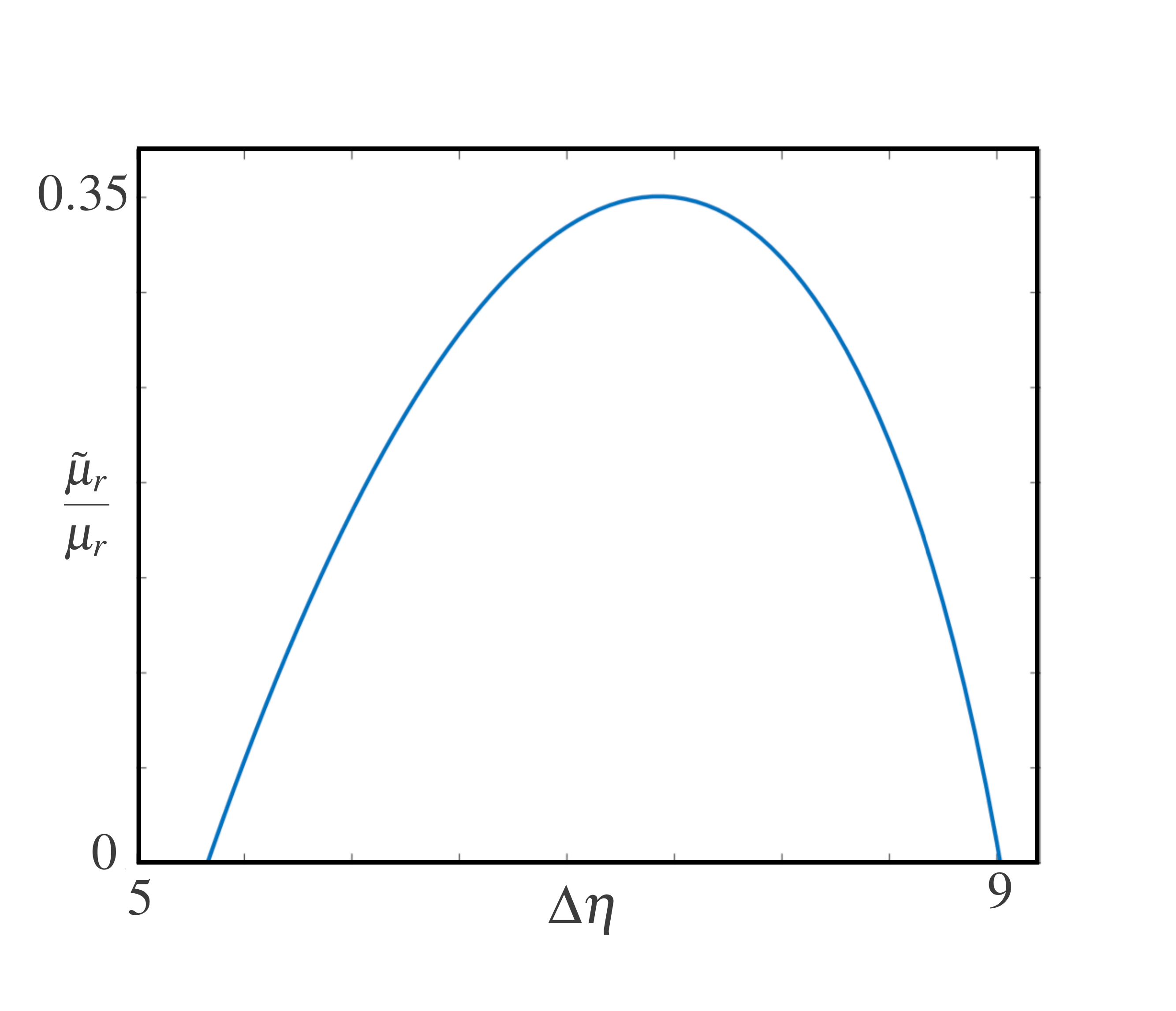}\\
  \caption{\label{Fig:rigidity_modulation} {\bf Dressed shear modulus:} Ratio of the dressed shear modulus~\eqref{eq:shear_modulus_dressed} and the original shear modulus~\eqref{eq:shear_modulus} of the parametrically-driven  crystal of $^{171}$Yb$^+$ ions, as a function of the relative modulation amplitude $\Delta\eta$ in Eq.~\eqref{eq:modulation}.}
\end{centering}
\end{figure}

Note that this parametric dressing can even change the sign of the tunneling~\cite{PhysRevB.34.3625}, inverting the band structure and changing the role of the zigzag mode. This can be   avoided in the present context, ensuring that the dressed rigidity $\mu_r>0$, by working with modulations $\Delta\eta\in(5.3, 9)$, as depicted in Fig.~\ref{Fig:rigidity_modulation}. This figure shows that the dressed rigidity can achieve much smaller values for $\Delta\eta\gtrsim5.3$, or $\Delta\eta\lesssim9$.
 Accordingly, both the transverse speed of sound and the Tomonaga-Luttinger parameter~\eqref{eq:mic_parameters}  can be reduced considerably, such that the ion crystal is less rigid and the effect of quantum fluctuations will become more important at the expense of a slower transverse  speed of sound. For instance, for $\Delta\eta=5.33$, we would get $K\approx3.2\cdot 10^3$, effectively reducing the Tomonaga-Luttinger parameter by  two orders of magnitude.

In contrast, in this regime $\hbar|\kappa_{i,j}^z|/4m_a\omega_{\textsf{z}}\ll\hbar\omega_{d}\neq\hbar\omega_{\textsf{z}}$, the bare mass of the scalar bosons  and the quartic coupling~\eqref{eq:coupling_strength_ions} would get no appreciable dressing as they are purely local terms. Accordingly, the effective range of the Ising spin-spin interactions in the harmonic limit of the crystal are still  described by Eq.~\eqref{eq:range_length}, and can thus be controlled by setting the frequency of state-dependent force closer or further from the zigzag mode, taking into account the flow of the classical critical point~\eqref{eq:flow_class_crictial} due to the quartic interactions. Since this shift scales with $\log K/K$, the dressing scheme will amplify the effects of the quartic interactions, making the corresponding change in $\xi_{\rm eff}$ more amenable to be measured experimentally. Note, however, that including the specific parametric modulation~\eqref{eq:modulation} may require the use of micro-fabricated trap electrodes in the spirit  of~\cite{PhysRevLett.123.100504}, or  global modulations, such as those stemming from ac-Stark shifts~\cite{PhysRevLett.107.150501,Bermudez_2012} that will lead to similar dressing effects in the long-wavelength description of the ion chain. Therefore, although the effect of the quartic interactions is amplified by reducing the rigidity of the crystal, the experimental complexity also increases.

\section{\bf Conclusions and outlook}

We have presented a new perspective on  a protocol for the interferometric sensing~\eqref{eq:parity_oscillations}  of  the generating functional of a scalar QFT~\cite{PhysRevX.7.041012},  which clarifies the role of the interactions between the quantum sensors mediated by the scalar bosons~\eqref{eq:spin_spin_unitary}, and the requirement of multipartite entanglement in the initial state. This   has allowed us to propose a new simplified sensing scheme, which uses unentangled probes that are coupled to the scalar field via always-on harmonic sources~\eqref{eq:sources_harmonic},  and which evolve under an effective   quantum Ising model with  long-range couplings~\eqref{eq:spin_spin_couplings} controlled by a dimensionally-reduced Euclidean propagator of the scalar field~\eqref{eq:Euclidean_propagator}. In the presence of self-interactions, these Ising spin-spin interactions are mediated by virtual $\lambda\phi^4$ bosons that can also scatter. This regime can be explored using functional-integral methods by means of a scalar-sigma model~\eqref{eq:sclalar_spin_action_final} and a diagrammatic expansion~\eqref{eq:spin_feynman}, which shows that this   additional scattering leads to renormalisations of the spin-spin couplings between the sensors, and  further $2n$-spin interactions~\eqref{eq:renormalised_transition_matrix}.
In the regime of interest, we have argued that the real-time dynamics is dominated by a renormalised quantum Ising  Ising model, where the pairwise interactions get additive and multiplicative renormalisations~\eqref{eq:mass_renormalisation}-\eqref{eq:int_renormalisation}.

These formal results find a natural application for $D=1+1$ dimensions in the physics of crystals of trapped ions  close to a structural phase transition, and  subjected to state-dependent forces. We have shown that a coarse-grained elasto-dynamical theory~\eqref{eq:scalar_qft_crystal} yields a transparent long-wavelength description of phonon-mediated spin-spin interactions in harmonic trapped-ion crystals~\eqref{eq:exp_tail}, and that additional and branch-cut contributions~\eqref{eq:dip_exp_tail} can be incorporated by considering the full lattice propagator~\eqref{eq:Euclidean_propagator_ions}. By approaching the structural phase transition, the quartic non-linearities start playing a role, and we have used RG predictions to estimate how the bare couplings flow~\eqref{eq:shift}, and how this will affect the range of the interactions in a regime where the rigidity of the ion crystal is very large, and the quantum fluctuations are largely suppressed. Finally, we have discussed a parametric driving scheme that can allow to reduce the shear modulus of the crystal~\eqref{eq:shear_modulus_dressed}, which should amplify the effect of the non-linearities in the effective Ising models.

Although the formalism hereby presented is specific to the zero-temperature $\lambda\phi^4$ QFT, the underlying ideas can be readily generalised to finite temperatures,  to other QFTs of interest, and other couplings between the Ising sensors and the corresponding field operators. This could include composite operators in the Ising-Schwinger sources~\eqref{eq:Ising_Z_source_coupling}, opening a route to probe fermionic  QFTs, or to connect to generating functionals of other types of correlations functions in bosonic QFTs. Likewise, although we have focused on Lorentz-invariant QFTs, the underlying idea can also be applied to long-wavelength descriptions of quantum many-body systems where Lorentz-invariant is not recovered at criticality.

Regarding the application to trapped-ion crystals, we have focused on $D=1+1$ dimensions and the linear-to-zigzag structural phase transition. It would be interesting to find the coarse-grained description of other situations, such as two-dimensional triangular crystals of ions  in Penning traps, or specific layouts of micro-fabricated Paul or Penning traps with other crystalline configurations. In addition, there are other types of structural phase transitions that can lead to other effective QFTs, and it would be interesting to understand the sensing scheme presented in our work for those situations. In addition to trapped-ion crystals, our results may also find applications in other quantum technologies, such as superconducting qubits coupled to Josephson-junction arrays, or other variants of transmission lines where non-linear effects can play a role. In the atomic realm, one could also explore the application of our formalism to atomic impurities trapped coupled to  ultracold quantum gases.

\acknowledgements
G.M-V. and A.B. acknowledge support from the Ram\'on y Cajal program RYC-2016-20066,  CAM/FEDER Project S2018/TCS-4342 (QUITEMAD-CM) and  Plan Nacional Generaci\'on de Conocimiento PGC2018-095862-B- C22. G.A. is supported by the UKRI Science and Technology Facilities Council (STFC) Consolidated Grant No. ST/T000813/1. M.M. acknowledges support by the ERC Starting Grant QNets Grant Number 804247 and the EU H2020-FETFLAG-2018-03 under Grant Agreement number 820495. 

\appendix
 \section{Generating functional of Klein-Gordon fields}
  \label{sec:KG_Z_0}

  In this appendix, we start with a basic description of the canonical quantisation  of a  scalar QFT~\cite{Klein1926,Gordon1926}, which appears in many textbooks, e.g.~\cite{Peskin:1995ev,fradkin_2021}. We then introduce the concept of the generating functional~\cite{Schwinger1951}, and make explicit derivations  using canonical quantisation instead of  functional integral methods. This derivation will serve to set the notation, and discuss the sensing schemes in the main text using  a QFT language that is closer to   quantum optics, preparing the ground for   the functional-integral formalism used at later stages.  As discussed in the main text, this derivation is useful to clarify key aspects of  the interferometric scheme to measure  such a generating functional by   incorporating Ising spins~\cite{PhysRevX.7.041012}, or the  simpler  sensing scheme that rely on the appearance of effective Ising models in specific dynamical situations.

The generating functional $\mathsf{Z}_0[J]$ depends on the so-called Schwinger sources $J(x)$~\cite{Schwinger1951} which, in the canonical formalism, are introduced as perturbations of  the Klein-Gordon Hamiltonian density~\eqref{eq:KG_field}  as follows  $\mathcal{H}_0\to \mathcal{H}_0+\mathcal{V}_J$, where 
\beq
\label{eq:V_J}
\mathcal{V}_J= -J(x)\phi(x),
\eeq
 and $\phi(x)$ is the scalar field operator described below Eq.~\eqref{eq:KG_field}. The sourced time-evolution operator $U(t_{\rm f},t_0)=U_{\mathcal{H}_0}U_{\mathcal{V}_J}$ can be expressed in terms of  two unitaries
 \beq
 \label{eq:unitaries}
  U_{ \mathcal{H}_0}=\ee^{-\ii\int\!{\rm d}^D\mathcal{H}_0},\hspace{2ex}
 U_{\mathcal{V}_J}=\mathsf{T}\left\{\ee^{+\ii\int\!{\rm d}^D J(x)\phi_H(x)}\right\}.
 \eeq
 Here, $\mathsf{T}\{\cdot\}$ is the time-ordering operator, and we have introduced the  fields $\phi_H(x)=(U_{\mathcal{H}_0})^\dagger \phi(x)U_{ \mathcal{H}_0}$ evolving in the Heisenberg picture with respect to the unsourced  Klein-Gordon Hamiltonian~\eqref{eq:KG_field}. In this picture, the vacuum persistence amplitude $\bra{0} U_{{\mathcal{V}_J}}\ket{0}$   can be  used to define the normalised generating functional, as it clearly encodes all the information about $n$-point  Feynman propagators of the field theory, $ G_0^{(n)}(x_1,\cdots, x_n)=\bra{0}\mathsf{T}\{\phi_H(x_1)\cdots\phi_H(x_n)\}\ket{0}$. These can be recovered  by taking functional derivatives
 \beq
 \label{eq:vpa}
 \mathsf{Z}_0[J]=\bra{0} \!U_{ \mathcal{V}_J}\!\!\ket{0}\!\!, \hspace{1.5ex} G_0^{(n)}(x_1,\cdots, x_n)=\left.\frac{(-\ii)^n\delta^n\mathsf{Z}_0[J]}{\delta J(x_1)\cdots\delta J(x_n)}\right|_{J=0}\!.
 \eeq
 
 In order to derive the explicit expression of the normalised generating functional, and the corresponding Feynman propagators, we use the canonical mode expansion 
  \beq
  \label{eq:fields_quantisation}
  \begin{split}
  \phi_H(x)&=\!\!\bigintssss\!\!\frac{{\rm d}^d\textsf{{k}}}{(2\pi)^d2\omega_{\textsf{\textbf{k}}}}\left.\left(\hspace{0.25ex}a^{\phantom{\dagger}}_{\textsf{\textbf{k}}}\ee^{-\ii kx}+a^\dagger_{\textsf{\textbf{k}}}\ee^{\ii kx}\hspace{0.25ex}\right)\right|_{\rm m.s.},\\
  \pi_H(x)&=\!\!\bigintssss\!\!\frac{{\rm d}^d\textsf{{k}}}{(2\pi)^d2}(\!-\ii)\!\!\! \left.\left(a^{\phantom{\dagger}}_{\textsf{\textbf{k}}}\ee^{-\ii kx}-a^\dagger_{\textsf{\textbf{k}}}\ee^{\ii kx}\right)\right|_{\rm m.s.},
  \end{split}
  \eeq     
  where $kx=k_\mu x^\mu=\omega_{\textsf{\textbf{k}}}t-\textsf{\textbf{k}}\cdot \textsf{\textbf{x}}$ is the  Minkowski product for $\mu\in\{0,1,\cdots,D-1\}$ on mass shell (m.s.) with 
  \beq
  \label{eq:dispersion}
  \omega_{\textsf{\textbf{k}}}=\sqrt{\textsf{\textbf{k}}^2+m_0^2}.
  \eeq
   This is the dispersion relation of the Klein-Gordon massive  bosons, which can be created or annihilated by the  covariant creation-annihilation operators $a_{\textsf{\textbf{k}}}^{{\dagger}}, a_{\textsf{\textbf{k}}}^{\phantom{\dagger}}$, which satisfy Lorentz-invariant commutation relations $\big[a_{\textsf{\textbf{k}}}^{\phantom{\dagger}},a_{\textsf{\textbf{k}}'}^{\dagger}\big]=(2\pi)^d2\omega_{\textsf{\textbf{k}}}\delta^d(\textsf{\textbf{k}}-\textsf{\textbf{k}}')$.
 Up to the zero-point energy, the Hamiltonian terms in their corresponding pictures become
 \beq
 H_0\!=\!\!\bigintssss\!\!\frac{{\rm d}^d\textsf{{k}}}{(2\pi)^d}\frac{\omega_{\textsf{\textbf{k}}}}{2\omega_{\textsf{\textbf{k}}}}a^{{\dagger}}_{\textsf{\textbf{k}}}a^{\phantom{\dagger}}_{\textsf{\textbf{k}}},\hspace{1ex} V_{\!J}\!(t)\!=-\!\!\bigintssss\!\!\frac{{\rm d}^d\textsf{{k}}}{(2\pi)^d}\frac{J^{\phantom{\dagger}}_{\textsf{\textbf{k}}}\!(t)}{2\omega_{\textsf{\textbf{k}}}}\!a^{{\dagger}}_{\textsf{\textbf{k}}}\ee^{\ii \omega_{\textsf{\textbf{k}}}t}+{\rm H.c.},
\eeq
where we have introduced the Fourier transform of the Schwinger sources $J^{\phantom{\dagger}}_{\textsf{\textbf{k}}}\!(t)=\int{\rm d}^{d}\textsf{{x}}\hspace{0.2ex}J(t,\textsf{\textbf{x}})\ee^{-\ii{\textsf{\textbf{k}}}\cdot \textsf{\textbf{x}}}$.

Let us note that the  contribution of the sources to the  evolution operator~\eqref{eq:unitaries} can be evaluated to the desired order of the coupling $J(x)$ by means of  the  Magnus expansion~\cite{https://doi.org/10.1002/cpa.3160070404,Blanes_2010}. Moreover, for free Klein-Gordon fields, this expansion can  be truncated at second order without any approximation
\beq
\label{eq:propagator}
\log U_{\mathcal{V}_J}=-\ii\int_{t_0}^{t_{\rm f}}\!\!\!{\rm d}t_1V_{J}(t_1)-\frac{1}{2}\int_{t_0}^{t_{\rm f}}\!\!\!{\rm d}t_1\!\!\int_{t_0}^{t_1}\!\!\!\!{\rm d}t_2\hspace{0.05ex}[V_{J}(t_1),V_{J}(t_2)].
\eeq 
 Using the Baker-Campbell-Haussdorf formula~\cite{Achilles2012} to normal-order the exponential of the first term, and considering the explicit time-ordering of the second one,  one  finds  that
 \beq
 \label{eq:int_picture_ev}
 U_{\mathcal{V}_J}=:\ee^{-\ii\!\!\bigintssss\!\!{\rm d}^Dx J(x)\phi_{\rm H}(x)}:\ee^{-\frac{1}{2}\!\!\bigintssss\!\!{\rm d}^Dx_1\!\!\bigintssss\!\!{\rm d}^Dx_2J(x_1)\Delta_{m_0}\!(x_1-x_2)J(x_2)},
 \eeq
 where $:\hspace{0.5ex}:$ is the normal-ordering symbol, and we have introduced the following spacetime function
 \beq
 \label{eq:F_propagator}
 \Delta_{m_0}\!(x)=\!\!\bigintssss\!\!\frac{{\rm d}^d\textsf{{k}}}{(2\pi)^d}\frac{1}{2\omega_{\textsf{\textbf{k}}}}\left.\left(\ee^{-\ii kx}\theta(t)+\ee^{+\ii kx}\theta(-t)\right)\right|_{\rm m.s}.
 \eeq
 
 As customary in canonical quantisation~\cite{Peskin:1995ev}, the mass-shell condition in Eq.~\eqref{eq:F_propagator} can be automatically imposed 
by introducing an extra contour integration over  $k^0$, such that 
  \beq
  \label{eq:feynman_propagator_app}
 \Delta_{m_0}\!(x)=\int_k \tilde{\Delta}_{m_0}\!(k)\ee^{-\ii kx},\hspace{2ex} \tilde{\Delta}_{m_0}\!(k)=\frac{\ii}{k^2-m_0^2+\ii\epsilon},
 \eeq
 where  one directly identifies Feynman's two-point propagator of the Klein-Gordon field for   $\epsilon\to 0^+$, and we have used the short-hand notation $\int_k=\int_{\phantom{k}}\!\!\! {\rm d}^Dk/(2\pi)^D$.
  Since the normal-ordered term in Eq.~\eqref{eq:int_picture_ev} will have a vanishing contribution when acting on the Klein-Gordon vacuum, one can easily see that the vacuum persistence amplitude in Eq.~\eqref{eq:vpa} yields 
    \beq
    \label{eq:free_Z_app}
  \mathsf{Z}_0[J]=\ee^{-\frac{1}{2}\!\!\bigintssss\!\!{\rm d}^Dx_1\!\!\bigintssss\!\!{\rm d}^Dx_2\hspace{0.2ex}J(x_1)\Delta_{m_0}\!(x_1-x_2)J(x_2)},
  \eeq
 which gives rise to any $n$-point propagator  upon functional differentiation~\eqref{eq:vpa}   in accordance to Wick's theorem, e.g. 
 \beq
 \label{eq:2_point_0}
 G_0^{(2)}\!(x_1,x_2)=-\left.\frac{\delta^2 \mathsf{Z}_0}{\delta J_{x_1}\delta J_{x_2}}\right|_{J=0}\!\!\!=\Delta_{m_0}\!(x_1-x_2)=:\Delta_{m_0,12},
 \eeq
  for the 2-point function, and the 4-point function
  \beq
   \label{eq:4_point_0}
  G_0^{(4)}\!(x_1,\cdots, x_4)\!=\!\Delta_{m_0,12}\Delta_{m_0,34}+\Delta_{m_0,13}\Delta_{m_0,24}+\Delta_{m_0,14}\Delta_{m_0,23}.
  \eeq
From now onwards we  refer to the normalised $\mathsf{Z}_0[J]$, or its interacting version $\mathsf{Z}[J]$, as simply the generating functional.

\section{Generating functional of  interacting $\lambda\phi^4$ fields}
 \label{sec:ren_gen_functional}

  In this appendix, we describe   the connection of the generating functional to the perturbative approach to QFTs based on Feynman diagrams, in which the   $\lambda\phi^4$ model plays a key role~\cite{ABERS19731}. Although this is discussed  in several textbooks, e.g.~\cite{Peskin:1995ev,fradkin_2021,ryder_1996}, this appendix    makes our work self-contained,  and present the results in a way that are useful  to   connect to the effective  Ising models,  making this material more amenable to the quantum-technology community not familiarised with  QFT functional methods.
  
   The time-evolution operator in the presence of classical Schwinger sources~\eqref{eq:V_J},  $U(t_{\rm f},t_0)=U_{  \!{\mathcal{H}}}U_{  \mathcal{V}_J}$, can again be  expressed in terms of  two unitaries
 \beq
 \label{eq:unitaries_interacting}
  U_{ \!\mathcal{H}}=\ee^{-\ii\!\!\bigintssss\!\!{\rm d}^Dx \mathcal{H}},\hspace{2ex}
 U_{ \mathcal{V}_J}=\mathsf{T}\left\{\ee^{+\ii\!\!\bigintssss\!\!{\rm d}^Dx J(x)\phi_H(x)}\right\}.
 \eeq
where  the  field operators $\phi_H(x)=U_{ \!\mathcal{H}}^\dagger \phi(x)U_{ \!\mathcal{H}}$ now evolve in the Heisenberg picture with respect to the full interacting $\lambda\phi^4$ model~\eqref{eq:int_V} in the absence of sources, thus accounting for possible scattering events. In this case, the 'vacuum' persistence amplitude will  lead to the full generating functional
 \beq
 \label{eq:vpa_interction}
 \mathsf{Z}[J]\!=\!\bra{0} U_{ \mathcal{V}_J}\!\ket{0}\!\!, \hspace{1.1ex} G^{(n)}\!(x_1,\cdots, x_n)=\left.\frac{(-\ii)^n\delta^n\mathsf{Z}[J]}{\delta J(x_1)\cdots\delta J(x_n)}\right|_{J=0}\!,
 \eeq
 where $\ket{0}$ is the groundstate of the interacting QFT, and the functional derivatives yield the  full $n$-point  propagators. Note that, depending on the bare parameters of the theory, this groundstate can support a non-zero particle content via spontaneous symmetry breaking~\cite{PhysRevD.7.1888,PhysRevD.9.1686}, although it is still  customary to refer to it as the 'vacuum'. In this article, we shall only be interested in  regimes with unbroken symmetry.
 
We note that the exact expression~\eqref{eq:propagator} for the source contribution to the evolution operator~\eqref{eq:unitaries_interacting} is no longer valid in presence of interactions, and  working with the canonical formalism gets   cumbersome rapidly. Hence, one typically    switches to functional integral methods. As customary, one uses the field/momentum representation  at fixed time $t=x^0$, $\phi(x)\ket{\{\varphi(x^0,\textsf{\textbf{x}})\}}=\varphi(x^0,\textsf{\textbf{x}})\ket{\{\varphi(x^0,\textsf{\textbf{x}})\}}$, $\pi(x)\ket{\{\Pi(x^0,\textsf{\textbf{x}})\}}=\Pi(x^0,\textsf{\textbf{x}})\ket{\{\Pi(x^0,\textsf{\textbf{x}})\}}$, where the eigenvalues are now classical fields, and the eigenvectors form a complete orthonormal basis~\cite{Peskin:1995ev,ryder_1996}. By splitting the time evolution in infinitesimal lapses,  one can introduce the resolution of the identity at  nearby fixed instants of time, alternating  the use of the field and momentum basis. Performing a Gaussian integral over the momentum fields $\Pi(x^0,\textsf{\textbf{x}})$, the generating functional~\eqref{eq:vpa_interction} can be expressed as the functional integral
  \beq
  \label{eq:path_integral_gen_functional}
   \mathsf{Z}[J]=\frac{1}{\mathsf{N}}\!\int\!{\rm D}\varphi\ee^{\ii S},\hspace{1ex} S=\!\int\!{\rm d}^Dx\big(\mathcal{L}_0-\mathcal{V}_{\rm int}(\varphi)-\mathcal{V}_J(\varphi)\big),
  \eeq
  where ${\rm D}\varphi$ is the functional-integral measure. Here,  the action is expressed in terms of the Klein-Gordon Lagrangian 
  \beq
  \label{eq:KG_lag}
 \mathcal{L}_0= \half\partial_\mu\varphi(x)\partial^\mu\varphi(x)-\half m_0^2\varphi^2(x),
  \eeq
 where $\partial_\mu=(\partial_t,\boldsymbol{\nabla})$, $\partial^\mu=(\partial_t,-\boldsymbol{\nabla})$,  and the source and  interaction parts correspond to Eqs.~\eqref{eq:V_J} and~\eqref{eq:int_V}, respectively, expressed in terms of the   basis fields $\phi(x)\to\varphi(x)$. In addition, we have introduced a normalization factor in terms of the unsourced action $\mathsf{N}=\!\int\!{\rm D}\varphi\ee^{\ii S}\big|_{J=0}$ such that $\mathsf{Z}[0]=1$.
  
  A central result in the perturbative approach to QFTs is that the full generating functional can be expressed in terms of the non-interacting one~\eqref{eq:free_Z} as follows
  \beq
  \label{eq:int_gen_functional}
     \mathsf{Z}[J]=\frac{\ee^{-\ii\!\!\bigintssss\!\!{\rm d}^Dx \mathcal{V}_{\rm int}\left(-\ii\delta_{J(x)}\right)}\mathsf{Z}_0[J]\hspace{3ex}}{\left.\ee^{-\ii\!\!\bigintssss\!\!{\rm d}^Dx \mathcal{V}_{\rm int}\left(-\ii\delta_{J(x)}\right)}\mathsf{Z}_0[J]\hspace{0.25ex}\right|_{J=0}},
  \eeq
  where we have introduced a short-hand notation for the functional derivatives $\delta_{J(x)}=\delta/\delta J(x)$. In perturbative treatments,  one typically expands the exponential of the interacting potential to the desired order of the interaction strength  $\lambda_0$, allowing for a graphical representation in terms of Feynman diagrams. In our case, keeping  second-order terms with at most  4 sources, we get the following diagrammatic representation
  \begin{widetext}
  \beq
   \label{eq:Z_interacting} 
 \mathsf{Z}[J]=\bigg( 1\hspace{-8ex}
 \setlength{\unitlength}{1cm}
\thicklines
\begin{picture}(19,0)
\put(1.25,0.0){$+\frac{\ii}{4}$}
\put(1.9,.1){\line(1,0){0.8}}
\put(1.75,0.01){$\color{gray}{\boldsymbol{\times}}$}
\put(2.3,0.3){\circle{0.4}}
\put(2.21,0.01){$\color{myred}\bullet$}
\put(2.84,0.){$+\frac{1}{8}$}
\put(2.55,0.01){$\color{gray}{\boldsymbol{\times}}$}
\put(3.45,0.01){$\color{gray}{\boldsymbol{\times}}$}
\put(3.6,0.1){\line(1,0){0.8}}
\put(4,0.3){\circle{0.4}}
\put(4,0.69){\circle{0.4}}
\put(3.91,0.01){$\color{myred}\bullet$}
\put(3.91,0.41){$\color{myred}{\bullet}$}
\put(4.25,0.01){$\color{gray}{\boldsymbol{\times}}$}
\put(4.65,0.){$+\frac{1}{8}$}
\put(5.3,0.1){\line(1,0){1.1}}
\put(5.17,0.01){$\color{gray}{\boldsymbol{\times}}$}
\put(5.6,0.30){\circle{0.4}}
\put(6.1,0.30){\circle{0.4}}
\put(5.51,0.01){$\color{myred}\bullet$}
\put(6.02,0.01){$\color{myred}\bullet$}
\put(6.25,0.01){$\color{gray}{\boldsymbol{\times}}$}
\put(6.6,0.){$+\frac{1}{12}$}
\put(7.4,0.1){\line(1,0){1}}
\put(7.25,0.01){$\color{gray}{\boldsymbol{\times}}$}
\put(7.9,0.1){\circle{0.55}}
\put(8.08,0.01){$\color{myred}\bullet$}
\put(7.54,0.01){$\color{myred}\bullet$}
    \put(8.25,0.01){$\color{gray}{\boldsymbol{\times}}$}
\put(8.5,0.0){$-\frac{\ii}{4!}$}
\put(9.22,0.1){\line(1,0){0.8}}
\put(9.61,-0.3){\line(0,1){0.8}}
\put(9.52,0.01){$\color{myred}\bullet$}
\put(9.1,0.01){$\color{gray}{\boldsymbol{\times}}$}
\put(9.86,0.01){$\color{gray}{\boldsymbol{\times}}$}
\put(9.465,0.41){$\color{gray}{\boldsymbol{\times}}$}
\put(9.465,-0.41){$\color{gray}{\boldsymbol{\times}}$}
\put(10.2,0.0){$-\frac{1}{12}$}
\put(10.95,0.1){\line(1,0){1.05}}
\put(11.61,-0.3){\line(0,1){0.8}}
\put(11.3,0.29){\circle{0.4}}
\put(11.52,0.01){$\color{myred}\bullet$}
\put(11.2,0.01){$\color{myred}\bullet$}
\put(10.8,0.01){$\color{gray}{\boldsymbol{\times}}$}
\put(11.85,0.01){$\color{gray}{\boldsymbol{\times}}$}
\put(11.465,0.41){$\color{gray}{\boldsymbol{\times}}$}
\put(11.465,-0.41){$\color{gray}{\boldsymbol{\times}}$}
\put(12.15,0.0){$-\frac{1}{32}$}
\put(12.9,.25){\line(1,0){0.8}}
\put(13.3,0.45){\circle{0.4}}
\put(13.21,0.16){$\color{myred}{\bullet}$}
\put(13.55,0.16){$\color{gray}{\boldsymbol{\times}}$}
\put(12.9,-.08){\line(1,0){0.8}}
\put(12.75,0.16){$\color{gray}{\boldsymbol{\times}}$}
\put(12.75,-0.165){$\color{gray}{\boldsymbol{\times}}$}
\put(13.3,-0.26){\circle{0.4}}
\put(13.21,-0.165){$\color{myred}{\bullet}$}
\put(13.55,-0.165){$\color{gray}{\boldsymbol{\times}}$}
\put(14.05,0.0){$-\frac{1}{16}$}
\put(14.9,.35){\line(1,0){0.8}}
\put(14.9,-.24){\line(1,0){0.8}}
\put(14.75,0.26){$\color{gray}{\boldsymbol{\times}}$}
\put(15.3,0.06){\circle{0.6}}
\put(15.21,0.26){$\color{myred}{\bullet}$}
\put(15.55,0.26){$\color{gray}{\boldsymbol{\times}}$}
\put(14.75,-0.33){$\color{gray}{\boldsymbol{\times}}$}
\put(15.21,-0.33){$\color{myred}{\bullet}$}
\put(15.55,-0.33){$\color{gray}{\boldsymbol{\times}}$}
\put(15.9,0.01){$+\cdots\bigg) \mathsf{Z}_0[J],$}
\end{picture}
\eeq
  \end{widetext}
which should be read as follows: crosses ${\color{gray}{\boldsymbol{\times}}}=J(x)$ represent the Schwinger sources acting at different spacetime locations $x$, and blobs ${\color{myred}\boldsymbol{\bullet}}=\lambda_0$ stand for interaction vertices with the bare quartic coupling. Solid lines that join  a cross and a blob should be translated into ${\color{gray}{\boldsymbol{\times}}}\boldsymbol{\!\!\!\!\!-\!\!\!\!-\!\!\!\!{\color{myred}{\bullet}}}=\Delta_{m_0}(x-z)$, and thus involve the free $d$-dimensional Feynman propagator~\eqref{eq:feynman_propagator} from the source at $x$ to  the  vertex at $z$. Likewise, solid lines  connected to the same blob stand for interaction loops that should be translated for  $\bigcirc\hspace{-2.5ex}{\color{myred}\bullet}\hspace{2ex}=\Delta_{m_0}(0)$, while those connecting two distant blobs must be substituted by the propagator between the corresponding points  $\boldsymbol{{\color{myred}\bullet}\!\!\!-\!\!\!\!-\!\!\!\!{\color{myred}\bullet}}=\Delta_{m_0}(z_1-z_2)$. For each of the above diagrams, we should integrate over all possible spacetime locations  of the sources $\int\!{\rm d}^Dx_i$, and those of the intermediate interaction vertices $\int\!{\rm d}^Dz_i$. 

This perturbative expression~\eqref{eq:Z_interacting}  suffices  to understand the main effects of the scalar-field self-interactions. The standard approach calculates how the 2- and 4-point functions $ G^{(2)}\!(x_1,x_2),
  G^{(4)}\!(x_1,\cdots,x_4)$, obtained by substituting Eq.~\eqref{eq:Z_interacting}  in  Eq.~\eqref{eq:vpa_interction}, change with respect to the free ones~\eqref{eq:2_point_0}-\eqref{eq:4_point_0} due to the self-interactions. This leads to a neat discussion of the appearance of divergences, the need to regularise the QFT, and renormalisation~\cite{Peskin:1995ev,fradkin_2021,ryder_1996}. We will  focus instead on rewriting this perturbative series as a renormalised  generating functional, as this  sets the stage for our calculation of the changes in the effective Ising model~\eqref{eq:eff_ising_lattice} when upgrading to  Ising-Schwinger sources, as described in the main text. For the moment, we  focus on the first 4  diagrams of Eq~\eqref{eq:Z_interacting}, which involve a pair of sources. The first 3  diagrams are combinations of the so-called tadpole, and one can see that a renormalised  functional  
\beq
\label{eq:ren_gen_function}
\mathsf{Z}_{ r}[J]=\ee^{-\frac{1}{2}\!\!\bigintssss\!\!{\rm d}^Dx_1\!\!\bigintssss\!\!{\rm d}^Dx_2J(x_1)\Delta_{m_{ r}}\!(x_1-x_2)J(x_2)},
\eeq
generates directly these terms to second order in   $\lambda_0$ if
 the propagator has 
  the following additive  mass renormalisation 
 \beq
 \label{eq:mass_renormalisation_app}
m_0^2\to m_{ r}^2=m_0^2+\mathsf{\Sigma}(0).
 \eeq
 Here, we have introduced   the so-called self-energy $\mathsf{\Sigma}(k)$, which relates the full and free propagators via the Dyson-Schwinger equation~\cite{PhysRev.75.1736,Schwinger452,Schwinger455}, namely
 \beq
 \label{eq:dyson}
 \ii \tilde{G}^{-1}(k)=\ii\tilde{\Delta}^{-1}_{m_0}(k)+\mathsf{\Sigma}(k), \hspace{1ex}\tilde{G}(k)=\frac{\ii}{k^2-m_0^2-\mathsf{\Sigma}(k)},
 \eeq
 such that the  renormalisation of the mass by the zero-momentum self energy~\eqref{eq:mass_renormalisation} becomes apparent.
 
  To show that the different powers of the Taylor series of the renormalised generating functional~\eqref{eq:ren_gen_function}  are  equivalent to the  first 3 diagrams, one must apply  $\int{\rm d}^Dk\ee^{-\ii k(x-y)}\!/\!(k^2-m_0^2+\ii\epsilon)^{2}=-(2\pi)^D\int{\rm d}^Dz \Delta_{m_0}\!(x-z)\Delta_{m_0}\!(z-y)$ repeatedly, 
and consider a mass renormalisation $\mathsf{\Sigma}(0)=\mathsf{\Sigma}_{m_0,\lambda_0}^{(1,{\rm td})}+\mathsf{\Sigma}^{\rm (2,td)}_{m_0,\lambda_0}$ with
 \beq
 \label{eq:bare_mass_renormalization_1_tadpole}
\mathsf{\Sigma}^{(1,{\rm td})}_{m_0,\lambda_0}= \frac{\lambda_0}{2}\!\!\bigintssss_{k_1}\!\!\tilde{\Delta}_{m_{ 0}}\!(k_1),
 \eeq
 for the single-tadpole diagram (i.e. first diagram in~\eqref{eq:Z_interacting}), and
 \beq 
 \label{eq:bare_mass_renormalization_2_tadpole}
\mathsf{\Sigma}^{(2,{\rm td})}_{m_0,\lambda_0}= -\ii\frac{\lambda_0^2}{4}\!\!\bigintssss_{k_1}\!\!\bigintssss_{k_2}\!\!\tilde{\Delta}_{m_{ 0}}^2\!(k_1)\tilde{\Delta}_{m_{ 0}}\!(k_2)
 \eeq
 for the   double-tadpole diagram (i.e second diagram in~\eqref{eq:Z_interacting}). Interestingly, the third diagram is directly generated by the Taylor expansion of the exponential, and does not contribute with an additional renormalisation of the mass. In the language of QFTs, the self-energy only has contributions from one-particle irreducible (1PI)  diagrams, namely those that cannot be split in two disconnected diagrams by cutting a single internal line/propagator. Since the third diagram in Eq.~\eqref{eq:Z_interacting} can indeed be split in a couple of disconnected tadpoles, it does not contribute to the self-energy as just shown.

 Finding the contribution of the fourth diagram in~\eqref{eq:Z_interacting}, the so-called sunrise diagram, to a renormalised generating functional like~\eqref{eq:ren_gen_function} is  slightly more involved. While the tadpole diagrams  involve  virtually-excited bosons with a momentum that is independent to that of the propagating bosons; this is not the case for the virtual bosons in the sunrise diagram. As a consequence,  its contribution to the self-energy depends on the external momentum, and  cannot be  simply recast as a  mass renormalisation~\eqref{eq:mass_renormalisation}. In fact, one finds that
 \beq
 \mathsf{\Sigma}^{(2,{\rm sr})}(k)=-\ii\frac{\lambda_0^2}{6}\!\!\bigintssss_{k_1}\!\!\bigintssss_{k_2}\tilde{\Delta}_{m_{ 0}}\!(k_1)\tilde{\Delta}_{m_{ 0}}\!(k_2)\tilde{\Delta}_{m_{ 0}}\!(k-k_1-k_2),
 \eeq
 which can be expanded in a power series of the external momentum as follows
 \beq
 \label{eq:expansion_sunrise}
  \mathsf{\Sigma}^{(2,{\rm sr})}(k)=\mathsf{\Sigma}^{(2,{\rm sr})}_{m_0,\lambda_0}+k^2\left.\frac{\partial\mathsf{\Sigma}^{(2,{\rm sr})}(k)}{\partial{k^2}}\right|_{k^2=0}+\cdots.
  \eeq
 
 To zero-th order in the external momentum, one  finds that the sunrise diagram  indeed contributes  to the renormalised  mass via $\mathsf{\Sigma}(0)=\mathsf{\Sigma}_{m_0,\lambda_0}^{(1,{\rm td})}+\mathsf{\Sigma}^{\rm (2,td)}_{m_0,\lambda_0}+\mathsf{\Sigma}^{\rm (2,sr)}_{m_0,\lambda_0}$, where
 \beq
 \mathsf{\Sigma}^{(2,{\rm sr})}_{m_0,\lambda_0}=-\ii\frac{\lambda_0^2}{6}\!\!\bigintssss_{k_1}\!\!\bigintssss_{k_2}\tilde{\Delta}_{m_{ 0}}\!(k_1)\tilde{\Delta}_{m_{ 0}}\!(k_2)\tilde{\Delta}_{m_{ 0}}\!(k_1+k_2).
 \eeq

 To  next-order in the external momentum, one  sees through the Dyson-Schwinger equation~\eqref{eq:dyson} that the sunrise diagram   changes the  propagator in  momentum space~\eqref{eq:feynman_propagator}  into
 \beq
 \label{eq:wv_renormalisation}
\tilde{\Delta}_{m_0}\!(k)\to\tilde{G}\!(k)=\frac{\ii}{z_{m_0,\lambda_0}^{\!-1}k^2-m_r^2+\ii\epsilon},
\eeq
where we have introduced the following parameter $z_{m_0,\lambda_0}^{\!-1}=1-\partial_{k^2}\mathsf{\Sigma}(k)|_{k^2=0}$ which, for the sunrise diagram reads
\beq
\label{eq:wavefunction_renormalisation}
z_{m_0,\lambda_0}^{\!-1}=1-\frac{\lambda_0^2}{6}\!\!\bigintssss_{k_1}\!\!\bigintssss_{k_2}\!\!\tilde{\Delta}_{m_{ 0}}\!(k_1)\tilde{\Delta}_{m_{ 0}}\!(k_2)\tilde{\Delta}^2_{m_{ 0}}\!(k_1+k_2).
\eeq
 This effect can be alternatively understood as a renormalisation of the derivative terms $\partial_\mu\varphi\partial^\mu\varphi\to z_{m_0,\lambda_0}^{-1}\partial_\mu\varphi\partial^\mu\varphi$ in the action~\eqref{eq:KG_lag}. This leads to the
 so-called wavefunction renormalisation~\cite{Peskin:1995ev,fradkin_2021}, where the fields must be rescaled $\varphi(x)\to\varphi(x)/\sqrt{z_{m_0,\lambda_0}}$ in order to recover the original form of the Klein-Gordon Lagrangian. As a consequence of this rescaling, one gets a multiplicative renormalisation of both the bare mass $m_0^2\to {m}^2_{r}$, and the Schwinger sources $  J(x)\to{J}_{ r}(x)$
 \beq
 \label{eq:source_ren_app}
 {m}_{ r}^2=\left(m_0^2+\mathsf{\Sigma}^{\rm (1,{\rm td})}_{m_0,\lambda_0}+\mathsf{\Sigma}^{\rm (2,sr)}_{m_0,\lambda_0}\right)z_{m_0,\lambda_0},\hspace{1ex} {J}_{ r}(x)=J(x)\sqrt{z_{m_0,\lambda_0}}.
 \eeq

It turns out that higher-order contributions in the external momentum to Eq.~\eqref{eq:expansion_sunrise} are irrelevant in the renormalisation-group sense disucssed in the main text~\cite{WILSON197475,hollowood_2013, RevModPhys.66.129}, and can thus be neglected at long wavelengths. Taking into account all these different renormalisations, the full generating functional of the  interacting QFT to second order is
 \beq
\label{eq:ren_gen_function_complete}
\mathsf{Z}_{ r}[{J}_r]=\ee^{-\frac{1}{2}\!\!\bigintssss\!\!{\rm d}^Dx_1\!\!\bigintssss\!\!{\rm d}^Dx_2{J}_r(x_1)\Delta_{{m}_{ r}}\!(x_1-x_2){J}_r(x_2)}.
\eeq
 
 Once all of the 2-source diagrams in Eq.~\eqref{eq:Z_interacting}  have been carefully accounted for in the renormalised generating functional~\eqref{eq:ren_gen_function_complete}, we shall focus on the remaining 4-source processes. We start with the tree-level vertex corresponding to the 5-th Feynman diagram, which can be recast as
   \beq
   \label{eq:4-source_correction}
\ee^{-\frac{\ii}{4!}\!\!\bigintssss\!\!{\rm d}^Dx_1\!\!\!\bigintssss\!\!{\rm d}^Dx_2\!\!\!\bigintssss\!\!{\rm d}^Dx_3\!\!\!\bigintssss\!\!\!{\rm d}^Dx_4{J}(x_1){J}(x_2){G}^{(4,{\rm c})}_{{m}_0,\lambda_0}\!\!(x_1,x_2,x_3,x_4){J}(x_3){J}(x_4)}\!\!.
\eeq
Here, we have introduced the connected 4-point propagator, which is consistent with the fact that only 1PI diagrams should be incorporated in the renormalised generating functional, as the reducible ones will be automatically generated by the power expansion of the exponential.  This connected  4-point function is obtained from Eq.~\eqref{eq:vpa_interction} via
 ${G}^{(4,{\rm c})}_{{m}_0,\lambda_0}(x_1,x_2,x_3,x_4)=G^{(4)}_{{m}_0,\lambda_0}(x_1\cdots x_4)-\Delta_{{m}_0,12}\Delta_{{m}_0,34}-\Delta_{{m}_0,13}\Delta_{{m}_0,24}-\Delta_{{m}_0,14}\Delta_{{m}_0,23}$. To this lowest-order in the interaction strength, we find
 \beq
 {G}^{(4,{\rm c})}_{{m}_0,\lambda_0}\!\!=\lambda_0\!\!\!\bigintssss\!\!\!{\rm d}^Dz\Delta_{{m}_0}\!(x_1-z)\Delta_{{m}_0}\!(x_2-z)\Delta_{{m}_0}\!(z-x_3)\Delta_{{m}_0}\!(z-x_4).
 \eeq
 The 6-th diagram in Eq.~\eqref{eq:Z_interacting} can be seen as a tadpole decoration of the previous one which, together with higher-order terms involving more tadpoles, can be accounted for by considering  the connected propagator with a renormalised mass. Note that the 7-th diagram, and higher-order disconnected diagrams of the like, are already accounted for by the expansion of the functional~\eqref{eq:ren_gen_function_complete} to the corresponding order. Had we considered $\mathcal{O}(\lambda_0^3)$ corrections, we would have also obtained a disconnected 4-source diagram decorated with a sunrise graph, which  is again  directly accounted for if we use the   the renormalised   functional in Eq.~\eqref{eq:ren_gen_function_complete}. Therefore, all of these decorations are accounted for by substituting  $ {G}^{(4,{\rm c})}_{{m}_0,\lambda_0}\to  {G}^{(4,{\rm c})}_{\tilde{m}_r,\lambda_0}$ in Eq.~\eqref{eq:4-source_correction} with the additive and multiplicative renormalisations of the mass and sources. 
 
 Once again, we have left the discussion about a diagram  leading to new effects, the 8-th diagram of Eq.~\eqref{eq:Z_interacting}, for the last part of this appendix. This term describes how the 4-point interaction can be mediated by a pair of  virtual bosons exchanged between a pair of propagating particles. Therefore, it will lead to a renormalisation of the coupling strength $\lambda_0\to{\lambda}_r$ 
 \beq
 \label{eq:int_strength_ren}
 {\lambda}_r=\lambda_0+\Gamma_{m_0,\lambda_0}^{(4)},\hspace{1ex}\Gamma_{m_0,\lambda_0}^{(4)}=-\ii\frac{3\lambda_0^2}{2}\!\!\bigintssss_{k_1}\!\!\tilde{\Delta}^2_{m_{ 0}}\!(k_1).
 \eeq
which should also be complemented with the wavefunction renormalisation
   \beq
 \tilde{\lambda}_r=\left(\lambda_0+\Gamma_{m_0,\lambda_0}^{(4)}\right)z^2_{m_0,\lambda_0}.
 \eeq
The combination of all these renormalisations must be included in the following renormalised generating functional
  \begin{widetext}  
      \beq
\label{eq:ren_gen_function_quartic}
\mathsf{Z}_{ r}[{J}_r]=\ee^{-\frac{1}{2}\!\!\bigintssss\!\!{\rm d}^Dx_1\!\!\bigintssss\!\!{\rm d}^Dx_2{J}_r(x_1)\Delta_{{m}_{ r}}\!(x_1-x_2){J}_r(x_2)-\frac{\ii}{4!}\!\!\bigintssss\!\!{\rm d}^Dx_1\!\!\bigintssss\!\!{\rm d}^Dx_2\!\!\bigintssss\!\!{\rm d}^Dx_3\!\!\bigintssss\!\!{\rm d}^Dx_4{J}_r(x_1){J}_r(x_2){G}^{(4,{\rm c})}_{{m}_r,{\lambda}_r}\!\!(x_1,x_2,x_3,x_4){J}_r(x_3){J}_r(x_4)}.
\eeq
\end{widetext}
We close this Appendix by noting that higher-order terms in the expansion will lead to $2n$-source contributions with $n=3,4,\cdots$ coupled to the corresponding renormalised connected $2n$-point propagator. In the limit of impulsive sources, one would obtain the $2n$-point propagator, including also the disconnected pieces, by evaluating the corresponding functional derivatives~\eqref{eq:vpa_interction} on  the renormalised  functional~\eqref{eq:ren_gen_function_quartic}. 

\section{Elastic crystals: compressibility, massless Klein-Gordon fields, and effective Ising models.}
\label{sec:elastodynamics_solids}

Following the multidisciplinary approach of this work, we include in this Appendix  a basic description of the long-wavelength theory of vibrations of crystalline solids, a topic that appears in several textbooks~\cite{goodstein_2017,ashcroft_2016}. This  serves to give a self-contained presentation, but also to present  key concepts in connection   to our previous  QFT approach, trying to make the   discussions of the main text more amenable to potential readers from the high-energy-physics community.  

We start by reviewing Debye's model for the specific heat of solids~\cite{debye_model}, which can be understood as the quantisation of a coarse-grained elasto-dynamical model~\cite{landau_elasticity,thorne_blandford_2017} that captures the   dynamics of the crystal at long wavelengths. A crystalline solid subjected to external forces can be modelled by a  displacement field $\boldsymbol{u}(t,\textsf{\textbf{x}})$ at a given instant  of time $t$ for each point of the crystal $\textsf{\textbf{x}}$. This accounts for a deformation of the solid  when its components support a non-zero gradient $\boldsymbol{\nabla}{u}_\alpha(t,\textsf{\textbf{x}})$, i.e. expansion/compression or shear strain. In the theory of elasticity, the response  to such strain  comes in the form of  stress forces  acting against the deformation, which have their microscopic origin in the  inter-atomic short-range interactions that try to restore  equilibrium  in the solid. For such a coarse-grained description, these forces are    local, and can be described in terms of pressure and shear stress. 

 To simplify the description, we shall consider the academic problem  of a  $d=1$-dimensional crystal. In this case, strain can only appear in the form of compression/expansion, and stress in the form of pressure. For elastic materials, Hook's law states that strain and stress are proportional. In particular, for this simple 1D situation, the  compression/expansion is modelled by the relative change of the length of the crystal $\delta L/L$ which, according to Hook's law, must  be proportional to the pressure $P$. This leads to $\delta L/L=-P/K_e$, where $K_e$ is the   
elastic or bulk modulus. This parameter can be  related to the inverse of  the  thermal compressibility $\beta_T=-\frac{1}{L}\frac{\partial L}{\partial P}\big|_T=K_e^{-1}$, and  thus quantifies  the  stiffness of the crystal to external forces, i.e. how it resists bulk changes in its length. 

For  homogeneous elastic solids, the dynamics of the coarse-grained field is described by a wave equation, the so-called Cauchy-Navier equation.  In this simple 1D case, setting the crystal along the $\mathsf{x}$-axis, this equation describes   vibrations of pressure in the form of a compressional wave evolving under 
\beq
\label{eq:long_waves}
 \left(\partial_t^2-c_{\ell}^2\partial_{\textsf{{x}}}^2\right){u}_{\mathsf{x}}(t,\textsf{{x}})=0,\hspace{2ex} c_\ell=\sqrt{\frac{K_e}{\rho}}.
\eeq 
Here, the sound speed $c_\ell$ is defined in terms of the elastic modulus and the  mass density of the crystal $\rho$, which must be homogeneous at this coarse-grained scale.

 At this point, one can rescale the displacement field to obtain a $D=(1+1)$-dimensional scalar field with the correct scaling natural dimension discussed below Eq.~\eqref{eq:spin_spin_couplings} in natural units $c_\ell=1=\hbar$. 
 Denoting by $a$   some characteristic microscopic length-scale of the crystal, one  finds that   the compressional wave~\eqref{eq:long_waves} is indeed governed by the massless Klein-Gordon Lagrangian~\eqref{eq:KG_lag} with the following identifications
\beq
\label{eq:debye_field}
\phi(x)\to \frac{1}{a}u_{\mathsf{x}}(t,\textsf{{x}}), \hspace{1.25ex}c\to c_\ell,\hspace{1.25ex}m_0^2\to 0.
\eeq
  Hence, if one  considers that the  speed of sound plays the role of the light speed, there is an effective Lorentz invariance  emerging at long wavelengths in this coarse-grained one-dimensional crystal. Although this symmetry is not strictly exact due to  the neglected microscopic short-wavelength details, or possible deviations from Hook's law, one expects it to dominate the low-energy behaviour of the  crystal, a deep and fundamental result that can be formalised through the use of the renormalisation group~\cite{WILSON197475,RevModPhys.66.129}. In fact, this emergence of symmetries at the coarse-grained level underlies contemporary  non-perturbative approaches to strongly-coupled QFTs, such as lattice quantum chromodynamics~\cite{PhysRevD.10.2445,gattringer_lang_2010}.

 In Debye's approach, one  proceeds by quantising this field by the introduction of operators~\eqref{eq:fields_quantisation} that create-annihilate quantised collective vibrations~\eqref{eq:long_waves}, the so-called phonons
 \beq
 \label{eq:phonons_crystal}
 H_p=\int{\rm }\!\!\frac{d{\textsf{{k}}}}{2\pi}\omega^{\phantom{\dagger}}_{\textsf{{k}}}a^\dagger_{\textsf{k}}a^{\phantom{\dagger}}_{\textsf{k}},\hspace{1.5ex} \omega^{\phantom{\dagger}}_{\textsf{{k}}}=|\textsf{k}|,
 \eeq
 and then calculates the partition function to extract any thermodynamic quantity, such as the specific heat~\cite{goodstein_2017}. Note that our 1D crystal would be unstable with respect to thermal fluctuations~\cite{PhysRev.158.383,PhysRevLett.17.1133,Coleman1973}, and this is why we categorised it as an academic problem. In more realistic situations, the crystal would likely consist of  weakly-coupled chains, and one  would  need to consider   also deformations with shear strain which, according to Hook's law, are proportional to the shear stress via a new coefficient: the rigidity or shear modulus $\mu_r$~\cite{landau_elasticity,thorne_blandford_2017}. These additional deformations change the longitudinal speed of sound and, moreover,  lead to additional transverse sound waves. For instance, for the $\mathsf{z}$-axis transverse to the weakly-coupled chains, the  wave equation is
\beq
\label{eq:transv_waves}
 \left(\partial_t^2-c_{\mathsf{t}}^2\partial_{\textsf{{x}}}^2\right){u}_{\mathsf{z}}(t,\textsf{{x}})=0,\hspace{2ex} c_{\mathsf{t}}=\sqrt{\frac{\mu_r}{\rho}}.
\eeq
This forbids the recovery of  Lorentz invariance even at long wavelengths, as massless fundamental particles should propagate within a unique light cone, but the longitudinal and transverse  waves propagate with different speeds of sound. In fact, both longitudinal and transverse waves play a key role in the usefulness of Debye's model to reproduce correctly the specific heat of a thermal solid. As  discussed in the main text, sound waves in laser-cooled crystals of trapped ions can be selectively excited, such that either the longitudinal or the transverse branch contributes to the dynamics, and there is again a Lorentz invariance emerging at long wavelengths.

It is interesting to note that, despite the absence of interactions and UV loop divergences, Debye's theory is  a long-wavelength description that requires a frequency cutoff to recover the correct temperature dependence of the specific heat~\cite{goodstein_2017}. This is the so-called Debye  frequency $\omega_{\rm D}$, which allows one to reduce the number of degrees of freedom from  the infinity of the coarse-grained field theory~\eqref{eq:long_waves} into  the large, yet finite, number  that characterises the elastic crystal. In $d=1$, this hard cutoff  would correspond to
\beq
\label{eq:Debye_cutoff}
\omega_{\textsf{k}}\leq \omega_{\rm D}=\pi\frac{ c_\ell N}{L},\hspace{2ex} |{\textsf{k}}|\leq\Lambda_{\rm c}=\pi \frac{N}{L},
\eeq
where $N$ is the number of  atoms in the solid. This equation becomes a  generic UV cutoff~\eqref{eq:UV_cutoff} using natural units $c_\ell=1$.

To carry this discussion further, let us now address the difficulties in including the Ising-Schwinger couplings~\eqref{eq:Ising_Z_source_coupling} in a solid-state crystal.  One may consider a collection of $n$  impurities  at positions $\textsf{\textbf{x}}_i={\textsf{x}}_i{\bf e}_{\textsf{x}}$, each of which can host  electrons  in a single orbital of energy $\omega_0$. These impurities are sufficiently far apart  that the electrons cannot tunnel between them, and thus remain localised. In the regime where all the electrons are spin polarised, e.g. using an external  magnetic field, they can be described by spinless fermion operators $c_i^{\phantom{\dagger}},c_i^\dagger$ with the following simple Hamiltonian
\beq
H_{e}=\sum_i\omega_0c_i^\dagger c_i^{\phantom{\dagger}}=\int{\rm d}{\textsf{x}}\delta\epsilon(\textsf{x})Q(t,\textsf{x}),
\eeq
where we have used the energy densities in Eq.~\eqref{eq:energies_lattice_spins},  and the Jordan-Wigner transformation $Z(t,\textsf{x}_i)=2c_i^\dagger \!c_i^{\phantom{\dagger}}-1$~\cite{Jordan1928}, such that the Ising projector is  $Q(t,{\textsf{x}})=(1+Z(t,{\textsf{x}}))/2$. In the context of electron-phonon coupling,  a so-called Holstein coupling~\cite{HOLSTEIN1959325} of  strength $g$, would then lead to
\beq
\label{eq:Holstein_coupling}
H_{e{\rm -}p}=\sum_i gc_i^\dagger c_i^{\phantom{\dagger}}u_{\textsf{x}}(t,{\textsf{x}}_i)=\int{\rm d}{\textsf{x}}\hspace{0.2ex}J(t,{\textsf{x}})Z(t,{\textsf{x}})\phi(t,{\textsf{x}}),
\eeq
where $J(t,{\rm x})$ is defined according to Eq.~\eqref{eq:lattice_sources} with $J_0=-2g$. In order to get the harmonic time-dependence of the sources, one would need to introduce  externally-controlled periodic modulations of the electron-phonon coupling, which does not sound  very realistic. In the main text, we  explore more realistic  alternatives based on trapped ions.

Let us close this Appendix by noting how Debye's cutoff enters in the discussion. In our high-energy physics model, we emphasised the importance of using a harmonic source with a frequency below the bare mass of the scalar field, avoiding in this way   possible  resonances and dissipative processes (see the discussion below Eq.~\eqref{coupling_density}). Clearly, in the absence of  a UV cutoff, the dispersion relation~\eqref{eq:dispersion} extends to arbitrarily-large energies, and the only possibility was to set $\omega_J^2\lesssim m_0^2$. However, in the presence of a physical cutoff, one can also set $\omega_J^2\gtrsim\Lambda_{\rm c}$ to avoid such resonances. For the massless scalar field associated to the longitudinal phonons~\eqref{eq:debye_field}, we would get effective Ising interactions with the exponentially-decaying couplings in Eq.~\eqref{eq:yukawa_coupling_1d}, controlled by  $m_{\rm eff}^2=\omega_{J}^2$. In this case, the spin-spin couplings read
 \beq
 \mathsf{J}_{ij}\propto{\ee^{-\omega_J|t_{\textsf{x}_i-\textsf{x}_j}|}}\cos\big(\textsf{k}_J(\textsf{x}_i-\textsf{x}_j)\big).
 \eeq
where $t_{{\textsf{x}}_i-{\textsf{x}}_j}=|\textsf{x}_i-{\textsf{x}}_j|/c_\ell$ is the time  the  sound wave takes to propagate  between the two corresponding spins.

Let us note that, in addition to  the limitations  already discussed, including the transverse field~\eqref{eq:eff_ising_lattice}, which makes the Ising model an archetype for the study of quantum phase transitions~\cite{sachdev_2011}, would also be unrealistic, as  the Jordan-Wigner transformation would  require highly non-local electronic terms. Last, but not least, the simplified sensing protocol discussed in Sec.~\ref{sec:KG_Ising_spins} would require to perform  local measurements of the electronic populations and, more critically, to initialise the  effective spins in a coherent superposition of the impurity being vacant or occupied by a single electron, which is also not very realistic. One could avoid this complication by considering  spinful fermions instead, but then the the problem would translate  into finding materials where the Holstein coupling~\eqref{eq:Holstein_coupling} is spin-dependent. In the main text, we  show that all these drawbacks can be overcome by moving from solid-state physics to trapped-ion quantum technologies.

\bibliographystyle{apsrev4-1}
\bibliography{bibliography}

\end{document}